\documentclass[a4paper,11pt]{article}
\pdfoutput=1 % if your are submitting a pdflatex (i.e. if you have
\usepackage{subfig}
\usepackage{float}
\usepackage{placeins}

\usepackage{jheppub} % for details on the use of the package, please
\setcitestyle{square}
\setcitestyle{citesep={,}}
\usepackage[usenames, dvipsnames]{xcolor}

\usepackage[]{todonotes}
\usepackage{verbatim}
\usepackage[utf8]{inputenc}
\usepackage{ulem}
\usepackage{mathrsfs}
\usepackage{cleveref}
\usepackage{amsmath}
\usepackage[shortlabels]{enumitem}
\usepackage{pgf}
\usepackage{tikz-cd}
\usetikzlibrary{shapes,arrows}
\usetikzlibrary{calc}
\usepackage{pgfplots}
\pgfplotsset{compat=1.16}
\usepackage{tikz-3dplot}
\usepackage{tabu}
\usepackage{fancyhdr}

\fancypagestyle{plain}{%
   \fancyhf{}
   \fancyfoot[C]{\iffloatpage{}{\thepage}}
   }
\usepackage{tikz}
\usepackage{lipsum}
\usetikzlibrary{matrix,arrows,calc}
\usepackage{bbm} 					%%% Blackboard Bold math symbols
\usepackage{slashed} 				%%% Feynman slash
\usepackage{graphicx}				%%% Graphics
\usepackage[font=footnotesize,labelfont=bf]{caption}
\usepackage{psfrag}				%%% Editing EPS files in TeX
\usepackage{tensor}				%%% The \indices command
\usepackage{fouridx}				%%% Arbitrary subscripts/superscripts
\usepackage{bm}					%%% Bold Math symbols
\usepackage{mdframed}				%%% Box for \aside
\usepackage{multirow}				%%% More flexible tables
\usepackage{soul}					%%% Strike-through text
\usepackage{bbold}				%%% Blackboard Bold
\usepackage{multicol}				%%% Pages with multiple columns
\usepackage{tikz-cd}
\usepackage{rotating}

\usetikzlibrary{arrows}

\tikzset{
commutative diagrams/.cd,
arrow style=tikz,
diagrams={>=latex}}

\newcolumntype{T}{>{\small}l}

\usepackage{amssymb}
\usepackage{amsfonts}
\usepackage{mathtools}

\usepackage{feynmf}
\usepackage{marvosym}

\usepackage{import}

\newcommand{\ie}{\textit{i.e.\ }}
\newcommand{\eg}{\textit{e.g.\ }}

\title{
Superradiance in String Theory}

\author[a]{Viraf M. Mehta,}
\author[b]{Mehmet Demirtas,}
\author[c]{Cody Long,}
\author[a]{David J. E. Marsh,}
\author[b]{\\Liam McAllister,}
\author[d]{and Matthew J. Stott}

\vspace{1cm}
\affiliation[a]{Institut f\"{u}r Astrophysik, Georg-August Universit\"{a}t, Friedrich-Hund-Platz 1, G\"{o}ttingen, Germany}
\affiliation[b]{Department of Physics, Cornell University, Ithaca, NY 14853, USA}
\affiliation[c]{Department of Physics and CMSA, Harvard University, Cambridge, MA 02138, USA}
\affiliation[d]{Department of Physics, Kings College London, Strand, London, WC2R, 2LS, United Kingdom}
\emailAdd{viraf.mehta@uni-goettingen.de}
\emailAdd{md775@cornell.edu}
\emailAdd{clong@g.harvard.edu}
\emailAdd{david.marsh@uni-goettingen.de}
\emailAdd{mcallister@cornell.edu}
\emailAdd{matt.stott.91@gmail.com}

\abstract{We perform an extensive analysis of the statistics of axion masses and interactions in compactifications of type IIB string theory, and we show that black hole superradiance excludes some regions of Calabi-Yau moduli space.  Regardless of the cosmological model, a theory with an axion whose mass falls in a superradiant band can be probed by the measured properties of astrophysical black holes, unless the axion self-interaction is large enough to disrupt formation of a condensate.  We study a large ensemble of compactifications on Calabi-Yau hypersurfaces, with $1 \le h^{1,1} \le 491$ closed string axions, and determine whether the superradiance conditions on the masses and self-interactions are fulfilled. The axion mass spectrum is largely determined by the K\"ahler parameters, for mild assumptions about the contributing instantons, and takes a nearly-universal form when $h^{1,1} \gg 1$.  When the K\"ahler moduli are taken at the tip of the stretched K\"ahler cone, the fraction of geometries excluded initially grows with $h^{1,1}$, to a maximum of $\approx 0.5$ at $h^{1,1} \approx 160$, and then falls for larger $h^{1,1}$.  Further inside the K\"ahler cone, the superradiance constraints are far weaker, but for $h^{1,1} \gg 100$ the decay constants are so small that these geometries may be in tension with astrophysical bounds, depending on the realization of the Standard Model.}

\begin{document}
\maketitle
\flushbottom
\normalem

\section{Introduction}\label{sec:intro}

String theory does not predict a unique low-energy effective theory in four-dimensional spacetime. Instead, compactifications of the ten-dimensional superstring theories on six-manifolds
generate a landscape of possible theories.  The ten-dimensional theories contain $p$-form fields, and reducing these on $p$-cycles of the internal space leads to a potentially large number of pseudoscalar axion-like fields\footnote{Throughout this work, we will refer to such fields as `axions'.} in the effective theory, a so-called \emph{axiverse}~\cite{Arvanitaki:2009fg}.
The properties of the axions are linked to the geometry of the compactification~\cite{Witten:1984dg,Svrcek:2006yi,Conlon:2006tq,Arvanitaki:2009fg,Marsh:2015xka,Demirtas:2018akl}, which suggests that observational constraints on axions could lead to constraints on the landscape.

However, despite the
impressive array of observational and experimental data constraining axion theories~\cite{Kim:1986ax,Marsh:2015xka,Bauer:2017ris}, it has proved difficult to translate such constraints into limits on string compactifications per se.  The challenge is that most constraints are highly model-dependent: they involve not just the Lagrangian for the axion itself, but also the couplings of the axion to the visible and hidden sectors, the history of our universe, or both.  It is therefore difficult to infer constraints on axions alone --- one can instead exclude axion theories paired with specific models of particle physics and cosmology. Furthermore, explicit constructions of the axion potential have until recently been limited to geometries with a small number of axions.

Terrestrial experimental constraints clearly rely on couplings between axions and the visible sector, while cosmology provides the alternative of constraining axion theories through gravitational couplings.
The most direct cosmological constraint is via the relic density. All axions can be produced by the vacuum realigment mechanism of coherent field evolution in the potential~\cite{1983PhLB..120..133A,1983PhLB..120..137D,1983PhLB..120..127P}. Heavy axions have the possibility to overclose the Universe and suffer from a form of the cosmological moduli problem~\cite{1983PhLB..131...59C}. Light axions may be stable, and could contribute to the dark matter density. The density of such stable ultralight axions is strongly constrained by cosmological structure formation~\cite{Hlozek:2014lca,Hlozek:2017zzf,Grin:2019mub}. However, computing the relic density relies on a number of assumptions about the initial state of the axions, and about the thermal history of the Universe.

Axions with suitable potentials can play an even more central role in the expansion history by
driving a phase of accelerated expansion, powering inflation in the very early universe~\cite{Freese:1990rb} (see \cite{Baumann:2014nda} for a review) or contributing to the present-day dark energy~\cite{Hlozek:2014lca,Cicoli:2018kdo,2006IJMPD..15.1753C}.
However, accelerating solutions are merely a possibility in such theories, not an inevitable outcome, and the actual expansion depends on the initial conditions.
In any case, even if one could prove that axion inflation and axion quintessence were impossible in a given effective theory, this would not rule out the theory as a description of our Universe, because there are also successful non-axion models of these phenomena.

In this work we will obtain limits on string compactifications that are \emph{independent} of the cosmological model, and are almost independent of the axion couplings to the visible sector, using \emph{black hole superradiance} (BHSR)~\cite{Penrose:1971uk,Press:1972zz,Arvanitaki:2009fg,Arvanitaki:2010sy,Brito:2014wla,Brito:2015oca}.
A brief summary of the BHSR mechanism, which we will review in detail in \S\ref{sec:BHSR}, is as follows.
A spinning astrophysical black hole (BH) grows an axion cloud from vacuum fluctuations via purely gravitational interactions.  The growth of the cloud extracts spin from the BH.  If the axion Compton wavelength is of order the size of the BH ergoregion then this process can be efficient enough to reduce the BH spin by an observable amount.  However, superradiance is shut off if the axion self-interactions
are large. The observation of spinning astrophysical BHs thus leads to limits on the allowed axion masses and self-interactions.

We will use BHSR limits on axion theories to obtain constraints on an ensemble of approximately 200,000 compactifications of type IIB string theory on orientifolds of Calabi-Yau threefold (CY$_3$)~\cite{Candelas:1985en} hypersurfaces in toric varieties.
We will find that part of the parameter space is indeed excluded by observations of astrophysical black holes.\footnote{As we will explain in \S\ref{sec:KS}, there is some fine print relating to modeling Euclidean D3-brane contributions to the axion potential, but we will present a series of tests that support the robustness of our findings.}

In the compactifications considered here, the number of axions resulting from the Ramond-Ramond
four-form $C_4$ is the Hodge number $h^{1,1}_+$ of the orientifold \cite{1987cup..bookR....G,2007stmt.book.....B}, which we write as $h^{1,1}$ for notational simplicity.
Upon including the scalar potential generated by instantons, the effective Lagrangian for the axion fields $\theta^i$, $i=1,\ldots, h^{1,1}$ takes the form
\begin{equation}\label{eq:full_lagrangian}
\mathcal{L}= - \frac{1}{8 \pi^2} M_{\mathrm{pl}}^2K_{ij}g^{\mu\nu}\partial_\mu\theta^i\partial_\nu\theta^j + \sum_{a=1}^\infty \Lambda_a^4 \left\{1 -  \cos \Biggl(\sum_i\mathcal{Q}^a_{i}\theta^i+\delta^a\Biggr)\right\}\, ,
\end{equation}
where $M_{\mathrm{pl}}$ is the reduced Planck mass, $K_{ij}$ is the K\"{a}hler metric, $g^{\mu\nu}$ is the inverse of the spacetime metric,
$\Lambda_a$ is a mass scale associated to
the $a$th instanton, $\mathcal{Q}^{a}_{i} \in \mathbb{Z}$ is the charge of the $a$th instanton under the $i$th axion shift symmetry, and $\delta^a$ is a $CP$ phase.

The data of $K_{ij}$, combined with $\{\Lambda_a, \mathcal{Q}^a_{i}, \delta^a\}$ for the finite set of instantons for which $\Lambda_a$ is not negligibly small,\footnote{Due to limitations of numerical precision we omit scales $\Lambda_a \lesssim 10^{-53}~\mathrm{eV}$, which in any case are so light as to be irrelevant for BHSR: see
\S\ref{sec:analysis}.} fully specifies the axion theory for the purpose of computing BHSR constraints.  The task is then to determine these parameters in an ensemble of compactifications.

Much prior work has made use of simple models for the distributions of axion parameters in string-inspired models.
The simplest model for the mass spectrum is log-flat, i.e.~uniform on a log scale. This relies on the observation that
the mass scale $\Lambda_a$ is related to the action $\mathcal{S}_a$
of the $a$th instanton as $\Lambda_a^4 \propto e^{-\lambda \mathcal{S}}$, where the order-one constant $\lambda$ is model-dependent.
A reasonably uniform distribution of actions $\mathcal{S}_a$ then leads to a log-flat mass spectrum~\cite{Arvanitaki:2009fg}. More involved models for the mass spectrum and for the K\"ahler metric eigenvalues can be motivated using random matrix theory~\cite{Easther:2005zr,Marsh:2011aa,Long:2014fba,Stott:2017hvl}.

In the present work we do not rely on this sort of modeling.  Instead, following \cite{Demirtas:2018akl}, we directly compute $K_{ij}$ and (up to an order-one prefactor) the $\Lambda_a$ for specific CY$_3$'s constructed from triangulations of four-dimensional reflexive polytopes~\cite{Kreuzer:2000xy}.  For the charge matrix $\mathcal{Q}^{a}_{i}$ we adopt a conservative model of the contributing instanton terms in the superpotential, and demonstrate that our results depend only weakly on this model. The basics of our construction are reveiwed in Section~\ref{sec:KS}. We then derive the masses directly from the Hessian eigenvalues of the resulting supergravity potential evaluated at its minimum.\footnote{Computing the phases $\delta^a$ is quite subtle --- see e.g.~\cite{Witten:1999eg,Stout:2020uaf}.  In Appendix \ref{app:tests} we consider two models, one with $\delta^a=0$ and the other with $\delta^a$ uniformly distributed in $[0,2\pi)$, and show that the results are indistinguishable.}

We are now in possession of our nut (the type IIB axion landscape), and our hammer (BHSR) with which to crack it. The process of cracking involves a great deal of numerical computation: minimizing axion potentials, and computing eigenvalues of large matrices and tensors involving a vast hierarchy of scales. This process, and the resulting statistics in the type IIB landscape, is presented in Section~\ref{sec:analysis}, and represents the main results of this paper. By computing the appropriate overlap between the axion parameter distributions and the available data on astrophysical BHs~\cite{Stott:2018opm,Stott:2020gjj}, we are able to exclude certain CY$_3$'s in certain regions in moduli space. A short summary of our main results has been presented already in \cite{Mehta:2020kwu}.

\section{The Kreuzer-Skarke Axiverse}\label{sec:KS}

In this section, following \cite{Demirtas:2018akl}, we describe an ensemble of axion effective theories
arising in compactifications of type IIB string theory on orientifolds of Calabi-Yau hypersurfaces in toric varieties.
These are the theories that we will constrain using BHSR in the remainder of this work.

\subsection{Calabi-Yau orientifolds}\label{sec:kinetic}

The Kreuzer-Skarke list \cite{Kreuzer:2000xy} of the 473,800,776 four-dimensional reflexive polytopes provides a starting point for generating an astronomically large set of solutions of string theory.  A fine, regular, star triangulation (FRST) of a polytope from the list defines a toric fourfold $V$ that contains a Calabi-Yau threefold (CY$_3$) hypersurface, $X$.  Computing the topology and the moduli space metric of $X$ is a combinatorial problem whose complexity grows rapidly with the number of K\"ahler moduli of $X$, corresponding to the Hodge number $h^{1,1}$.  Recent progress \cite{Demirtas:2020dbm,cytools} allows for very efficient computation for any favorable\footnote{See \eg\cite{Demirtas:2018akl} for the definition of a favorable polytope.} polytope in the Kreuzer-Skarke list, including at the maximal value $h^{1,1}=491$.

Compactifying type IIB string theory on such a CY$_3$ hypersurface $X$ yields a four-dimensional theory with $\mathcal{N}=2$ supersymmetry.  In order to allow for realistic models of particle physics and cosmology, we would like, instead, to preserve $\mathcal{N} \le 1$ supersymmetry.  To this end we consider compactification of type IIB string theory on an O3/O7 orientifold, $X/\mathcal{O}$, with $\mathcal{O}$ an orientifold involution, arriving at a theory with
$\mathcal{N}=1$ supersymmetry in four dimensions.  Systematic enumeration of all such involutions is not yet possible for the Kreuzer-Skarke list (although such a classification has been achieved for a related ensemble \cite{Carta:2020ohw}), and our approach will be to work with the data of $X$.  We do not fully specify $\mathcal{O}$, stipulating only that $h^{1,1}_-(X/\mathcal{O})=0$, so that $h^{1,1}_+(X/\mathcal{O})=h^{1,1}(X) \equiv h^{1,1}$.  In passing from $X$ to $X/\mathcal{O}$ for a specific $\mathcal{O}$, some of the data that will enter our analysis of superradiance would change: for example, some cycles have their volumes reduced by a factor of two.  We expect that our conclusions, expressed as averages over an ensemble of geometries, are robust against such changes, and thus against the choice of $\mathcal{O}$.  Even so, the axion spectrum of any individual geometry would depend to some degree on the particular involution $\mathcal{O}$ that is considered.

Let us now consider a compactification of type IIB string theory on such an orientifold of a Calabi-Yau threefold hypersurface $X$.
The four-dimensional theory contains $h^{1,1}$ axions from reduction of the Ramond-Ramond four-form $C_4$.
In terms of a basis $\{D^i\}$, $i=1,\ldots,h^{1,1}$ for $H_4(X,\mathbb{Z})$, we define
\begin{align}
\theta^i := \int_{D^i} C_4
\end{align}
to be the corresponding dimensionless axions.  We take the $D^i$ to be a set of $h^{1,1}$ irreducible toric divisors, and we term the resulting basis the\footnote{Strictly speaking, there are other toric bases corresponding to other choices of $h^{1,1}$ irreducible toric divisors, but we will work with just one choice throughout.  Our results are of course independent of this choice.} \emph{toric basis}.

The K\"ahler coordinates on K\"ahler moduli space are the complexified divisor volumes
\begin{align}
T^i := \tau^i + i \theta^i\,,
\end{align} with $\tau^i = \text{Vol}(D^i)$.
The volume of $X$ is given by
\begin{align}\label{eq:xvol}
  \mathcal{V} = \frac{1}{6}\kappa_{ijk} t^i t^j t^k\,,
\end{align}
where $\kappa_{ijk}$ are the triple intersection numbers of the $D^i$, and the $t^i$ are the K\"ahler parameters, corresponding to the volumes of curves.  The curve and divisor volumes are related by $\tau^i = \frac{\partial}{\partial t^i} \mathcal{V}$, and we note that indices are raised using $\delta^{ij}$.

\subsection{Stretched K\"ahler cone}

Throughout this work we will aim to consider only the region of K\"ahler moduli space in which the $\alpha'$ expansion is well-controlled.
This ensures that the kinetic term for the axions is accurately described by the metric on the K\"ahler moduli space of $X/\mathcal{O}$:
\begin{align}\label{eq:klagrangian}
  \mathcal{L} = -\frac{M_{\mathrm{Pl}}^2}{8 \pi^2} K_{ij}\partial^{\mu}\theta^i\partial_{\mu}\theta^j \,,
\end{align}
where the K\"ahler metric $K_{ij}$ is obtained from the K\"ahler potential $\mathscr{K}=-2\,\log \mathcal{V}$ by $K_{ij}=\partial^2 \mathscr{K}/\partial\tau^i\partial\tau^j$. We denote the square roots of the eigenvalues of $K_{ij}$ as $f_K$. As $\mathcal{V}$ is determined by the K\"ahler parameters and the intersection numbers of $X$, the axion kinetic term \eqref{eq:klagrangian} is readily computable in terms of the data of a triangulated polytope and a specification of the vevs of the K\"ahler moduli (saxions) $\tau^i$.
Background for these results is given in \eg\cite{Demirtas:2018akl, Bachlechner:2014gfa, Long:2016jvd}.

The $\alpha'$ expansion is well-controlled if all curvatures are small in units of the string scale $\ell_s \equiv 2\pi\sqrt{\alpha'}$, so we would like to identify the region in K\"ahler moduli space
in which this condition holds. We call this region the \textit{stretched} K\"ahler cone of $X$, denoted by $\widetilde{\mathcal{K}}(X)$.
A standard approach, which we will adopt, is to require that the volume of $X$ itself, and of all divisors and holomorphic curves within $X$, are at least one in units of $\ell_s$.  This condition, which amounts to imposing a restriction on all submanifolds that are calibrated by the K\"ahler form, is actually not necessary --- for example, certain holomorphic curves can be small in string units without giving large corrections to the effective action --- and it is also not sufficient, because the curvature of the metric on $X/\mathcal{O}$ depends on the complex structure moduli as well as the K\"ahler moduli.

With increasing $h^{1,1}$, the number of constraints defining $\widetilde{\mathcal{K}}(X)$ grows rapidly. As a result, $\widetilde{\mathcal{K}}(X)$ becomes very narrow, and gets pushed away from the origin. This translates into large hierarchies between the K\"ahler parameters $t^i$ and the divisor volumes $\tau^i$ \cite{Demirtas:2018akl}. This is the dominant effect that controls the scaling of divisor volumes with $h^{1,1}$. In turn, the divisor volumes determine the actions of Euclidean D3-branes and ultimately the instanton mass scales $\Lambda_a$.

Given a toric variety $V$ with a Calabi-Yau hypersurface $X \subset V$, constructing the stretched K\"ahler cone requires identifying the homology classes of holomorphic curves, which generate a cone in $H_2(X,\mathbb{Z})$ called the Mori cone of $X$.
Computing the Mori cone $X$ is rather challenging, and there is no general algorithm available. However, the Mori cone of the ambient variety $V$ is easily computed using toric geometry.  This cone \textit{contains} the Mori cone of the hypersurface $X$: any holomorphic curve in $X$ is also a holomorphic curve in $V$. Then, one can approximate $\widetilde{\mathcal{K}}(X)$ by imposing restrictions on the volumes of all holomorphic curves in $V$. The resulting region in the K\"ahler moduli space, $\widetilde{\mathcal{K}}(V)$, is contained inside $\widetilde{\mathcal{K}}(X)$. In our main datasets, we set the K\"ahler moduli at the tip of $\widetilde{\mathcal{K}}(V)$, which we denote $\mathcal{K}^V_1$ (because the minimal curve volume in units of $\ell_s$ is one).
This point in the moduli space is relatively easy to identify for all $h^{1,1}$, and the volumes of holomorphic curves and divisors in $\widetilde{\mathcal{K}}(V)$ scale similarly with $h^{1,1}$ as those in $\widetilde{\mathcal{K}}(X)$ \cite{Demirtas:2018akl}. Next, we define another point further inside the K\"ahler cone, $\mathcal{K}^V_{25}$, by uniformly scaling $\mathcal{K}^V_1$ so that the smallest holomorphic divisor volume $\tau_\mathrm{min}=25=1/\alpha_\mathrm{GUT}$ \cite{Buras:1977yy, Dimopoulos:1981yj}. These two points mark the boundaries of the region in the K\"ahler cone where the $\alpha'$ expansion is well controlled and a visible sector with a realistic grand unified gauge coupling $\alpha_\mathrm{GUT}$ can be realized via D7-branes wrapping holomorphic divisors.

\begin{figure}[ht!]
\begin{center}
  \includegraphics[width=0.7\textwidth]{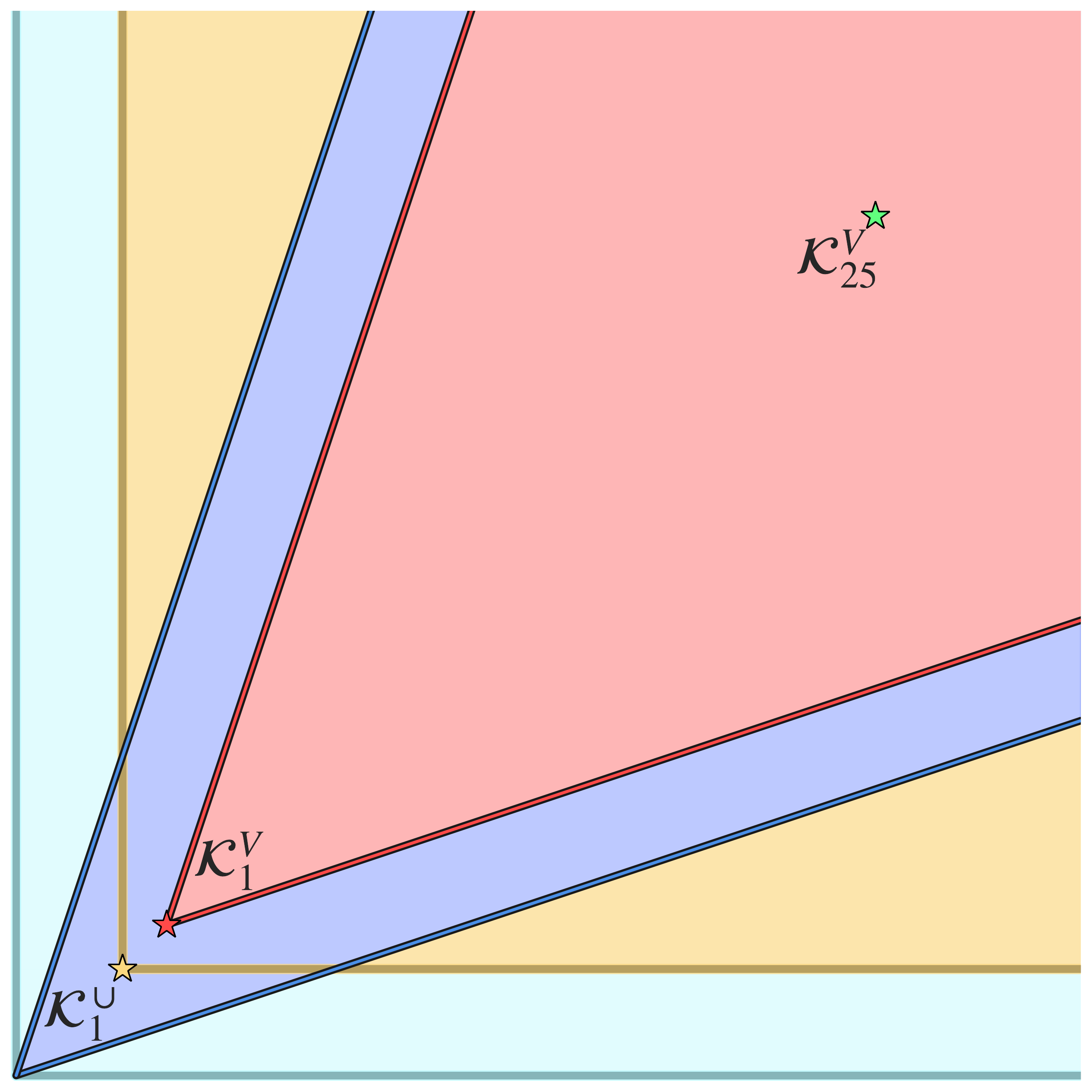}
  \caption{Cartoon of the K\"ahler cones and stretched K\"ahler cones for $h^{1,1}=2$.
  The three adjacent cones whose boundaries are cyan or blue rays are the K\"ahler cones $\mathcal{K}(V_1)$, $\mathcal{K}(V_2)$, and $\mathcal{K}(V_3)$ of three ambient toric varieties $V_1$, $V_2$, and $V_3$.  The union of these cones, which here is the first quadrant, is denoted $\mathcal{K}_\cup(V)$.  The gold region is the stretched K\"ahler cone $\widetilde{\mathcal{K}_\cup}(V)$, defined such that no curve that is in all three of $V_1$, $V_2$, and $V_3$ has volume $< \ell_s$.  The tip of this cone, marked with a gold star, is denoted $\mathcal{K}^{\cup}_1$.
  The blue cone is $\mathcal{K}(V_2)$, and the pink region  the stretched K\"ahler cone $\widetilde{\mathcal{K}}(V_2)$, defined such that no curve that is in $V_2$ has volume $< \ell_s$.  The tip of this cone, marked with a red star, is denoted $\mathcal{K}^{V}_1$.  (Strictly speaking this point should be called $\mathcal{K}^{V_2}_1$, but the numbering of $V_1$, $V_2$, and $V_3$ is arbitrary and we will select just one of them.) The point $\mathcal{K}^{V}_{25}$ is obtained by uniformly scaling $\mathcal{K}^{V}_{1}$ such that the smallest holomorphic divisor volume $\tau_\mathrm{min}=25=1/\alpha_\mathrm{GUT}$.}
\label{fig:cones}
\end{center}
\end{figure}

A better but more computationally expensive approximation to the stretched K\"ahler cone can be obtained by constraining the volumes of curves in $X$ that are holomorphic in \textit{any} ambient variety $V \supset X$ that arises from the same reflexive polytope. We denote this region by $\widetilde{\mathcal{K}_\cup}(V)$, and we have
$\widetilde{\mathcal{K}}(V) \subseteq \widetilde{\mathcal{K}_\cup}(V) \subseteq \widetilde{\mathcal{K}}(X)$.
As the number of such ambient varieties increases exponentially with $h^{1,1}$, constructing them via brute force becomes unfeasible when $h^{1,1} \gg 1$. Instead, one can construct $\widetilde{\mathcal{K}_\cup}(V)$ by considering the relevant circuits on the reflexive polytope.\footnote{We thank Andres Rios Tascon for this algorithm \cite{AndresUnpublished}.} Even then, the resulting cones become prohibitively complicated when $h^{1,1} \gtrsim 100$. To quantify the improvement achieved by computing $\widetilde{\mathcal{K}_\cup}(V)$, we constructed a smaller dataset with $1 \leq h^{1,1} \leq 100$ and found that the axions in the resulting sample of theories were relatively heavier and the decay constants were larger than in theories where the K\"ahler moduli are set at the tip of $\widetilde{\mathcal{K}}(V)$.  We will describe these tests in detail in Section \ref{sec:analysis} and Appendix \ref{app:tests}.

In this work we will not consider the dynamics of the saxions $\tau_i$.
Such moduli are famously problematic in cosmology, from the inflationary era to the present --- see \cite{Baumann:2014nda} for an overview --- and understanding their stabilization is one of the central problems in string compactifications.
Our interest here is in \emph{excluding}
certain geometries based on BHSR.  The effects of saxions could make our considerations irrelevant by disrupting cosmology so much that black holes, and the galaxies that contain them, never form, but this would only exclude the corresponding geometry.
Provided that the saxions are heavier than the axions and as such do not contribute to the decay rate of the cloud (see Section \ref{sec:BHSR}), it is implausible that saxion dynamics could \emph{rule in} a geometry that is excluded here based on superradiance.  So it will suffice to suppose that the saxions are stabilized at a point inside the K\"ahler cone, for example by perturbative corrections to the K\"ahler potential.  Where  exactly  they are stabilized is pivotal, as we shall see.
We will find that, for some geometries, the axion effective theory arising in certain regions in K\"ahler moduli space is ruled out by BHSR.
Thus, if the physics of moduli stabilization leads the saxions to be stabilized in such a region, the corresponding model is excluded.

\subsection{Instanton potential}\label{sec:instantons}

The axions $\theta^i$ enjoy all-orders shift symmetries $\theta^i \to \theta^i + \mathrm{const.}$ that are broken by nonperturbative effects.
Specifically, a Euclidean D3-brane wrapping a holomorphic four-cycle $\Sigma^a = q^a_{~i} D^i$, with $q^a_{~i} \in \mathbb{Z}$, and with
$i=1,\ldots,h^{1,1}$, contributes a superpotential term
\begin{align} \label{eq:terminw}
  W \supset W_a \equiv \mathcal{A}_a\, \mathrm{exp}\Bigl(- 2 \pi q^a_{~i} \left(\tau^i+i\theta^i\right)\Bigr)\,.
\end{align}
Here $\mathcal{A}$ is a moduli-dependent Pfaffian that we will set to unity in the following, and the index $a$ is a label that runs over all the holomorphic four-cycles wrapped by Euclidean D3-branes.  In general this is an infinite set, but only some finite number $P$ of terms
will be important for our purposes.
The scalar potential then takes the form
\begin{align}\label{eq:vlagrangian}
  V = \sum_{a=1}^P \Lambda_a^4 \biggl\{1 - \cos \left(\mathcal{Q}^{a}_{~i}\theta^i+\delta^a\right)\biggr\}\,,
\end{align} where summation over the repeated index $i$ is understood.
Here we have defined the rectangular \emph{charge matrix}
\begin{align} \label{eq:Qq}
  \mathcal{Q}^a_{~i} = \begin{pmatrix} q^\alpha_{~i} \\ q^\beta_{~i} - q^\gamma_{~i} \end{pmatrix},
\end{align}
in which $\alpha=1,\ldots,p$, $a=1,\ldots, p(p+1)/2 \equiv P$, and $p$ is the number of nonperturbative contributions to the superpotential. The rows involving
$q^\beta_{~i} - q^\gamma_{~i}$ arise from cross terms in the F-term potential.\footnote{A classical flux superpotential $W_0$, if present, would also contribute to the axion masses via cross terms with the non-perturbative superpotential, but the effect on the axion masses is subleading compared to the exponential dependence on the $\tau^i$.  The same holds for the Pfaffian prefactors $\mathcal{A}_a$ that we have set to unity.}  The mass scales $\Lambda_a$ in \eqref{eq:vlagrangian} are determined by the instanton actions, and the exponential dependence is given by
\begin{align} \label{eq:slamb}
  \Lambda_a \sim \, \mathrm{exp}\Bigl(-2\pi \mathcal{Q}^a_{~i} \tau^i\Bigr)\,.
\end{align}
As a result, there are typically exponential hierarchies among the $\Lambda_a$, as shown in Appendix~\ref{app:tests}.

For a given geometry, one may, in principle, determine which four-cycles $\Sigma^a$ support Euclidean D3-brane superpotential terms --- and so, using \eqref{eq:Qq}, compute the axion charge matrix $\mathcal{Q}_a^{~i}$ --- by counting the  zero modes of suitable Dirac operators on the cycles $\Sigma^a$~\cite{Witten:1996bn}.
Some progress in this direction was made in \eg \cite{Marsano:2008py,Blumenhagen:2009qh,Blumenhagen:2010ja,Donagi:2010pd,Cvetic:2011gp,Grimm:2011dj,Martucci:2015oaa,Braun:2017nhi}, but a complete understanding remains out of reach.  At the very least, one would have to specify the orientifold involution $\mathcal{O}$, which projects out certain zero modes, and even then no effective algorithm is currently available to count zero modes for $h^{1,1} \gg 1$.  Finally, even if a certain four-cycle $\Sigma^a$ has too many zero modes to support a superpotential term from a \emph{non-magnetized} Euclidean D3-brane, once one sums over possible magnetizations (\ie possible quantized worldvolume fluxes) and accounts for the effect of bulk closed string fluxes, the result is often a nonvanishing superpotential.  These effects of fluxes are again largely understood in principle, but explicit counting in an ensemble with $h^{1,1} \gg 1$ is not yet feasible.

We will nevertheless be able to arrive at a plausible and fairly robust model of the axion mass spectrum.
To describe this model, we consider the $h^{1,1}+4$ prime toric divisors $D^A$, which correspond to the vanishing loci of the toric coordinates $x_A$.  The $D^A$ are irreducible holomorphic hypersurfaces with the property that any holomorphic hypersurface (\ie any effective divisor)
can be written as a nonnegative integer linear combination of the $D^A$.  As Euclidean D3-brane superpotential terms can only arise from effective divisors, we may view the $D^A$ as a generating set of all possible Euclidean D3-brane superpotential terms.

Our model is just that: we suppose that there is a Euclidean D3-brane superpotential term for each of the $h^{1,1}+4$ prime toric divisors $D^A$.
Prime toric divisors often support Euclidean D3-branes even before accounting for the possibility of magnetization \cite{Long:2016jvd}, and so it is reasonable to anticipate that most of the $D^A$ will contribute superpotential terms once magnetization and closed-string fluxes are incorporated.
Indeed, \emph{completeness conjectures}~\cite{Polchinski:2003bq,Heidenreich:2016aqi,Andriolo:2018lvp} applied to these instantons would imply that all these divisors --- or in the case of the sub-Lattice Weak Gravity Conjecture \cite{Heidenreich:2016aqi}\footnote{See also \cite{Montero:2016tif}.} at least enough to fill out a sublattice --- give non-negligible contributions.

For the purposes of our analysis, what matters is that the axion mass spectrum obtained from our simplified model of the contributing instantons --- \ie of the instanton charge matrix --- gives a good approximation to the axion mass spectrum that would result from a complete computation.  To this end we have carried out a series of tests for varying models of the charge matrix.
Performing these tests on the entire dataset would be prohibitively expensive, so we used a smaller dataset consisting of $1000$ geometries at each of $h^{1,1}=10$, $30$, and $50$. For every geometry in this set, we constructed effective theories using different models of the instanton charge matrix and computed the key physical parameters: the axion masses, decay constants, and quartic interactions. We then performed Kolmogorov-Smirnov 2-sample tests on these distributions to compare them to our original dataset.

The models of the charge matrix that we analyzed are as follows.  First of all, what if other effective divisors besides the $D^A$ support Euclidean D3-branes?  Because all effective divisors are nonnegative linear combinations of the $D^A$, such an extension only introduces terms subdominant to those already captured by the $D^A$.  The Kolmogorov-Smirnov 2-sample test for this model gave $p \sim 1$, i.e.~the effect on the physical parameters is insignificant.

Second, if only some number $K<h^{1,1}$ of the divisors support Euclidean D3-branes, then the axion mass matrix necessarily has one or more zero eigenvalues. We have verified that when $h^{1,1}-K \ll h^{1,1}$, the effect of this change on the nonzero eigenvalues, \ie those interesting for BHSR, is negligible, with $p \sim 1$.  On the other hand, for sufficiently small $K$ the impact is substantial.  Without an explicit computation we cannot exclude the possibility that $K \ll \mathcal{O}(h^{1,1})$, but we find this unlikely in view of the evidence from examples \cite{Long:2016jvd}.

Third, what if non-holomorphic divisors support Euclidean D3-brane contributions to the K\"ahler potential?  We computed the effects of such terms from Euclidean D3-branes taken to wrap piecewise-calibrated representatives (cf.~\cite{Demirtas:2019lfi}) of all classes $D^A-D^B$ for $A\neq B$, as well as $10^3$ random charges of the form $\sum_A \alpha_A D^A$ where $\alpha_A \in \{-1,0,1\}$. We found the effect of these additional contributions to be negligible, with the Kolmogorov-Smirnov 2-sample test giving $p \sim 1$.

Finally, for any non-effective divisor class, there may exist a non-holomorphic representative that is smaller than the piecewise-calibrated representative. In this case, contributions from the corresponding instantons could be dominant. In general, computing volumes of non-holomorphic cycles and the actions of the associated Eulidean D3-branes is a difficult problem (see \eg \cite{Demirtas:2019lfi} and references therein). However, the instantonic form of the (sub-)Lattice Weak Gravity Conjecture puts conjectural upper bounds on the volumes of non-holomorphic cycles,\footnote{There are cases where these upper bounds are smaller than the volume of any piecewise-calibrated representative: see \cite{Demirtas:2019lfi} for details.} corresponding to lower bounds on the importance of the resulting terms in the effective action.  We used the Lattice Weak Gravity Conjecture bounds as estimates for the volumes of non-holomorphic cycles, and found that the effect of the corresponding non-holomorphic instantons on the axion masses, decay constants, and quartic interactions is negligible, with $p \sim 1$, especially for $h^{1,1}\gg 1$.  Thus, even the strongest form of the Weak Gravity Conjecture does not require non-holomorphic instantons to be relevant for our purposes.

In summary, we compared the axion mass spectrum resulting from our model of the charge matrix, in which all prime toric divisors support Euclidean D3-brane superpotential terms, to a range of related models, representing plausible corrections and refinements in which instantons wrapping various divisors are added or removed.  The mass spectra we found were practically indistinguishable, with $p \simeq 1$, in all cases \emph{except} that in which there are $\ll h^{1,1}$ independent instantons, such that the axion mass matrix has many exactly zero eigenvalues.  We argued that this final case is unlikely, though we did not prove it is excluded.  Thus, for the remainder of this work we will adopt the basic model in which instantons occur on prime toric divisors.

\section{Black Hole Superradiance}\label{sec:BHSR}

In this section we provide a brief summary of BHSR --- see \cite{Arvanitaki:2010sy,Brito:2014wla,Brito:2015oca} for more extensive reviews.  Our quantitative model and our treatment of BH data follow \cite{Stott:2018opm,Stott:2020gjj}.

\subsection{Preliminaries}

The Klein-Gordon equation on a Kerr spacetime background \cite{PhysRevD.22.2323,Dolan:2007mj} possesses bound states with complex eigenvalues, signalling that the occupation number of such a state can grow or decay without any external influence. These states are hydrogenic in nature, and are parameterised by the principal ($n$), orbital angular momentum ($l$), and azimuthal angular momentum ($m$) quantum numbers.
Extraction of angular momentum from the BH supplies energy to increase the occupation number of the bound states. Since the Klein-Gordon equation on a Schwarzschild background possesses no such classical instability, the process switches off in the limit that the BH spin goes to zero.

\begin{figure}[!ht]
\begin{center}
 \includegraphics[width=0.75\textwidth]{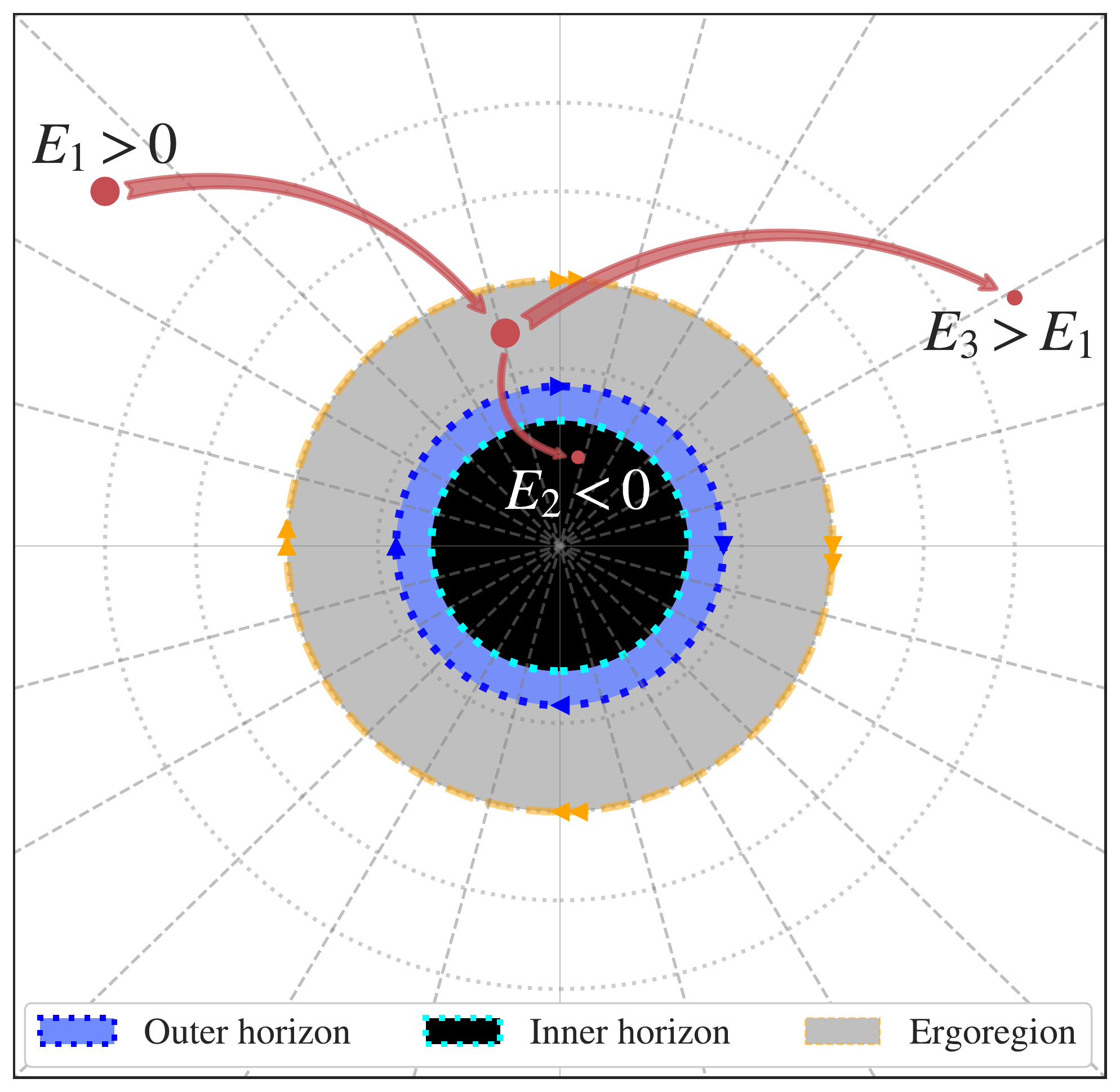}
\caption{\textbf{The Penrose Process}. The figure shows the ergogregion and the horizons of the Kerr metric from a view looking down the spin axis.  A particle enters the ergoregion and decays as $1\longrightarrow 2+3$, with particle 2 falling into the event horizon. If particle 2 has angular momentum in the opposite orientation to the BH itself, then the energy $E_2$ can be negative and particle 3 can escape to infinity with $E_3>E_1$, thus extracting angular momentum and energy from the BH. If the BH is surrounded by a mirror, the process can repeat. A bosonic field living on the Kerr background creates this scenario naturally. The hydrogenic bound states with radius inside the ergoregion can partially tunnel into the event horizon, with the remaining bound part increasing in energy by sapping the rotational energy of the BH.}
\label{fig:Penrose_Cartoon}
\end{center}
\end{figure}

The existence of the instability can be understood qualitatively by appeal to the Penrose process~\cite{Penrose:1971uk} and the scenario of the ``Black Hole Bomb''~\cite{Press:1972zz,Cardoso:2004nk}. Consider a physical object entering the ergoregion of the Kerr spacetime. Within the ergosphere, the Killing vector field associated to time translations at spatial infinity becomes spacelike.
Although all timelike paths must corotate with respect to observers at spatial infinity, an infalling object with sufficiently negative angular momentum has negative energy, and can reduce the energy of the black hole upon falling through the event horizon.
Thus, an infalling object that splits in two in a process $1\longrightarrow 2+3$, where the part entering the event horizon has $E_2<0$,
can lead to an emerging component with more energy than it went in with, as a consequence of energy conservation, \emph{i.e.}~$E_3=E_1+|E_2|$. See Fig.~\ref{fig:Penrose_Cartoon} for a cartoon depiction of this process.  This signals the presence of an instability, since placing a ``mirror'' around the BH allows this process to repeat.

Furthermore, such a process occurs naturally for a massive boson. Bound state solutions exist in the ergoregion, where the effective potential has a large angular momentum barrier separating this region from the event horizon, and a gravitational potential energy barrier preventing escape to infinity, providing the ``mirror''. The superradiance process occurs for frequencies that satisfy the superradiance condition $0<\omega<m\Omega_{\rm H}$, where $\Omega_{\rm H}$ is the angular velocity of the event horizon and $m$ is the magnetic quantum number.
The bound state wavefunction allows a small probability of tunnelling through the angular momentum barrier to the event horizon, allowing some fraction of the particles to fall into the event horizon, while the reverse process is (locally) forbidden due to the one-way nature of the horizon. Efficient tunnelling for the Penrose process requires significant overlap of the wavefunction with the region inside the event horizon. Superradiance is maximal when the Compton wavelength is of order the radius of the event horizon, $\lambda_{\rm C}=1/m_a\sim R_H\sim G_N M_{\rm BH}$.  When $\lambda_{\rm C}\ll R_H$ (i.e.~high particle mass) the would-be bound states are inside the event horizon, for small quantum numbers (and are thus unstable), while for large quantum numbers the states are sharply localised in radius and cannot effectively tunnel through the potential barrier. When $\lambda_{\rm C}\gg R_H$ (i.e.~low particle mass) there is significant tunnelling out of the gravitational potential well, and the states are unbound. The instability is thus a narrow-band resonance when the effective coupling of the hydrogen-like bound states satisfies $\alpha=G_NM_{\rm BH}m_a\sim 1$.

The value of the dimensionless product $\alpha$ parameterises the efficiency of the energy extraction process. For non-interacting bosons the occupation number grows according to the rate equation $\frac{dN}{dt}=N\Gamma_{\rm SR}$, until the angular momentum of the BH is significantly depleted, at \cite{Arvanitaki:2014wva}
\begin{equation}\label{eqnmax}
N = N_{\mathrm{max}} \approx 10^{76} \times \left(\frac{\Delta a_{\star}}{0.1}\right)\Bigl(\frac{M_{\rm BH}}{10M_{\odot}}\Bigr)^2\,,
\end{equation} where $\Delta a_{\star}$ is the change in spin.
The quasibound state growth rate $\Gamma_{\rm SR}$ is determined by the largest unstable eigenvalue, and defined by the imaginary component of the angular frequency $\omega$, which can be decomposed as $\omega = \omega_{\rm Re}+i\Gamma_{\rm SR}$, where $\omega_{\rm Re}$ represents the leading order hydrogenic mode eigenfrequencies.

Scalar perturbations on the Kerr spacetime in Boyer-Lindquist coordinates are fully separable in the Teukolsky formalism \cite{1973ApJ...185..635T,PhysRevLett.29.1114} as a product of one-dimensional functions, according to the general ansatz $\Phi(t,\bm{r})=e^{-i\omega t+im\phi} R(r)S(\theta)$ \cite{PhysRevD.5.1913}. This separability ansatz permits solvable spheroidal wavefunctions and a confluent Heun radial equation that leads to non-singular solutions satisfying certain boundary conditions. For scalars, the superradiance timescale can be found analytically in the non-relativistic limit ($\alpha \ll 1$) using a Taylor expansion. In the range of validity of the radial equation, the imaginary component of the frequency is \cite{PhysRevD.22.2323,Baumann:2019eav},
\begin{align}
\Gamma_{\rm SR} &= 2m_{a}r_+(m\Omega_{\rm H}-\omega)\alpha^{4l+4}\mathcal{A}_{nl}\prod^{l}_{k=1}\bigl(k^2(1-a_\star^2)+4r_+^2(m\omega-m_a)^2\bigr)\ , \\
\mathcal{A}_{nl} &= \frac{2^{4l+2}(2l+n+1)!}{(l+n+1)^{2l+4}n!}\left(\frac{l!}{(2l)!(2l+1)!}\right)^2\ , 	
\end{align}
where $r_+$ is the radius of the outer horizon.\footnote{Numerical solutions \cite{Dolan:2007mj} have been found to agree reasonably well when in the limit $\alpha \lesssim \mathcal{O}(0.1)$. See also \cite{Zouros:1979iw} for results in the strong coupling regime.}

The most dominant state is the nodeless corotating mode with $n=2,l=m=1$, which has the approximate growth rate $\Gamma_{\rm SR}\sim 24^{-1}a_{\star}\alpha^8m_a$. In order for SR superradiance to be the dominant process affecting the BH, $\Gamma_{\rm SR}$ must be larger than all other rates that could possibly affect the BH mass and spin. The superradiance rate decreases as the dimensionless spin $a_\star$ decreases, leading to a saturation bound when $\Gamma_{\rm SR}(a_{\star,{\rm crit}})=\Gamma_{\rm sat}$, with $\Gamma_{\rm sat}$ the largest non-superradiance rate. For definiteness, we use the Salpeter characteristic timescale
$\tau_{\rm Sal} = 4.5\times 10^{7}$ yrs for compact objects radiating at their Eddington limit \cite{1964ApJ...140..796S}
to fix the saturation rate, $\Gamma_{\rm sat}^{-1}\equiv \tau_{\rm BH} = \tau_{\rm Sal}$.
This saturation defines the \emph{Regge trajectories} $a_{\star,{\rm crit}}(M_{\rm BH},m_a)$ for astrophysical BHs.  BHs above the Regge trajectory are efficiently spun down by a boson until they reach the trajectory, at which point superradiance becomes too slow and other processes take over the evolution. Thus, the observation of BHs above the Regge trajectory can be used to exclude the existence of certain massive bosons.

Due to the instability, any unstable state will become populated from vacuum fluctuations alone, and hence the BHSR process is \emph{independent of the cosmological model}, in particular on the details of inflation or the axion relic density. This process leads to non-trivial macroscopic phenomena (depletion of BH spin) from initially small quantum perturbations.
Moreover, BHSR relies only on gravitational interactions of a minimally coupled massive boson, and does not require as a necessary condition any additional particle content or interactions. However, as we will discuss below, BHSR can be disrupted by non-gravitational interactions, such as axion decays to other sectors, or, more importantly, axion self-interactions.

\subsection{Axion self-interactions}\label{sec:BHSRconsl}

Self-interactions are one of several non-linear characteristics present in the hydrogenic boson-BH system.  Such features can allow the exponential amplification of the dominant state to be quenched before the instability extracts maximal spin within a characteristic e-folding timescale.\footnote{The other key example is the interactions between growing and decaying eigenmodes, through both perturbation level-mixing \cite{Arvanitaki:2010sy,Arvanitaki:2014wva} and resonances from inspiral orbital dynamics \cite{Baumann:2018vus,Zhang:2019eid,PhysRevD.99.064018,Baumann:2019ztm,Berti:2019wnn,Kavic:2019cgk,Cardoso:2020hca}.} The case of a single axion field has been well studied for the canonical cosine potential~\cite{Yoshino:2012kn,Arvanitaki:2010sy,Arvanitaki:2014wva,Yoshino:2015nsa,Mocanu:2012fd}. In this case, the self-interactions are attractive and lead to collapse of the cloud in a ``Bosenova'' process.  The critical occupation number beyond which a Bosenova occurs is
\begin{equation}
N_{\rm Bose} = \frac{10^{78}c_{\rm Bose}n^4}{\alpha^3}\left(\frac{M_{\rm BH}}{10M_{\odot}}\right)^2\left(\frac{f_{\rm pert}}{M_{\rm pl}}\right)^2	\,,
\label{eqn:bosenova_limit}
\end{equation}
where here $f_{\rm pert}$ is defined in terms of the coefficient $\lambda$ of the quartic self-interaction (for a multi-axion system, care is needed in the definition, see Appendix~\ref{app:basis}), and $c_{\rm Bose}$ is an $\mathcal{O}(1)$ constant to be specified. This process was studied numerically in \cite{Yoshino:2012kn} with the critical coupling fitting the estimate in \eqref{eqn:bosenova_limit} with $c_{\rm Bose}\approx 5$~\cite{Arvanitaki:2014wva}.

After a Bosenova occurs, the axion cloud grows back and the process can repeat.  However, the rate at which superradiant growth punctuated by Bosenovae depletes the angular momentum of the BH is slower by a factor $\sim N_{\rm{Bose}}/N_{\mathrm{Max}}$ than the rate in the absence of axion self-interactions \cite{Arvanitaki:2014wva}.  Sufficiently strong self-interactions therefore cause the depletion of BH spin through superradiance to be slower than other dynamical processes, rendering superradiance effectively irrelevant in determining the BH spin.  Thus, axions with sufficiently small $f_{\rm pert}$ cannot be excluded by measurements of BH spin.\footnote{However, as we will discuss in \S\ref{sec:conc}, strengthened interactions with the visible sector at small $f_{\rm pert}$ may allow for exclusions based on astrophysical bounds not involving BHs.}

\begin{figure}
\begin{centering}
\includegraphics[width=0.8\textwidth]{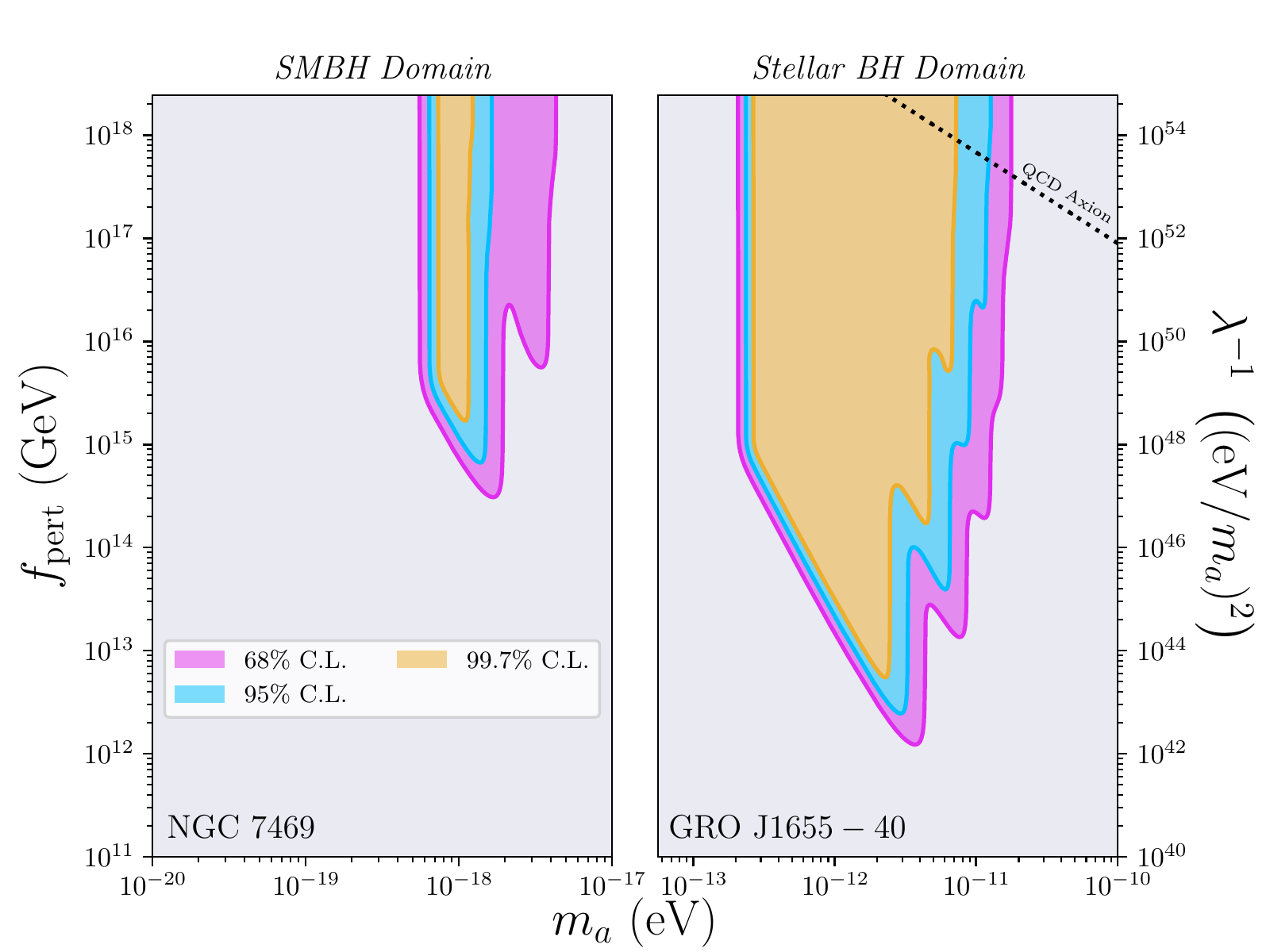}
\caption{Example exclusion functions for two typical BHs. The left panel shows a supermassive BH, and the right panel a stellar mass BH. BHSR operates over a resonant region in $m_a$, and is shut off by the Bosenova process at large values of the quartic coupling, parameterised by $f_{\rm pert}^{-1}$.}
\label{fig:BH_exclusion_example}
\end{centering}
\end{figure}
%%%%%%%%%%%%%

The Bosenova limit \eqref{eqn:bosenova_limit} was used in \cite{Arvanitaki:2014wva} to compute limits in the $(m_a,\lambda)$ plane, or equivalently $(m_a,f_{\rm pert})$, where there is now a maximum $\lambda$ (minimum $f_{\rm pert})$ above (below) which superradiance constraints no longer apply. These constraints were also computed in \cite{Stott:2020gjj}, where the exclusion probability, $P_{\rm ex}$, in the plane $(m_a,f_{\rm pert})$ was computed for a given BH, illustrated in Fig.~\ref{fig:BH_exclusion_example} for two typical BHs. Rather than repeat this analysis for every point in our database, we precompute the exclusion probabilities, and use them as a likelihood for $(m_a,f_{\rm pert})$. This approach to the superradiance exclusions is somewhat crude, and does not account in detail for all the different possible processes affecting the evolution of the cloud~\cite{Baryakhtar:2020gao}. However, it provides the correct broad picture in the plane $(m_a,f_{\rm pert})$, and is sufficient for our statistical exclusions on the landscape: slight shifts in the exclusion probabilities will not change our conclusions based on large samples of geometries. The reason for the correct broad picture can be justified by a simple argument.

The limit \eqref{eqn:bosenova_limit} can be estimated by appeal to the action for a massive field with self-interactions in a Kerr spacetime. Self interactions can shut off superradiance when the self-coupling $\lambda$ is large enough to play a role in the evolution. We anticipate this to happen when the corresponding term in the action, $\lambda\phi^4/4!$, becomes of the same order as the mass term, $m_a^2 \phi^2 / 2$, that drives the instability.\footnote{The superradiance condition itself can be estimated by equating $m_a^2\phi^2/2$ to the Ricci curvature, $R\sim 1/2(G_N M_{\rm BH})^2$ and taking $\phi\sim M_{pl}$ by dimensional analysis.} We have:
\begin{align}
	M_\text{cloud} &= N m_a \, , \\
	V_\text{cloud} &= \frac{4\pi}{3} R_\text{cloud}^3 \sim \frac{4\pi}{3} r_\text{ergo}^3 = \frac{32\pi}{3} (G_N M_{\rm BH})^3 \, ,
\end{align}
where $N$ is the occupation number of the cloud.  By equating $m^2 \phi^2 / 2 \sim M_\text{cloud}/V_\text{cloud}$, we find that
\begin{equation}
	\phi^2 \sim \frac{3}{16\pi} \frac{N}{(G_N M_{\rm BH})^3 m_a} \, . \label{eq:p3.4:phi}
\end{equation}
using this value of $\phi^2$ we equate the terms in the potential to find the critical value of $\lambda$:
\begin{equation}
    \phi^2 \sim \frac{12m_a^2}{\lambda} \Rightarrow N \sim 64\pi \frac{(G_N M_{\rm BH} m_a)^3}{\lambda} \sim \frac{64\pi}{\lambda} \, ,
\end{equation}
where in the final step we used that superradiance occurs approximately when $\alpha = G_N M_{\rm BH} m_a=1$. Using $\lambda \equiv m^2/f_{\rm pert}^2$ and $\alpha = G_N M_{\rm BH} m_a=1$ gives:
\begin{align}
    N &\sim \frac{64\pi}{\lambda} \approx 7\times 10^{78} \left(\frac{M_{\rm BH}}{10\,M_\odot}\right)^2 \left(\frac{f_{\rm pert}}{M_{\rm pl}}\right)^2\, ,
\label{eqn:bosenova_estimate}
\end{align}
which is remarkably close to \eqref{eqn:bosenova_limit} with $c_{\rm Bose}=5$ and $\alpha=n=1$. The utility of the estimate \eqref{eqn:bosenova_estimate} is that it is easily generalised to arbitrary interactions, and applies whether or not a Bosenova proper (ejection of large quantities of relativistic axions) occurs.

As we describe shortly, in our database of compactifications the four-axion interaction $V_{\rm int}=\lambda_{ijkl}\theta^i\theta^j\theta^k\theta^l/4!$ has most entries nonzero, but the dominant effect comes from the self-interaction $\lambda_{iiii}$.  Furthermore, the self-interaction term $\lambda_{iiii}$ can be positive or negative (see \S\ref{sec:analysis}), corresponding to attractive and repulsive self-interactions, respectively. Repulsive interactions are not present in the canonical single cosine potential studied numerically by \cite{Yoshino:2012kn}, and may or may not lead to collapse of the cloud in a Bosenova. However, as our argument above was based only on the magnitude of the action \eqref{eqn:bosenova_estimate} and thus \eqref{eqn:bosenova_limit} can be used with $|\lambda_{iiii}|$ for either case. Furthermore, as discussed in detail in \cite{Baryakhtar:2020gao}, the perturbative evolution of the cloud depends on the tree level matrix element, which is proportional to $\lambda^2$. Thus the point where scattering dominates over superradiance is independent of the sign of $\lambda$. Lastly, we approximate the superradiance rates in the non-relativistic limit, where the two body potential is attractive regardless of the sign of $\lambda$.

In a multi-axion system, axion flavour-changing processes contribute to the rate via the off-diagonal terms in $\lambda_{ijkl}$. Assuming that the scattering process involves two fields in the cloud annihilating into one that remains in the cloud, and one that escapes, then the $\lambda$ in \eqref{eqn:bosenova_limit} should simply be replaced by the sum of all terms in $\lambda_{iiij}$ where $m_j<m_i$ so that the process is kinematically allowed. It turns out that the off-diagonal terms in $\lambda_{ijkl}$ are hierarchically smaller than those on the diagonal, and so we neglect these processes, but they could be included in principle.

Finally, a typical critical point in a multi-axion potential need not satisfy a $\mathbb{Z}_2$ symmetry for the mass eigenbasis directions, i.e.~cubic terms will in general be present in the Taylor expansion of the potential, although we have not investigated this in our database. Repeating the above argument for a cubic interaction $\mathcal{L}=\eta \phi^3/3! \equiv (m^2/f_3)\phi^3/3!$ leads to the critical occupation number:
\begin{equation}
N_3\sim 192\pi \left(\frac{m}{\eta}\right)^2 \approx 2\times 10^{79} \left(\frac{M_{\rm BH}}{10\,M_\odot}\right)^2 \left(\frac{f_3}{M_{\rm pl}}\right)^2\, .
\end{equation}
If $f_3\sim f_{\rm pert}$, \emph{i.e.}~the same scale controls all the perturbative interactions (as we show in \S\ref{sec:analysis}, $f_{\rm pert}$ is strongly correlated with the K\"{a}hler metric eigenvalues, $f_K$, and thus we expect the same for $f_3$), then the critical occupation number for cubic interactions is sub-dominant to that for quartic interactions. Cubic interactions are also discussed in detail in \cite{Baryakhtar:2020gao}, where it is argued that cubic-interaction processes that compete with superradiance are relativistic, and are sub-dominant to non-relativistic effects of the quartic interactions.

\subsection{Single-field constraints}\label{sec:BHSRconstraints}

%%%%%%%%%%%%%%%%%%%%%%%%%%%%%%%%%%%
\begin{table}[!htp]
\begin{tabular}{ |T||T|T|T|T|  }
 \hline
 \multicolumn{5}{|c|}{Stellar Mass Black Holes} \\
 \hline
 Black Hole&$M_{\rm BH}\ \left(M_\odot\right)$&$a_*$&Refs.&{\normalsize Axion Mass Constraint (68\% CL)}\\
 \hline
GRO J1655-40  & $6.30^{+0.5}_{-0.5}$ & $0.7^{+0.1}_{-0.1}$ & \cite{1538-4357-636-2-L113}/\cite{Greene:2001wd} &$2.3 \times 10^{-13} {\rm eV} \leq m_{a} \leq 1.6 \times 10^{-11} {\rm eV}$ \\
A 0620-00   & $6.61^{+0.25}_{-0.25}$ & $0.12^{+0.19}_{-0.19}$ & \cite{2010ApJ...710.1127C}/\cite{2010ApJ...718L.122G} &\multicolumn{1}{c|}{$-$}  \\
LMC X-3  & $6.98^{+0.56}_{-0.56}$ & $0.25^{+0.13}_{-0.16}$ & \cite{2014ApJ...794..154O}/\cite{Steiner:2014zha} & $2.6 \times 10^{-13} {\rm eV} \leq m_{a} \leq 8.7 \times 10^{-13} {\rm eV}$  \\
XTE J1550-564  & $9.10^{+0.61}_{-0.61}$ &$0.34^{+0.37}_{-0.34}$ & \cite{2011ApJ...730...75O}/\cite{2011MNRAS.416..941S} & $2.0 \times 10^{-13} {\rm eV} \leq m_{a} \leq 6.4 \times 10^{-13} {\rm eV}$ \\
4U 1543-475  & $9.40^{+1.0}_{-1.0}$ & $0.8^{+0.1}_{-0.1}$ & \cite{Orosz67}/\cite{Shafee:2005ef} & $1.6 \times 10^{-13} {\rm eV} \leq m_{a} \leq 1.7 \times 10^{-11} {\rm eV}$  \\
LMC X-1 & $10.91^{+1.41}_{-1.41}$ & $0.92^{+0.05}_{-0.07}$ & \cite{Orosz:2008kk}/\cite{2009ApJ...701.1076G}& $1.4 \times 10^{-13} {\rm eV} \leq m_{a} \leq 1.9 \times 10^{-11} {\rm eV}$  \\
GRS 1915+105 & $10.10^{+0.6}_{-0.6}$ & $\geq 0.95$ & \cite{Steeghs:2013ksa}/\cite{McClintock:2006xd}  & $1.5 \times 10^{-13} {\rm eV} \leq m_{a} \leq 2.2 \times 10^{-11} {\rm eV}$\\
Cygnus X-1 & $14.80^{+1.0}_{-1.0}$ & $\geq 0.983$ & \cite{2011ApJ...742...84O}/\cite{Gou:2013dna} & $1.0 \times 10^{-13} {\rm eV} \leq m_{a}\leq 1.8 \times 10^{-11} {\rm eV}$ \\
M33 X-7 & $15.65^{+1.45}_{-1.45}$ &$0.84^{+0.05}_{-0.05}$ &\cite{Orosz:2007ng}/\cite{1538-4357-679-1-L37}& $1.0 \times 10^{-13} {\rm eV} \leq m_{a} \leq 1.1 \times 10^{-11} {\rm eV}$\\
\hline
GW150914 (1) & $35.6^{+4.7}_{-3.1}$ & $0.28^{+0.57}_{-0.25}$ & \cite{LIGOScientific:2018mvr}/\cite{LIGOScientific:2018mvr}& \multicolumn{1}{c|}{$-$}   \\
GW150914 (2) & $30.6^{+3.0}_{-4.4}$ & $0.34^{+0.53}_{-0.30}$ &\cite{LIGOScientific:2018mvr}/\cite{LIGOScientific:2018mvr}& $8.4 \times 10^{-14} {\rm eV} \leq m_{a} \leq 9.7 \times 10^{-14} {\rm eV}$ \\
GW151226 (1) & $13.7^{+8.8}_{-3.2}$ & $0.57^{+0.36}_{-0.43}$ &\cite{LIGOScientific:2018mvr}/\cite{LIGOScientific:2018mvr} & $1.3 \times 10^{-13} {\rm eV} \leq m_{a} \leq 1.8 \times 10^{-12} {\rm eV}$ \\
GW151226 (2)& $7.7^{+2.2}_{-2.5}$ & $0.51^{+0.44}_{-0.45}$ &\cite{LIGOScientific:2018mvr}/\cite{LIGOScientific:2018mvr}& $2.2 \times 10^{-13} {\rm eV} \leq m_{a}\leq 2.7 \times 10^{-12} {\rm eV}$\\
GW170104 (1)& $30.8^{+7.3}_{-5.6}$ & $0.34^{+0.52}_{-0.30}$ & \cite{LIGOScientific:2018mvr}/\cite{LIGOScientific:2018mvr}& $7.9 \times 10^{-14} {\rm eV} \leq m_{a} \leq 1.0 \times 10^{-13} {\rm eV}$\\
GW170104  (2)&  $20.0^{+4.9}_{-4.6}$ & $0.43^{+0.48}_{-0.38}$ & \cite{LIGOScientific:2018mvr}/\cite{LIGOScientific:2018mvr}&$9.7 \times 10^{-14} {\rm eV} \leq m_{a} \leq 3.5 \times 10^{-13} {\rm eV}$ \\
GW190521 (1) & $85.0^{+21.0}_{-14.0}$ & $0.69^{+0.27}_{-0.62}$ & \cite{Abbott:2020tfl}/\cite{Abbott:2020tfl}& $2.6 \times 10^{-14} {\rm eV} \leq m_{a} \leq 3.3 \times 10^{-13} {\rm eV}$\\
GW190521 (2) & $66.0^{+17.0}_{-18.0}$ & $0.73^{+0.24}_{-0.64}$ &\cite{Abbott:2020tfl}/\cite{Abbott:2020tfl}& $3.1 \times 10^{-14} {\rm eV} \leq m_{a} \leq 6.8 \times 10^{-13} {\rm eV}$\\
  \hline
 \multicolumn{5}{|c|}{Supermassive Black Holes $\left(M_{\rm BH}\ \left(10^6M_\odot\right)\right)$} \\
 \hline
Fairall 9  & $255.0^{+56.0}_{-56.0}$ & $0.52^{+0.19}_{-0.15}$ & \cite{Peterson:2004nu}/\cite{2012ApJ...758...67L} & \multicolumn{1}{c|}{$-$}    \\
Mrk 79   & $52.40^{+14.40}_{-14.40}$ & $0.70^{+0.1}_{-0.1}$ & \cite{Peterson:2004nu}/\cite{2011MNRAS.411..607G} & $1.8 \times 10^{-19} {\rm eV} \leq m_{a}\leq 8.8 \times 10^{-19} {\rm eV}$ \\
NGC 3783   & $29.80^{+5.40}_{-5.40}$ & $\geq 0.98$ & \cite{Peterson:2004nu}/\cite{2011Brenneman} & $2.8 \times 10^{-19} {\rm eV} \leq m_{a} \leq 5.0 \times 10^{-18} {\rm eV}$ \\
Mrk 335  & $14.20^{+3.70}_{-3.70}$ & $0.83^{+0.09}_{-0.13}$ & \cite{Peterson:2004nu}/\cite{2013MNRAS.428.2901W} & $5.8 \times 10^{-19} {\rm eV} \leq \mu_{0} \leq 4.2 \times 10^{-18} {\rm eV}$ \\
MCG-6-30-15   & $2.90^{+1.80}_{-1.60}$ & $\geq 0.98$ & \cite{McHardy:2005ut}/\cite{Brenneman:2006hw} & $2.5 \times 10^{-18} {\rm eV} \leq m_{a} \leq 5.9 \times 10^{-17} {\rm eV}$  \\
Mrk 110 & $25.10^{+6.10}_{-6.10}$ & $\geq 0.89$ & \cite{Peterson:2004nu}/\cite{2013MNRAS.428.2901W} & $3.4 \times 10^{-19} {\rm eV} \leq m_{a} \leq 2.6 \times 10^{-18} {\rm eV}$ \\
NGC 7469 & $12.20^{+1.40}_{-1.40}$  & $0.69^{+0.09}_{-0.09}$ & \cite{Peterson:2004nu}/\cite{doi:10.1111/j.1365-2966.2011.19224.x} & $6.5 \times 10^{-19} {\rm eV} \leq m_{a} \leq 4.0 \times 10^{-18} {\rm eV}$   \\
Ark 120 & $150.0^{+19.0}_{-19.0}$ & $0.64^{+0.19}_{-0.11}$ & \cite{Peterson:2004nu}/\cite{2013MNRAS.428.2901W} & $7.5 \times 10^{-20} {\rm eV} \leq m_{a} \leq 1.4 \times 10^{-19} {\rm eV}$   \\
NGC 4051 & $1.91^{+0.78}_{-0.78}$ & $\geq 0.99$ & \cite{Peterson:2004nu}/\cite{8175999} & $3.3 \times 10^{-18} {\rm eV} \leq m_{a} \leq 9.6 \times 10^{-17} {\rm eV}$ \\
M87* & $6500.0^{+700.0}_{-700.0}$ & $0.9^{+0.1}_{-0.1}$ &\cite{Akiyama:2019fyp}/\cite{Tamburini:2019vrf} & $2.6 \times 10^{-21} {\rm eV} \leq m_{a} \leq 1.2 \times 10^{-20} {\rm eV}$  \\
\hline
\end{tabular}
\caption{Mass and dimensionless spin parameter measurements, with corresponding literature references, for the full set of BHs used in this work. Mass values are quoted up to $1\sigma$ confidence, whereas the dimensionless spin values are quoted at the 90\% confidence level. The value of the mass constraint represents the excluded axion mass values from BHSR for each BH at the 68\% confidence level. For the gravitational wave LIGO observations the designations (1) and (2) denote the primary and secondary merger components of the binary system measurements. \\}\label{bh_table}
\end{table}

%%%%%%%%%%%%%%%%%%%%%%%%%%%%%%%%%%%

Our BH data set is summarised in Table~\ref{bh_table}, along with the axion mass ranges excluded by each individual BH, for which we adopt the statistical method of \cite{Stott:2018opm,Stott:2020gjj}. The exclusion probability in this method is computed as the probability that a given BH lies within the region of the $(M_{\rm BH},a_\star)$ plane forbidden by superradiance for a given value of $(m_a,f_{\rm pert})$, which can be found from the area overlap of the error ellipse with the Regge trajectories. The method approximates the $(M_{\rm BH},a_\star)$ errors as Gaussian and uncorrelated when computing the overlap. The method is also Bayesian. Overall this leads to a slightly wider exclusion on the axion mass by an $\mathcal{O}(1)$ factor on either end of the 95\% region compared to the statistical method employed in \cite{Arvanitaki:2014wva}.

For the self-interactions, following~\cite{Stott:2020gjj}, we use the approximation based on $N_{\rm Bose}$ consistent with \cite{Arvanitaki:2014wva}, and neglect cubic interactions. The analysis in \cite{Stott:2020gjj} shows that, in the stellar mass domain, the free field mass constraints hold for $f_{\rm pert}\gtrsim 10^{14}\text{ GeV}$, and disappear completely for $f_{\rm pert}\lesssim 10^{12}\text{ GeV}$. In the supermassive domain, the free field mass constraints hold for $f_{\rm pert}\gtrsim 10^{16}\text{ GeV}$, and disappear completely for $f_{\rm pert}\lesssim 10^{15}\text{ GeV}$. This treatment of the self-interactions is cruder than the detailed evolution considered in \cite{Baryakhtar:2020gao}, and in terms of bounds on $f_{\rm pert}$ it gives weaker exclusions.  Applying the approach of \cite{Baryakhtar:2020gao} to the Kreuzer-Skarke axiverse would be worthwhile, but is beyond the scope of this work.

Measurements of SMBHs are subject to larger uncertainty than those of stellar mass BHs. Even when a high spin is confirmed there can be large errors on the mass. As we will show, SMBHs have comparably little effect on our main conclusions due to the relatively small values of $f_{\rm pert}$ across most of our datasets.

A separate issue concerns the LIGO BH spin measurements, which are taken over a short period of time shortly before merger, and thus have a large spin uncertainty from fitting the waveforms without a significant inspiral period. As a default, we include the LIGO BHs in our dataset, taking the quoted errors from waveform fitting at face value.  In \cite{Arvanitaki:2014wva,Baryakhtar:2020gao}, precise measurements of both the BH mass and spin are demanded, whereas our Bayesian method allows for the inclusion of BHs with imprecisely measured values of $(M_{\rm BH},a_\star)$, thus enlarging the dataset.~\footnote{The authors of \cite{Baryakhtar:2020gao} object to the inclusion of spins with large errors on the grounds that, using our method, an imprecise measurement still leads to an exclusion. This is not a problem of the method. Take the example in \cite{Baryakhtar:2020gao} where the poor measurement implies an almost flat \emph{posterior} on the BH spin. This indeed implies that a BH has 95\% probability of $a_\star>0.05$, and thus will lead to an exclusion on the axion mass using the information from the measurement to update the prior. Furthermore, if one had a \emph{physically motivated} prior on $a_\star$ (whatever that was), a Bayesian analysis would also allow exclusions on the axion mass for BHs with no spin measurement at all. A Bayesian analysis demands an assumption on the prior, and prior dependence in the case of poor measurements is unavoidable.}

Nevertheless, to be complete, in the following we consider constraints both including and excluding the LIGO BHs (i.e.~only using cases where the likelihood dominates the prior on the spin). As is evident from Table~\ref{bh_table}, the LIGO BHs only extend the range of excluded masses by around one order of magnitude overall, and thus including or excluding them also only has a minor effect on our later conclusions.

\subsection{Gauge interactions}

Interactions between axions and other fields not included in our model could also outcompete superradiance if the interactions are strong enough~\cite{Boskovic:2018lkj,PhysRevLett.122.081101,Boskovic:2019qao}. The most dangerous are those between axions and massless fields. For the case of photons and hidden photons, such interactions occur via the Chern-Simons term, $g\phi F\tilde{F}/4$, where $F$ is the photon or hidden photon field strength, $g_{a\gamma}$ is the coupling strength, and $\phi$ is the canonically normalised axion field. For closed string axions we expect that $g \sim \alpha/(\pi f_K)$, where $f_K$ is defined below \eqref{eq:klagrangian}, and $\alpha$ is the gauge coupling constant. The limit on $g$ for this process to dominate over superradiance can also be estimated by appeal to the relevant terms in the action. We do not include this possibility in our model, and leave the study of the axion interactions with gauge fields to future work. We suspect that if axion-gauge field interactions were large enough to compete with superradiance, then they might be large enough to be in conflict with astrophysical constraints on the axion-photon coupling, as in the large $h^{1,1}$ models studied in~\cite{Halverson:2019cmy}; this possibility is discussed in \S\ref{sec:conc}.

\section{Axion Parameters and Constraints}\label{sec:analysis}

Having detailed the constraints arising from BHSR, we can now apply these bounds to an ensemble of axion effective theories obtained from type IIB string compactifications.  The construction of the ensemble has been outlined in \S\ref{sec:KS}.  Here we first present numerous statistics of the ensemble in \S\ref{sec:KSstats}, and then derive the relevant couplings of the resulting axion effective theories in terms of the geometric data in \S\ref{sec:axdata}. We finally compute exclusions on geometries in \S\ref{sec:exclusions}

\subsection{Ensemble of geometries}\label{sec:KSstats}

We generated 210,379 fine, regular, star triangulations (FRSTs), for the full range of $h^{1,1}$ in the Kreuzer-Skarke database, \ie for $1\leq h^{1,1}\leq 491$.  At each $h^{1,1}$ we picked up to 1000 random favorable reflexive polytopes from the database and obtained at least one FRST per polytope.  For $177 \le h^{1,1} < 238$ the number $N(h^{1,1})$ of favorable polytopes in the database falls between 100 and 1000, and in such cases we included only $N(h^{1,1})$ polytopes.  For $h^{1,1} \ge 238$ one has $N(h^{1,1}) < 100$, and in such cases we computed more than one FRST per polytope, in order to have at least 100 distinct geometries at each $h^{1,1}$ for which $N(h^{1,1}) \neq 0$.  We calculated the relevant topological data of the resulting $\mathrm{CY}_3$ hypersurfaces using \texttt{CYTools} \cite{cytools}.  We then considered the moduli at the three fixed positions $\mathcal{K}^V_1$, $\mathcal{K}^\cup_1$, and $\mathcal{K}^V_{25}$, as defined in Figure \ref{fig:cones}. 
%\footnote{As mentioned previously, due to the computational intensity, the dataset $\mathcal{K}_1^\cup$ consists of 100 geometries per $h^{1,1}\leq 100$.}

The statistics of the geometric data of the CY$_3$'s depends on the algorithm used to obtain the FRSTs. In this work, we obtained the FRSTs by lifting the points of the polytope into one higher dimension by a set of heights
and constructing the convex hull of the resulting point set --- see Section 2 of \cite{Demirtas:2020dbm} for a detailed discussion. In the following we will refer to this as the \textit{fast} sampling algorithm. While this algorithm is robust and efficient, the set of FRSTs obtained this way may not be a \textit{fair} random sample of all FRSTs, \ie with each FRST sampled with equal probability.  To construct a sample that is plausibly closer to being representative of the total set, one can use the random sampling algorithm of \cite{Demirtas:2020dbm}. As this method is computationally expensive, we used it to prepare a smaller dataset of 1000 CY$_3$'s at each of $h^{1,1}=10,30,50$ and quantified its effect on the statistics of axion masses and decay constants. We found that the sample of CY$_3$'s obtained by random sampling give rise to theories where the axions are lighter and the decay constants are smaller.

Our aim in this work is to consider the region of K\"ahler moduli space in which the $\alpha'$ expansion is well-controlled. We call this region the stretched K\"ahler cone of $X$, and denote it by $\widetilde{\mathcal{K}}(X)$. In our main datasets, we approximate this region by the stretched K\"ahler cone of the ambient variety, $\widetilde{\mathcal{K}}(V)$. An even better approximation can be obtained by computing $\widetilde{\mathcal{K}_\cup}(V)$, defined in \S\ref{sec:KS}. This results in theories where the axions are heavier and the decay constants are larger.

Given that the statistics of axion masses and decay constants depend on the way we sample CY hypersurfaces and approximate the streched K\"ahler cones, we prepared a small dataset where we used both the fair sampling algorithm of \cite{Demirtas:2020dbm} and computed $\widetilde{\mathcal{K}_\cup}(V)$ for each geometry. As these computations are expensive both in CPU time and memory, we picked 100 random favorable reflexive polytopes and obtained 10 FRSTs per polytope at each $1 \leq h^{1,1} \leq 100$ and analyzed the resulting theories.  We found that the effects of using these two algorithms mostly cancel out to give theories where axions are only slightly lighter and decay constants only slightly larger  -- see Appendix \ref{app:tests}.

We wish to emphasize that, regardless of the sampling method and selection of a point within the stretched K\"ahler cone, all of the geometries considered are valid string compactifications in which the resulting effective theory is expected to be under good theoretical control: that is, all known perturbative and non-perturbative corrections are small.  Thus, our analysis will rule out portions of the string landscape via BHSR constraints, and the only question we are addressing by tests of our sampling method is how \emph{representative} these constraints are of manifolds specified by a given $h^{1,1}$.

\subsection{Computing axion data}\label{sec:axdata}

As outlined in Section \ref{sec:KS}, we construct the axion Lagrangian using the underlying geometric quantities arising in our compactifications.  The data we extract to construct these is, thus, in a particular basis: \textit{the lattice basis}. We must therefore transform these axion systems into a \textit{physical} basis -- see Appendix \ref{app:basis} --  and derive the various physical quantities -- \ie $m_a, f_a,\lambda_{ijkl}$ -- in order to apply the constraints from Section \ref{sec:BHSRconstraints}.  We describe the methods using examples from the $\mathcal{K}^V_1$ dataset for illustration, however they are applied to each of the 3 datasets: $\mathcal{K}^V_1, \mathcal{K}^V_{25}$, and $\mathcal{K}^\cup_1$.

\subsubsection{Cycle volume hierarchies}\label{sec:Lambdas}

As shown in Appendix~\ref{app:tests}, the hierarchies between elements of $\Lambda_a$ are extremely large, and inhibit automated computations significantly due to numerical precision issues.\footnote{Unfortunately, the hierarchies are also large enough to preclude the use of the excellent methods developed in \cite{Bachlechner:2017zpb} which analyse and find the global minima of multiaxion systems.}
Indeed, there are elements of $\Lambda_a$ that are so small they would be inaccessible using floating point-precision limited scientific packages such as \texttt{scipy}.  To handle such hierarchies, one can use multiprecision packages such as \texttt{mpmath}~\cite{mpmath} or proprietary software such as \texttt{Mathematica}.  With an increase in precision, large hierarchies slow computations \textit{significantly}, and has prevented analysis of minima in geometries with $h^{1,1}\gg 100$.
Despite the difficulties, we were still able to optimize a sample of our potentials, using a variety of optimization techniques available in \texttt{pygmo2}~\cite{Biscani2020} in order to compute an ensemble of local minima and critical points\footnote{We only include critical points for which every tachyonic direction has $m_a^2 > - H_0^2$, and so is stable on cosmological timescales.}

Working with data type \texttt{float64}, commonly known as \textit{double precision}, in \texttt{pygmo2} we can reach a numerical precision of
\begin{equation}\label{eq:deflambdanum}
\Lambda_{\tt{num}} :=  10^{-80} M_{\mathrm{pl}} \sim 10^{-53}\,\mathrm{eV}\,,
\end{equation}
and so we impose $\Lambda_{\tt{num}}$ as a cutoff, setting to zero any $\Lambda_a < \Lambda_{\tt{num}}$.
Such extremely small terms in $\Lambda_a$ would, in any case, contribute axion mass contributions well below the Hubble scale, and therefore outside of any observational possibility, rendering these directions effectively massless.  We now turn to the massive directions.

%%%%%%%%%%%%%%%%%%%%%
\subsubsection{Axion masses}\label{sec:axm}
%%%%%%%%%%%%%%%%%%%
\begin{figure}[ht!]
\begin{center}
  \includegraphics[width=\textwidth]{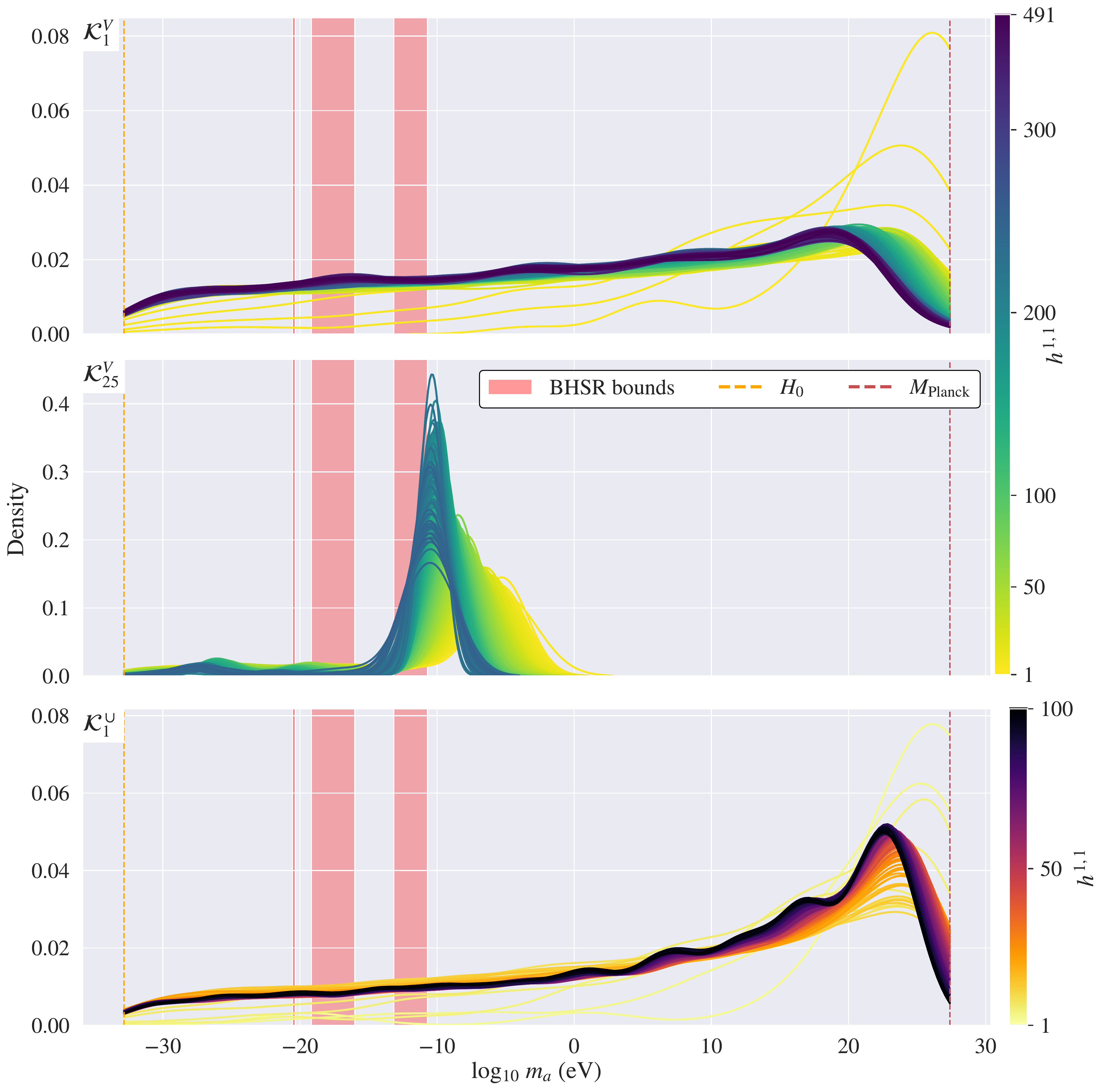}
  \caption{Distributions of $m_a$ for the datasets we studied, evaluated at the origin of the axion field space. Results for different triangulations are combined to a single distribution at each $h^{1,1}$.  Note that we clip the mass distributions at $H_0$ and $M_{\mathrm{Pl}}$ -- though there are no $m_a\gtrsim M_{\mathrm{Pl}}$.}
\label{fig:masses}
\end{center}
\end{figure}
%%%%%%%%%%%%%%%%%%%%

Superradiance occurs in the vacuum. As such, the axion masses and quartic self interactions should be evaluated at points in the potential that are stable on cosmological time scales. The only absolutely stable point is the global minimum. Local minima are stable if the tunnelling rate~\cite{Coleman:1980aw} is longer than the age of the Universe, while critical points are stable if the most tachyonic mass satisfies $|m^2|<H_0$.

Due to the dimensionality of the field space, finding global minima is not feasible.
We used the method of \textit{differential evolution} (DE), a genetic algorithm-like optimization routine, to find ensembles of local minima and critical points.  Importantly, for the case with vanishing phases, $\delta^a=0$,  the origin $\vec{\theta}=\vec{0}$ is always a critical point.  In Appendix~\ref{app:tests} we summarize our computation of axion parameters at all these locations in the potential (critical points, local minima, at the origin, and both with and without phases), as well as our comparison of the distributions. Remarkably, we find that our conclusions are robust across these possibilities, and the axion mass and decay constant spectra for a given point in moduli space are largely independent of where they are evaluated in the axion potential.\footnote{This can be attributed to the large hierarchies in $\Lambda_a$: only a handful of particularly large elements will contribute to the axion parameters.} For computational simplicity, we present the subsequent high statistics results computed at the origin.

Once a minimum or critical point is found we can compute the masses: these are just the eigenvalues of the Hessian matrix transformed to the canonical basis,
\begin{align}\label{eq:Hess}
  \mathcal{H}_{ij}=\frac{\partial^2V(\theta)}{\partial\theta^i\partial\theta^j}.
\end{align}
The results for the datasets are shown in Fig. \ref{fig:masses}. For presentation, we bin these results to present a single distribution at each value of $h^{1,1}$. Each individual triangulation, \ie each individual CY$_3$ makes a draw from this distribution, with the value of the mass \textit{independent} of the critical point in the potential (see Appendix~\ref{app:tests}).
%%%%%%%%%%%%%%%%%%%%%%%%%%
\begin{figure}[ht!]
\begin{center}
  \includegraphics[width=\textwidth]{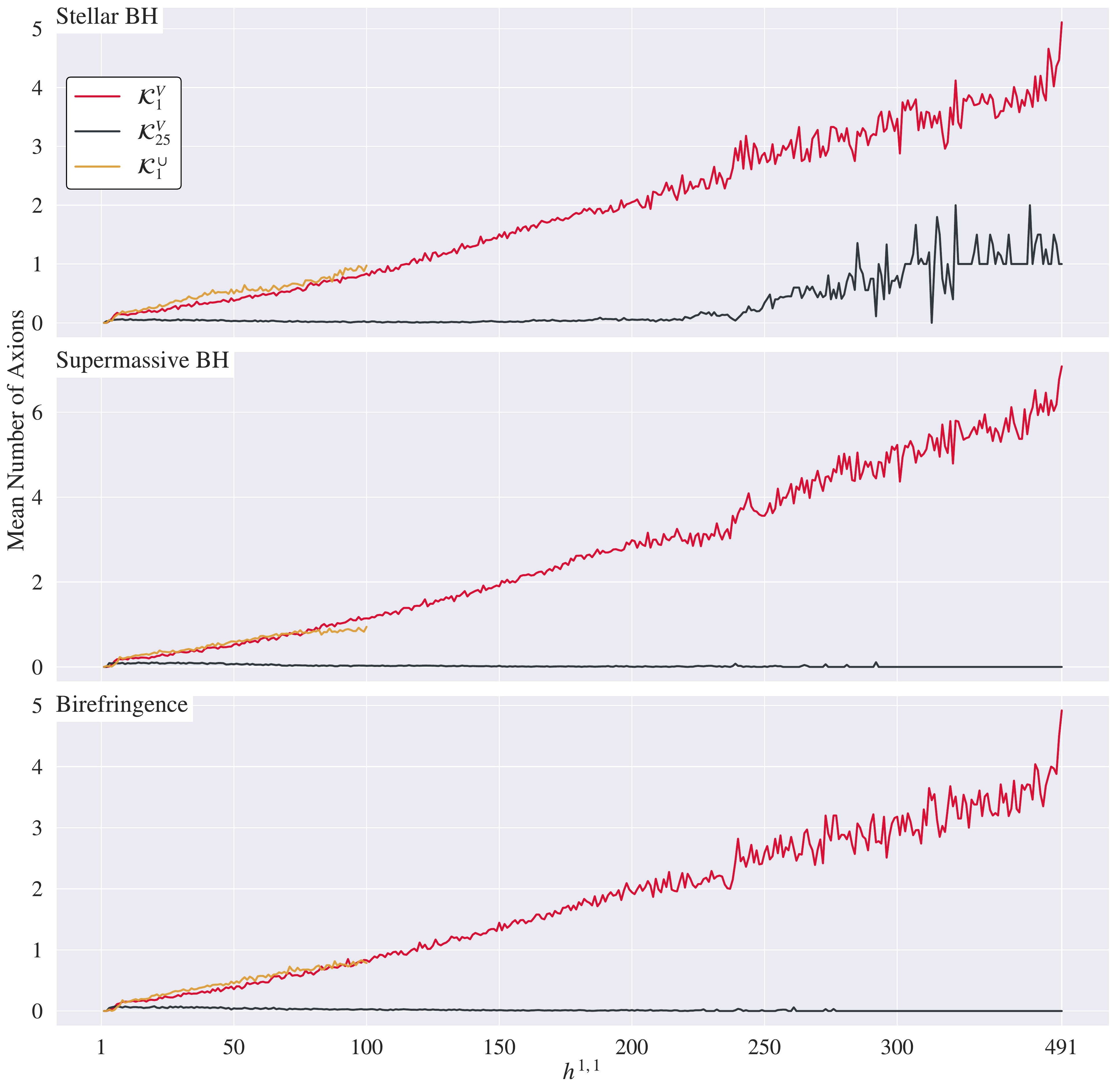}
  \caption{Mean number of axions per geometry that fall within different interesting physical mass windows. The noise at large $h^{1,1}$ is caused by low number statistics in our dataset at these values.}
\label{fig:windows}
\end{center}
\end{figure}
%%%%%%%%%%%%%%%%%%%%%%%%%%%

We observe that at the tip of the K\"{a}hler cone the mass distribution converges to an almost universal shape at large $h^{1,1}$. The shape is almost log-flat in the tails, with a broad peak below the Planck scale. The peak moves slowly to smaller masses as $h^{1,1}$ increases and the average volume grows. Consistent with the convergent nature of the distributions, we find an almost fixed fraction of massless axions, i.e. axions with $m_a<H_0$ (see Figure \ref{fig:tachyon}). Inside the K\"{a}hler cone, we find a similar convergent behaviour in the fraction of effectively-massless fields. Here the average volumes are larger, and thus the peak of the distribution is far lower, near $10^{-10}\text{ eV}$. Due to the cutoff imposed at $H_0$, the peak appears narrower in this case, and the movement of the peak as $h^{1,1}$ increases is more pronounced.

Having the mass spectra to hand, as a first investigation we look at the number of axions that fall in interesting mass windows, as shown in Fig. \ref{fig:windows}. As we see, in fact, a non-zero number of axions lie within the excluded BHSR windows at \textit{each} $h^{1,1}$. We also show for reference the ``birefringence'' window, $10^{-33}\text{ eV}\leq m_a\leq 10^{-28}\text{ eV}$~\cite{Arvanitaki:2009fg} (see discussion in Sec.~\ref{sec:conc}). For the two datasets near the tip of the stretched K\"{a}hler cone, we observe that the number of axions in any fixed mass window increases linearly with $h^{1,1}$, showing that a fixed fraction of all the axions in \emph{any} of these geometries reside in these windows. This is consistent with our assertion that the tails of these distributions at large $h^{1,1}$ have converged to a universal form away from the evolving peak near the Planck scale. On the other hand, for our dataset inside the stretched K\"{a}hler cone, we observe a sharp increase in the number of axions in the stellar mass BHSR region at large $h^{1,1}$. This is caused by the evolution of the peak of this distribution into the BHSR window (see Fig.~\ref{fig:masses}).

\subsubsection{Quartic interactions}\label{sec:axq}
%%%%%%%%%%%%%%%%%%%
\begin{figure}[ht!]
\begin{center}
  \includegraphics[width=\textwidth]{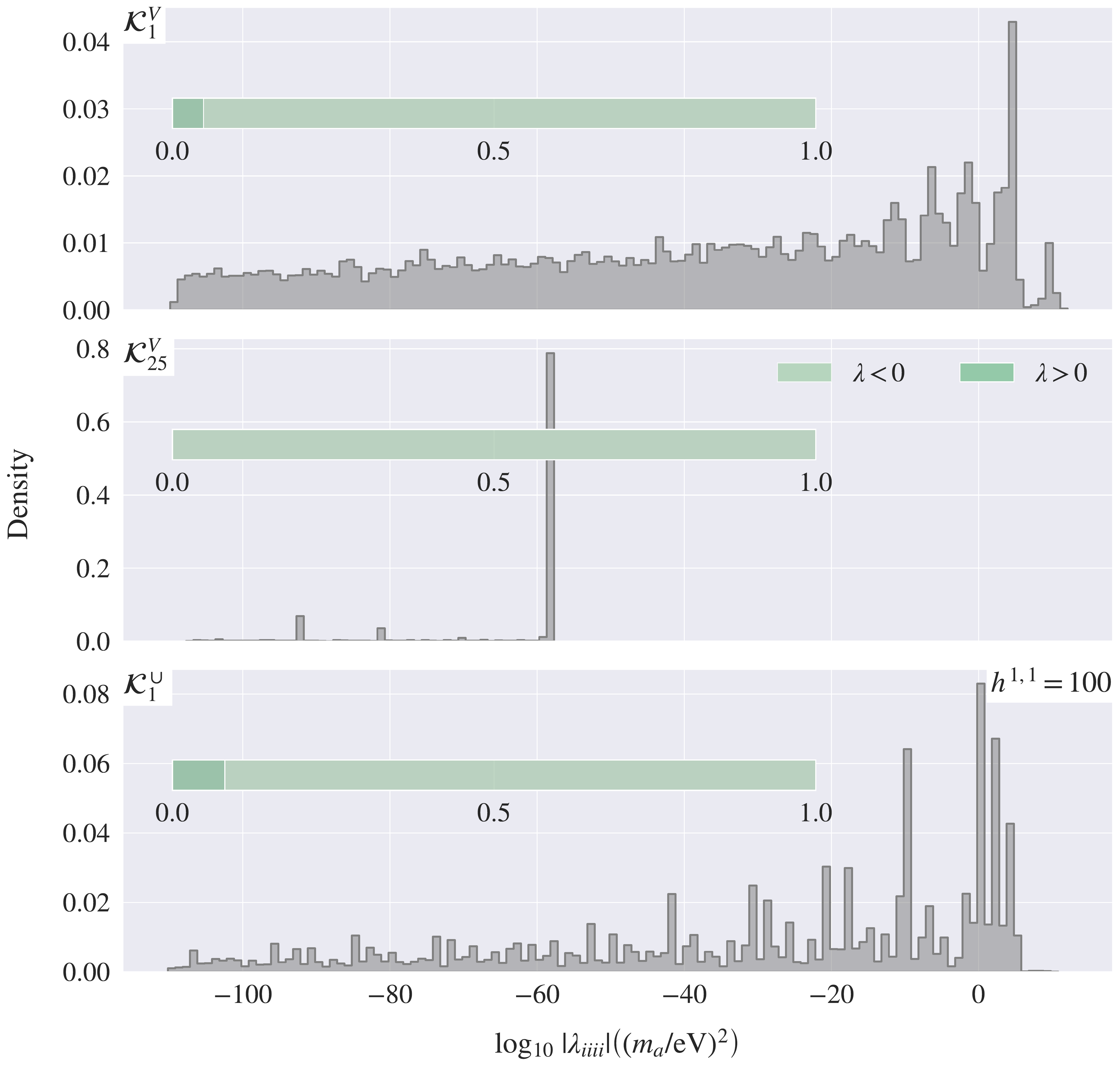}
  \caption{Distributions of $|\lambda_{iiii}|$ for the datasets we studied, evaluated at the origin of the axion field space (see Appendix~\ref{app:tests} for other locations) for $h^{1,1}=100$. The green bars indicate the distribution of the signs of $\lambda_{iiii}$ in the complete dataset.}
\label{fig:lambdas}
\end{center}
\end{figure}
%%%%%%%%%%%%%%%%%%%%

With access to the full axion potentials, we can Taylor expand a critical point to fourth order and extract the full four-axion interaction tensor, $\lambda_{ijkl}$ in the mass eigenbasis (see Appendix~\ref{app:basis}). We did this for all $1\leq h^{1,1}\leq 20$ and the diagonal components only for the remainder of the data. We found off-diagonal elements of $\lambda_{ijkl}$ are hierarchically smaller than the diagonal elements $\lambda_{iiii}$.

The distribution of the on-diagonal terms is shown in Fig.~\ref{fig:lambdas} for $h^{1,1}=100$. Similarly to the mass, a convergent and almost log-flat distirbution is found at large $h^{1,1}$. We also noticed a strong positive correlation between $m_a$ and $\lambda_{iiii}$, \ie $\rho_s\left( m_a,\lambda_{iiii} \right)\sim 1$, where $\rho_s$ is the Spearman rank coefficient.
The correlations $\rho_s\left( m_a,f_{\mathrm{pert}} \right)$ and $\rho_s\left( \lambda_{iiii} \right)$, however, were found to be only weakly positive at $\simeq 0.5$ (see Appendix~\ref{app:tests} for details).

In addition, we examined the distributions of signs of $\lambda_{iiii}$, where $\lambda<0$ is attractive and $\lambda>0$ is repulsive. The distribution of the signs at the origin of field space is shown in Fig.~\ref{fig:lambdas}, where we see a preference for attractive interactions, but a number of directions have repulsive interactions. In the case of the signs, we found some dependence on the location of the critical point in the potential: at the origin, the self-interactions were mostly attractive, while at generic local minima and critical points, they were found to be both repulsive and attractive (see Appendix~\ref{app:tests} for details).

\begin{figure}[ht!]
\begin{center}
  \includegraphics[width=\textwidth]{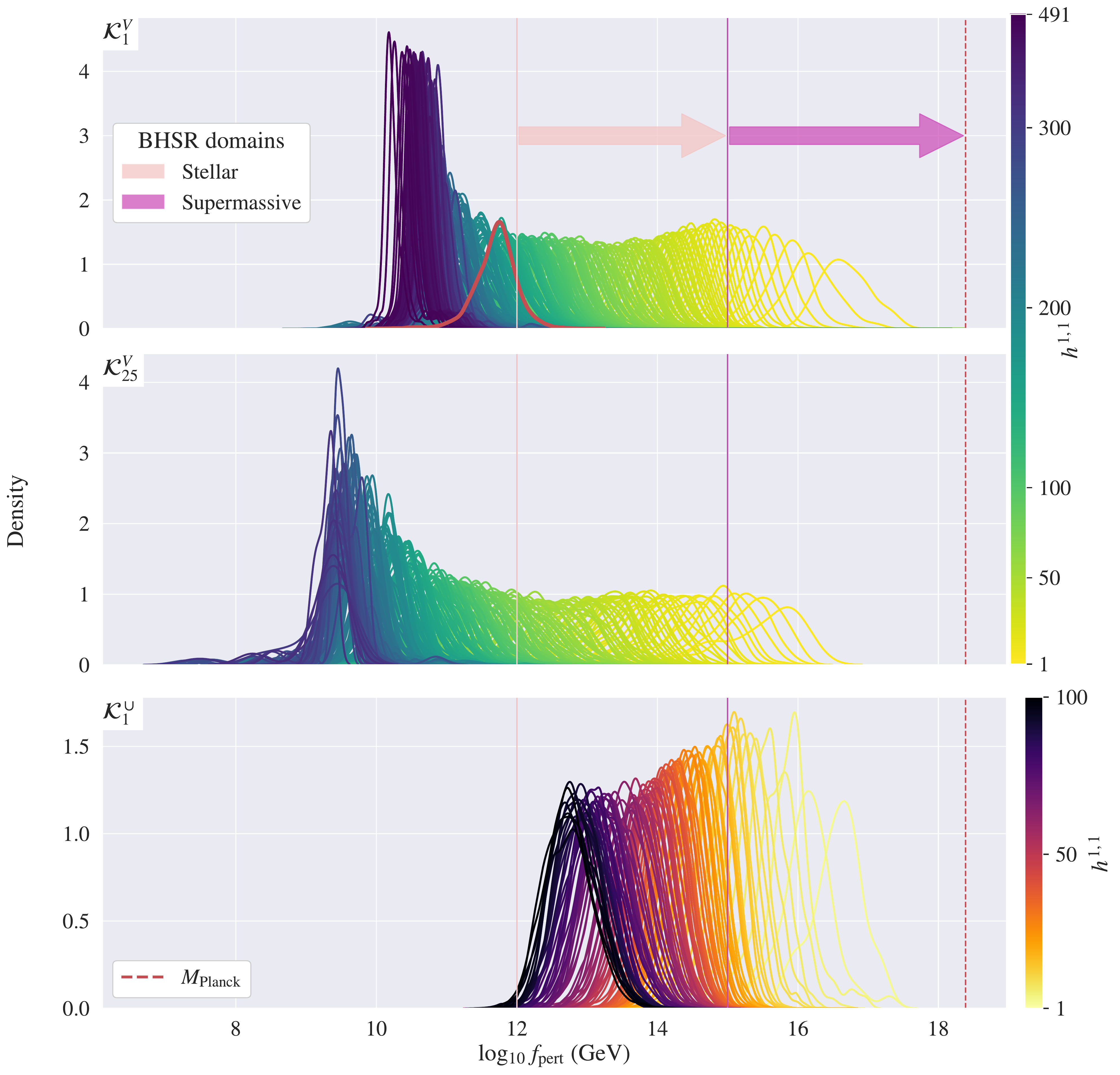}
  \caption{Probability distributions of $f_{\mathrm{pert}}$ for the three datasets we considered. Marked in arrows are the approximate minimum values of $f_{\mathrm{pert}}$ required for BHSR to operate for the stellar and supermassive cases. In the top panel, the distribution for $h^{1,1}=157$ (where, as we will see, the probability of exclusion peaks) is shown in red.}
\label{fig:fpert}
\end{center}
\end{figure}

From the diagonal terms in the interaction tensor we extract the perturbative decay constant:
\begin{align}
  f_{\mathrm{pert}}\equiv \sqrt{\frac{m^2}{\lambda_{iiii}}}.
\end{align}
We computed $f_{\mathrm{pert}}$ for all of the geometries in our three datasets: the distributions of $f_{\mathrm{pert}}$ are shown in Fig.~\ref{fig:fpert}. We observe a log-normal distribution of values when binned over $h^{1,1}$.  The log-normal distribution can be understood loosely from the product distributions of $m$ and $\lambda$ with tight, but not exact, correlations. The mean of the log-normal distribution shows a clear decreasing trend as $h^{1,1}$ increases. For $\mathcal{K}_{25}^V$, the mean is typically two orders of magnitude smaller than for $\mathcal{K}_{1}^{V}$. Fig.~\ref{fig:fpert} also indicates the approximate values of $f_{\rm pert}$ above which superradiance can occur in the stellar and supermassive mass windows.
%%%%%%%%%%%%%%%
%%%%%%%%%%%%%%%

\begin{figure}[ht!]
\begin{center}
  \includegraphics[width=\textwidth]{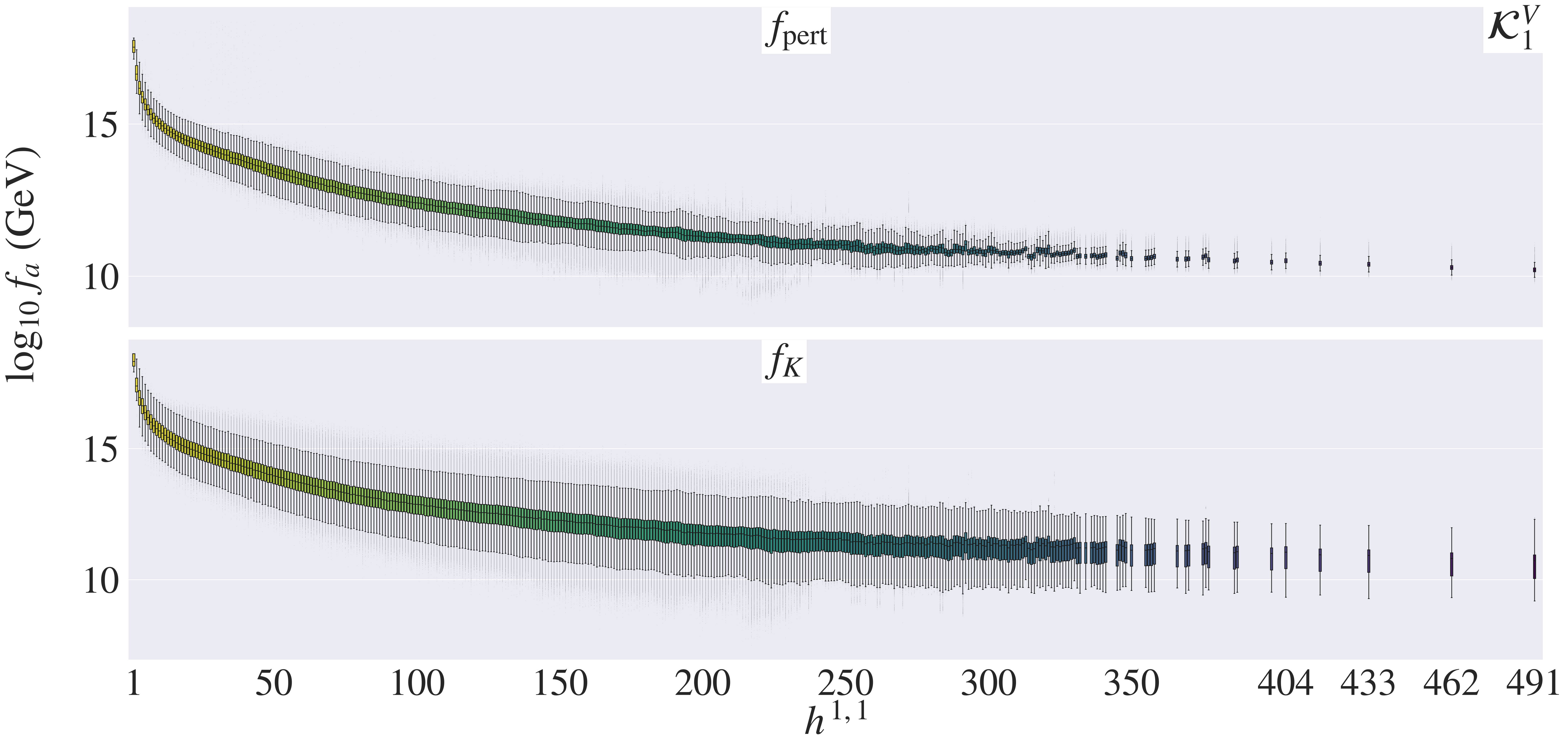}
  \caption{A comparison of the $f_{K}$ and $f_{\mathrm{pert}}$ distributions.}
\label{fig:fk_fpert}
\end{center}
\end{figure}

For all three datasets and for every geometry, we also computed the square roots of the K\"{a}hler metric eigenvalues $f_K$, see equation \eqref{eq:klagrangian}.\footnote{The $f_K$ are related to `axion decay constants', but this term is used to refer to several quantities that are in general inequivalent.  To avoid ambiguity we work with the precisely-defined quantities $f_K$ and $f_{{\rm pert}}$.}
Importantly, $f_K$ are never larger than $M_{\mathrm{Pl}}$ in our dataset. Just as with $f_{\mathrm{pert}}$, the $f_K$ distributions are log-normal, with the mean reducing as $h^{1,1}$ increases. In fact, the distributions of $f_K$ and $f_{\rm pert}$ are statistically similar. The distributions are displayed side-by-side for $\mathcal{K}_{1}^V$ in Fig.~\ref{fig:fk_fpert}.

Due to the computational cost of calculating $f_{\mathrm{pert}}$ at large $h^{1,1}$, only those $f_{\mathrm{pert}}$ in $\phi_{m>H_0}$ directions were computed.  This leads to a narrower log-normal distribution than the $f_K$, distributions -- as seen in Fig. \ref{fig:fk_fpert} -- with a higher peak value, though a Kolmogorov-Smirnov 2-sample test comparing the $N_{m>H_0}\ f_{\mathrm{pert}}$ values with the largest $N_{m>H_0}\ f_K$ values shows they represent very similar distributions.  Thus, we may conclude that the easily attainable $f_K$ could be used as a good statistical proxy for $f_{\mathrm{pert}}$.

%%%%%%%%%%%%%%%%%%%%%%%%%%%%%%%%%%%%
\subsection{Exclusions on triangulations from BHSR}\label{sec:exclusions}

We are now in a position to apply the exclusions outlined in Section \ref{sec:BHSR} to our axion spectra. Using the mass spectra and decay constants, we compute the probability of exclusion for each triangulation (i.e. each CY$_3$) in our database. The methodology is outlined in Appendix B of \cite{Stott:2018opm} and we summarise here for convenience.

The probability of excluding a model, $\mathcal{A}=\{m_a,f_{\rm pert}\}$, with a single axion given a dataset of BHs, $\left\{ d_i \right\}$, is given by $P_{\mathrm{ex}}\left( \left. \mathcal{A}\right|\left\{ d_i \right\} \right)$. The total exclusion probability is equivalent to the probability that any \emph{single} BH lies in the excluded superradiance region of the $(M_{\rm BH}, a_*)$. To compute this, we first note that the probability is normalised such that
\begin{align}\label{eq:Pex}
  P_{\mathrm{ex}}\left( \left. \mathcal{A}\right|\left\{ d_i \right\} \right) = 1 -  P_{\mathrm{al}}\left( \left. \mathcal{A}\right|\left\{ d_i \right\} \right)
\end{align}
where $ P_{\mathrm{al}}\left( \left. \mathcal{A}\right|\left\{ d_i \right\} \right)$ is the probability that the model is allowed. Using a simple probability tree we note that $ P_{\mathrm{al}}\left( \left. \mathcal{A}\right|\left\{ d_i \right\} \right)$ is the cumulative probability that \textit{all} BH points simultaneously fall outside the excluded region. Thus:
\begin{align}\label{eq:Pal}
  P_{\mathrm{al}}\left( \left. \mathcal{A}\right|\left\{ d_i \right\} \right) = \prod_i  P_{\mathrm{al}}\left( \left. \mathcal{A}\right| d_i \right),
\end{align}
where $P_{\mathrm{al}}\left( \left. \mathcal{A}\right| d_i \right)$ is the probability that the model is allowed given the single BH data point $d_i$ in the dataset.
%%%%%%%%%%%%%%%%%%%%%
\begin{table}[ht!]
\noindent
{\small
\begin{center}
{\tabulinesep=1.2mm
  \begin{tabu}{|m{2cm}|m{1.5cm}|m{1.5cm}|m{1.5cm}|m{1.5cm}|m{1.5cm}|m{1,5cm}|}
\hline
\hline
\multirow{3}{*}{Point in MS}&  \multicolumn{4}{c|}{Stellar}  &  \multicolumn{2}{c|}{Supermassive}  \\
\cline{2-7}  & \multicolumn{2}{c|}{With LIGO} & \multicolumn{2}{c|}{Without LIGO} & \multirow{2}{*}{$h^{1,1}$} & \multirow{2}{*}{$P_\mathrm{ex}\left( \mathrm{CY}_3 \right)$}\\ \cline{2-5}
&  $h^{1,1}$ & $P_\mathrm{ex}\left( \mathrm{CY}_3 \right)$  & $h^{1,1}$ & $P_\mathrm{ex}\left( \mathrm{CY}_3 \right)$  &  &   \\
\hline
{$\centering \mathcal{K}^V_1$} &    157 &  0.541 & 157 & 0.474 &  6 &  0.128 \\ \cline{1-7}
{$\centering \mathcal{K}^V_{25}$}&    11 &  0.067 &   11& 0.058& 3 &  0.062 \\ \cline{1-7}
{$\centering \mathcal{K}^\cup_1$ }&  42 & 0.55 & 92 & 0.44&  13 & 0.18 \\

\hline
\hline
\end{tabu}}
\end{center}
}
\caption{\label{table:maxPex}
$h^{1,1}$ for largest fraction of geometries excluded -- at 95\% C.L. -- by stellar and supermassive BH data at different points in the moduli space (extracted from Fig.~\ref{fig:moneyplot}).}
\end{table}
%%%%%%%%%%%%%%%%%%%%%%

Now let us consider the axiverse. Each triangulation, with a choice of a point in moduli space, is a model, $\mathcal{M}=\{\mathcal{A}_i\}$ containing $N_{\mathrm{ax}}$ axions. Since any one axion in the excluded region is enough to falsify that model, the probability that the model is allowed is given by the probability that every axion in the model is allowed, i.e.
\begin{align}\label{eq:PalNax}
  P_{\mathrm{al}}\left(\left. \mathcal{M}\right|\left\{ d_i \right\} \right) = \prod_i P_{\mathrm{al}}\left( \left. \mathcal{A}_i\right|\left\{ d_i \right\} \right).
\end{align}
Thus, the exclusion probability for the triangulation is:
\begin{align}\label{eq:PexNax}
  P_{\mathrm{ex}}\left(\left. \mathcal{M}\right|\left\{ d_i \right\} \right) = 1 -  P_{\mathrm{al}}\left(\left. \mathcal{M}\right|\left\{ d_i \right\} \right).
\end{align}

\begin{figure}[ht!]
\begin{center}
  \includegraphics[width=0.82\textwidth]{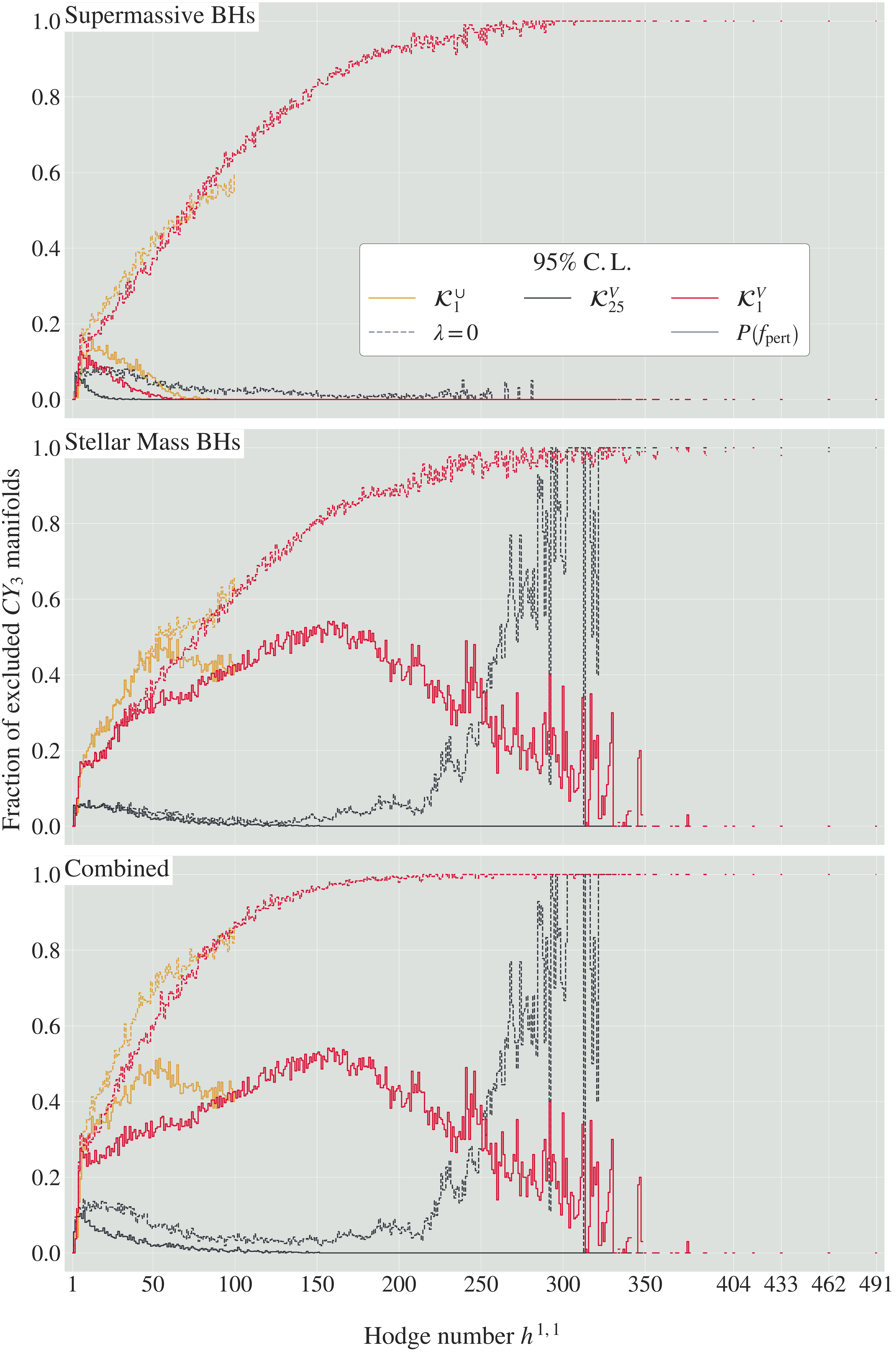}
  \caption{Fractions of geometries excluded by BHSR as a function of $h^{1,1}$.  We note that our overall exclusions are dominated by bounds from stellar mass BHs, and that the results are largely unchanged if the LIGO BHs with poorly measured spin are omitted (see Appendix~\ref{app:tests}).  The noise at large $h^{1,1}$ is caused by low number statistics in our dataset at these values.}
\label{fig:moneyplot}
\end{center}
\end{figure}

We now compute the probability of exclusion for each geometry individually. In most cases a geometry is either excluded at high probability (if its axions lie in the BHSR window), or it is not, \ie there are relatively few cases with low but not vanishing $P_{\rm ex}$. In Fig. \ref{fig:moneyplot} we present the fraction of excluded CY$_3$'s -- with  ${P_{\mathrm{ex}}\left(\left. \mathcal{M}\right|\left\{ d_i \right\} \right)>0.9545}$ -- for each $h^{1,1}$ for all three points in the moduli space. We summarise our results in Table \ref{table:maxPex}, stating the $h^{1,1}$s with the peak fraction of excluded CY$_3$'s -- \ie before the curve turns over -- and the excluded fraction at this value.
%%%%%%%%%%%%%%%%%%%%
%%%%%%%%%%%%%%%%%%%%

Our results show that the exclusion probability depends greatly on both the position in K\"ahler moduli space, \ie the volumes of the cycles, as well as the number of axions. In particular, the exclusions are stronger for $\mathcal{K}^V_1$ than for $\mathcal{K}^V_{25}$.  Initially, all constraints become stronger with increasing $h^{1,1}$, but the effect of self-interactions, which increases in magnitude as a function of $h^{1,1}$ due to the shrinking average value of $f_{\mathrm{pert}}$ is significant in suppressing the constraints compared to the case where self-interactions are neglected. Self-interactions cause the constraints derived from supermassive BHs to be irrelevant in determining exclusions for $h^{1,1} \gtrsim 15$. For $\mathcal{K}^V_1$, the constraints peak around $h^{1,1} \approx 160$, where we exclude more than 50\% of CY$_3$'s at greater than 95\% C.L. This result is easy to understand from the $f_{\rm pert }$ distributions shown in Fig.~\ref{fig:fpert}, where we observe that for $h^{1,1}=157$ the distribution is already moving to values outside the region allowing for superradiance. At $\mathcal{K}_{25}^V$, the constraints peak near $h^{1,1} = 10$, where we exclude around 7\% of CY$_3$'s at greater than 95\% C.L. At $\mathcal{K}_{25}^V$ the fraction of excluded CY$_3$'s falls to almost zero at $h^{1,1} \gtrsim 100$.

Note that the downward trend of the exclusion probabilities, as $h^{1,1}$ increases beyond the turnover in $P_{\mathrm{ex}}\left( \mathrm{CY}_3 \right)$, would drop off due to disruption of the condensate by self-interactions, once every axion has $f_{\mathrm{pert}}\lesssim 10^{12}\mathrm{\,GeV}$: see Fig.~\ref{fig:fpert}.  We also note that both the $m_a$ and $f_a$ for the FRSTs combine to give overall PDFs per $h^{1,1}$ -- shown in Figs. \ref{fig:masses} and \ref{fig:fpert} -- assuming that the FRSTs generated for each $h^{1,1}$ are representative of the remaining geometries in each category.

\section{Conclusions}\label{sec:conc}

This paper has two principal results: a systematic study of the axion mass spectrum and quartic self-interactions in more than $2\cdot 10^5$ compactifications of type IIB string theory at three different points in moduli space, and an analysis of the parameter space excluded by black hole superradiance in these theories. We summarise these in turn.

\begin{figure}
\includegraphics[width=\textwidth]{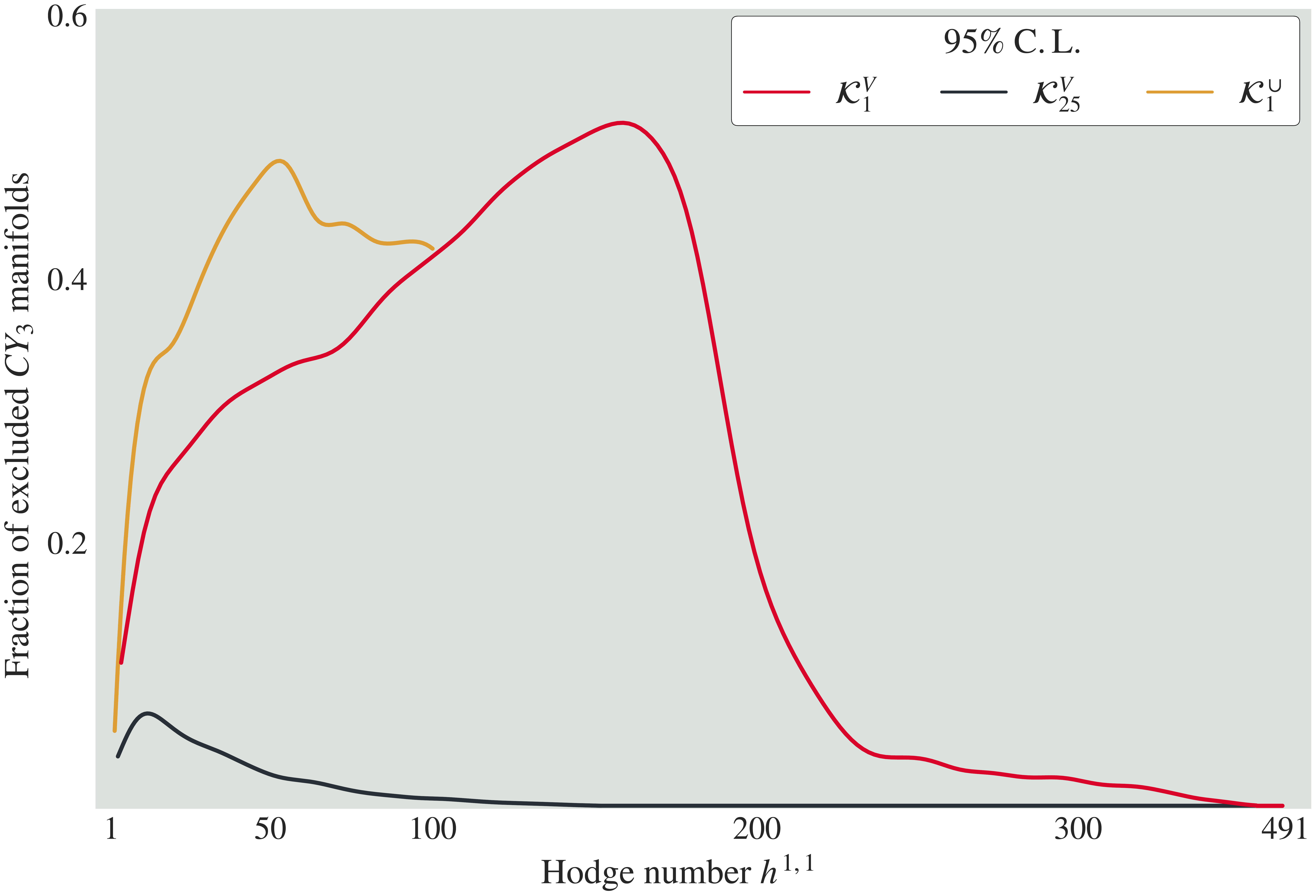}
\caption{{\bf Summary of constraints} Fraction of Calabi-Yau manifolds in our database excluded by black hole superradiance, as a function of the Hodge number $h^{1,1}$, at three different locations in moduli space (see Fig.~\ref{fig:cones}).  The data is from Fig.~\ref{fig:moneyplot}, including axion self-interactions, and has been binned and smoothed.}
\label{fig:Pex_kde}
\end{figure}

\subsection{Summary of results}

The geometries studied in this work are CY$_3$ hypersurfaces in toric varieties obtained from triangulations of reflexive polytopes, spanning the full range of Hodge numbers in the Kreuzer-Skarke database, namely $1\leq h^{1,1}\leq 491$.

Computing the instanton contributions from first principles by counting fermion zero modes is not currently feasible for an ensemble of this size.  To analyze the scalar potential generated by instantons, we instead used a model in which every prime toric divisor supports a Euclidean D3-brane superpotential term.  We performed a series of tests to assess how much the properties of the scalar potential in the ensemble would differ if the true set of contributing instantons were different from those in our model.
The results are as follows:
\begin{itemize}
\item Our conclusions are not appreciably changed by including arbitrarily many additional superpotential contributions from Euclidean D3-branes wrapping effective divisors that are linear combinations of prime toric divisors.
\item Our conclusions are not appreciably changed by including arbitrarily many K\"ahler potential contributions from four-cycles that are either piecewise-calibrated or else saturate the bound dictated by the axionic form of the Lattice Weak Gravity Conjecture.
\item Our conclusions change qualitatively if there are $\ll h^{1,1}$ independent four-cycles supporting superpotential terms, or if there are sufficiently many K\"ahler potential instantons manifesting strong recombination.  We argued that these situations are implausible, but we did not exclude them.
\end{itemize}

The K\"{a}hler metric and the scalar potential generated by instantons were evaluated at three locations in moduli space: at the tip, denoted $\mathcal{K}^V_1$, of the stretched K\"{a}hler cone $\widetilde{K}_V$ associated to an ambient toric variety $V$ containing the CY$_3$; at the tip, denoted  $\mathcal{K}^\cup_1$, of the stretched K\"{a}hler cone $\widetilde{\mathcal{K}_\cup}(V)$ associated to the union of all such ambient toric varieties; and at a location inside $\widetilde{K}_V$, which we denoted $\mathcal{K}^V_{25}$, that is an extremal point for realising the Standard Model gauge couplings (see Fig.~\ref{fig:cones}).  The numerical challenge of our study came from the vast hierarchies present in the instanton scales, which necessitated the use of extremely high precision calculations to achieve convergent results.

We computed the axion mass and quartic interaction spectrum at the origin of the axion field space (which, with zero $CP$ phase $\delta^a$, is a critical point of the potential) for all geometries in our ensemble. We found that statistically similar results apply including a random $CP$ phase, and at generic local minima and critical points.

The axion mass spectra displayed the following properties:
\begin{itemize}
\item There is a broad peak encompassing the most massive axions, the location of which moves to smaller masses as $h^{1,1}$ increases. The low mass tails of the distribution converge to a uniform shape at large $h^{1,1}$.
\item The fraction of axions that are effectively massless, with $m_a<H_0$, approaches a constant value as $h^{1,1}$ increases. At $\mathcal{K}^V_1$ this fraction is 75\%, while at $\mathcal{K}^V_{25}$ it is greater than 95\%.
\item At $\mathcal{K}^V_1$, the distribution of axions in the tail of the distribution appears to converge to nearly log-uniform across a wide range of phenomenologically interesting values.
\item At $\mathcal{K}^V_{25}$, the peak of the distribution covers astrophysically interesting regions from $10^{-5}\text{ eV}$ (at low $h^{1,1}$) to $10^{-10}\text{ eV}$ (at high $h^{1,1}$).
\end{itemize}

We computed the axion self-interaction tensor, $\lambda_{ijkl}$, and used this to define the perturbative decay constant $f_{\rm pert}$ from the diagonal elements $\lambda_{iiii}$. In all cases, the distribution of $f_{\rm pert}$ was found to be log-normal, with the mean decreasing with $h^{1,1}$. The distribution of $f_{\rm pert}$ is controlled by the K\"ahler metric eigenvalues and is correlated with the axion mass in a well-defined way. We found that axion self interactions can be either repulsive or attractive around generic minima and critical points, and flavour-changing interactions are weak. The trend of decreasing axion masses and decay constants with increasing $h^{1,1}$ is readily understood from the increase in volumes associated more topologically complex compactifications.

We then used our distributions to compute the exclusion probability of a CY$_3$ based on the observed masses and spins of astrophysical black holes. Our results are summarised as follows:
\begin{itemize}
\item For small $h^{1,1}$, the probability of exclusion increases as $h^{1,1}$ increases.
\item In all cases, the effect of self-interactions is significant in suppressing the constraints compared to the case where self-interactions are neglected.
\item Self-interactions cause the constraints derived from supermassive BHs to be irrelevant in determining exclusions for $h^{1,1}\gtrsim 15$.
\item At $\mathcal{K}^V_1$, the constraints peak around  $h^{1,1}\simeq 160$, where we exclude more than 50\% of CY$_3$'s at greater than 95\% C.L.
\item At $\mathcal{K}^V_{25}$, the constraints peak near $h^{1,1}=10$, where we exclude around 7\% of CY$_3$'s at greater than 95\% C.L. In this case, the fraction of excluded CY$_3$'s falls to almost zero for $h^{1,1}\gtrsim 100$.
\end{itemize}

In summary, this work represents the first extensive and systematic study of the properties of light axions in an ensemble of explicit string compactifications sampling the entire range of the Kreuzer-Skarke database. We have uncovered a rich phenomenology, and an intriguing level of regularity that is ripe for deeper explanations and further exploration. We have also shown how this vast landscape of possible models can be effectively cut down by appeal to astrophysical observations. Black hole superradiance provides some very clear boundaries on what are observationally acceptable compactifications, and our work paves the way to bringing yet more data to bear on the landscape.

\subsection{Outlook: axion couplings to the visible sector at large $h^{1,1}$}

We made use of significant recent advances in computing and analysing triangulations \cite{cytools}, and deployed an array of high-precision tools for computation and minimisation of the axion potential.
Even so, our most detailed comparative analysis across all three datasets focused on the range $1 \le h^{1,1} \le 100$.
However, we did include at least one CY$_3$ for every $h^{1,1}$ in the Kreuzer-Skarke database, both inside and at the tip of the stretched K\"ahler cone, and can roughly characterize the limits on theories with many axions. Here we offer some speculations about possible future constraints on such theories, taking $h^{1,1}=491$ as an illustration.

Our results for $f_{\mathrm{pert}}$ and $f_K$ for the CY$_3$ we studied with $h^{1,1}=491$ follow the general trend expected for smaller values of $h^{1,1}$, with a log-normal distribution, and mean $\langle f_\mathrm{pert}\rangle\approx \langle f_K\rangle\approx 10^{10}\text{ GeV}$ at $\mathcal{K}^V_1$, and $\langle f_{\mathrm{pert}}\rangle\approx \langle f_K\rangle\approx 10^{8}\text{ GeV}$ at $\mathcal{K}^V_{25}$. The low value of $\langle f_\mathrm{pert}\rangle$ makes superradiance almost irrelevant in this case.

The majority of the axions are expected to be massless. There are strong constraints on massless axions arising from visible sector couplings, in particular to electromagnetism, $\mathcal{L} = g_{a\gamma}\phi F_{\mu\nu}\tilde{F}^{\mu\nu}/4$, where $\phi$ is the canonically normalised axion field. The coupling is expected to be $g_{a\gamma}\sim \alpha_{\rm em}/(\pi f_K)$ with $\alpha_{\rm em}$ the electromagnetic fine structure constant. In the case of multiple axions and multiple $U(1)$ sectors, such a coupling is generated by kinetic mixing of the $U(1)$ fields, but will in general be suppressed by a mixing angle. The allowed value of $g_{a\gamma}$ is constrained by astrophysics.

One such constraint comes from the galactic supernova SN1987A.  Axions are produced inside the supernova by the Primakoff process with $E_a\sim T$ and they free stream out of the supernova. Subsequently, these axions convert into photons in the galactic magnetic field. If this process occurred, a gamma ray burst would have been observed coincident with SN1987A, and it was not. This excludes the existence of massless axions with $g_{a\gamma}\gtrsim 5\times 10^{-12}\text{ GeV}^{-1}$~\cite{Payez:2014xsa} implying $f_K\lesssim 4\times 10^{10}\text{ GeV}$. The gamma ray rate scales like $g_{a\gamma}^4$, thus if there are $N$ massless axions the constraint on $f_K$ increases with $N^{1/4}$.\footnote{The weaker bound from cooling of horizontal branch stars, $g_{a\gamma}\gtrsim 5\times 10^{-11}\text{ GeV}^{-1}$~\cite{Hoof:2018ieb}, excludes $f_K\lesssim 4\times 10^7\text{ GeV}$. This bound does not rely on reconversion of axions to photons, and so scales like $N^{1/2}$.}

A second astrophysical constraint of interest arises from the X-ray spectra of distant quasars. If there are massless axions with non-zero $g_{a\gamma}$, then the X-ray photons can convert into axions in the intergalactic magnetic field. Since there are a large number of different coherence volumes for the magnetic field, this leads a random modulation of the quasar spectrum caused by conversion and reconversion. Such modulations are not observed, and consequently exclude values of $g_{a\gamma}\gtrsim 6-8\times 10^{-13}\text{ GeV}^{-1}$~\cite{Day:2018ckv,Reynolds:2019uqt} excluding $f_K\lesssim 4\times 10^9\text{ GeV}$. This bound also increases like $N^{1/4}$.

The astrophysical bounds would appear to strongly exclude a typical triangulation at $h^{1,1}=491$ at $\mathcal{K}^V_{25}$, even allowing for some mixing angle suppression. At $\mathcal{K}^V_1$ there is some tension if a significant proportion of the $\mathcal{O}(400)$ massless axions have large visible sector couplings. On the other hand, the huge number of possible triangulations for the polytope with $h^{1,1}=491$, possibly as many as $10^{428}$ \cite{Demirtas:2020dbm}, means that at either point in moduli space there likely exist a vast number of outliers that are not excluded by these considerations. A systematic exploration of $h^{1,1}=491$ will require the use of more advanced numerical tools than we have developed here, possibly benefitting from machine learning. Our results suggest that such an exploration would be very fruitful if astrophysical constraints are taken into consideration. This conclusion holds in general: when superradiance bounds are weaker due to a small $f_{\rm pert}$ and stronger self-interactions, astrophysical constraints are likely to be stronger due to larger visible sector couplings controlled by $f_K$.

Recently, Minami and Komatsu have reported a measurement of \emph{birefringence} in the polarization spectra of the cosmic microwave background (CMB)~\cite{PhysRevLett.125.221301}. By using the frequency dependence of the polarized galactic foreground, Minami and Komatsu calibrated the absolute polarization angle in the \emph{Planck} CMB data. They could then measure the correlation between $E$ (divergence)-type and $B$ (curl)-type polarization, which is expected to be generated only by a rotation of polarization (i.e. birefringence) taking place between the surface of last scattering (redshift $z_{\rm LSS}=1100$, where $E$ type polarization is generated by Compton scattering) and the present day. The inferred angle of rotation is $\beta=0.35\pm 0.14^\circ$, i.e.~a possible detection of birefringence at greater than 2$\sigma$ confidence.
While caution is warranted in the absence of independent confirmation of this finding, it is nevertheless interesting to take the claim of \cite{PhysRevLett.125.221301} at face value, and ask whether the axions in our dataset could naturally produce such a signal.

Cosmic birefringence can be generated by the parity violating coupling between an axion and electromagnetism~\cite{Carroll:1989vb,Arvanitaki:2009fg,Pospelov:2008gg,Caldwell:2011pu,Zhao:2014yna,Fedderke:2019ajk,Fujita:2020ecn,Takahashi:2020tqv}. The large scale, uniform rotation (as compared to an angle-dependent rotation) inferred by Minami and Komatsu can be generated by the slow rolling of the axion field between the surface of last scattering and today, which corresponds to axion masses in the range
\begin{equation}
H_0<m_a<H(z_{\rm LSS}) \Rightarrow 10^{-33}\text{ eV}\lesssim m_a\lesssim 10^{-28}\text{ eV}\, .
\label{eqn:birefringence_window}
\end{equation}
The angle of rotation depends only on $\theta$, not the canonically normalised axion field, and thus it is independent of the axion decay constant. For a single axion with initial displacement $\theta_{\rm init}\in [-\pi,\pi]$ the rotation angle is $\beta \approx |(\theta_{\rm init}/\pi)|\times 0.21^\circ$ radians. Thus the necessary condition to explain the Minami and Komatsu result is to have just a single axion in the mass window \eqref{eqn:birefringence_window}, provided that it has $\mathcal{O}(1)$ mixing angle and initial displacement.

The amount of birefringence depends on the \emph{integrated} $\Delta \theta$, the total change in the coefficient of $F_{\mu\nu}\tilde{F}^{\mu\nu}$, and can thus increase if multiple axions couple to electromagnetism. If the displacements $\vec{\theta}_{\rm init}$ are aligned, then the increase scales with $N$, while if the displacements are random, it scales with $\sqrt{N}$ as a random walk. Thus a typical mixing angle suppression 0.1 can be overcome with ten aligned axions, or 100 random displacements.

We have computed the number of axions in the birefringence window for all $h^{1,1}$ in our database, and the result is displayed in Fig.~\ref{fig:windows}. The necessary condition for birefringence occurs on average at $h^{1,1}\approx 100$ for $\mathcal{K}^V_1$ and $\mathcal{K}^\cup_1$. The behaviour of $\mathcal{K}^\cup_1$ is, as expected, close to $\mathcal{K}^V_1$ for the covered values of $h^{1,1}$.  At larger $h^{1,1}$ we find an average of at most six axions in the birefringent window for $\mathcal{K}^V_1$.
The maximum number of axions in the birefringent window is 13 for $\mathcal{K}^V_1$, and five for $\mathcal{K}^V_{25}$. Thus all three points in moduli space could explain the observation \cite{PhysRevLett.125.221301}, and for aligned initial conditions can overcome moderate amounts of mixing angle suppression using multiple axions. The observation is explained most naturally close to the origin of moduli space and for large $h^{1,1}$.

In closing, we remark that advances in computing the effective theories resulting from string compactifications, combined with continuing progress on an array of experimental tests of axion theories, present the tantalizing prospect of probing swaths of the landscape.
To realize the full potential of this connection, it will be important to move beyond the comparatively well-understood realm of weakly-curved type IIB orientifolds.  Likewise, it will be worthwhile to better characterize the axion couplings in solutions of string theory with realistic visible sectors, so that constraints from BHSR can be used alongside constraints from non-gravitational interactions.

\vspace{1cm}
\emph{Acknowledgements:} We are grateful to Mona Dentler, Naomi Gendler, Sebastian Hoof, Manki Kim, Jakob Moritz, Andres Rios Tascon, and Fuminobu Takahashi.
The work of DJEM and VMM was supported by the Alexander von Humboldt Foundation and the German Federal Ministry of Education and Research.  The work of MD and LM was supported in part by NSF grant PHY-1719877. The work of CL was supported in part by the Alfred P. Sloan Foundation Grant No. G-2019-12504 and by DOE Grant DE-SC0013607. The work of MJS was supported by funding from the UK Science and Technology Facilities Council (STFC).  We made use of the open source packages \textsc{numpy}~\cite{numpy}, \textsc{matplotlib}~\cite{matplotlib}, \textsc{scipy}~\cite{scipy}, \textsc{seaborn}~\cite{waskom2020seaborn}, and \textsc{pandas}~\cite{mckinney-proc-scipy-2010,reback2020pandas}.

\FloatBarrier
\newpage
\appendix

\section{Changing Basis}\label{app:basis}

From the data of a triangulation we reach an axion effective theory that takes the form
\begin{align}\label{applagrangian}
\mathcal{L}= - \frac{1}{8 \pi^2} M_{\mathrm{pl}}^2K_{ij}g^{\mu\nu}\partial_\mu\theta^i\partial_\nu\theta^j + \sum_{a=1}^P \Lambda_a^4 \left\{1 -  \cos \biggl(\sum_i\mathcal{Q}^{a}_{i}\theta^i\biggr)\right\}\, ,
\end{align}
where from an a priori infinite sum over instantons, we have retained the finite number $P$ of terms for which $\Lambda_a \lesssim \Lambda_{\mathtt{num}}$ (see \S\ref{sec:analysis}), and in this appendix we set to zero the phases $\delta^{a}$ in each cosine, cf.~\eqref{eq:full_lagrangian}.  The Lagrangian \eqref{applagrangian} is expressed in a particular basis of integral homology defined by $h^{1,1}$ prime toric divisors $\{D_i\}$.  We call this the lattice basis.

Transforming to the basis where the indices on $\phi^i$ are contracted with the identity matrix, \ie
\begin{align}
  \frac{M_{\mathrm{Pl}}^2}{2}K_{ij}\partial^\mu\theta^i\partial_\mu\theta^j\to \frac 12\partial^\mu\phi^i\partial_\mu\phi_i\,,
\end{align}
gives,
\begin{align}
  \phi^i = M_{\mathrm{Pl}}\,\mathrm{diag}\left(f_K\right)^i{}_k U^k{}_{j}\theta^j
\end{align}
where $U^i{}_{j}$ is an orthogonal transformation matrix, which corresponds to the matrix of eigenvectors of $K_{ij}$, and $f_K = \sqrt{\mathrm{eig\,}(K_{ij})}$ in Planck units. The transformation takes the constant K\"{a}hler metric to its canonical form by means of a rotation and a rescaling.

This gives the canonical Lagrangian,
\begin{align}
  \mathcal{L} = -\frac{1}{2} \partial_\mu\phi^i\partial^\mu\phi_i - \tilde{V}(\phi)
\end{align}
where $\tilde{V}(\phi)$ is the transformed potential,
\begin{align}
  \tilde{V}(\phi) = \sum_{a=1}^{P}\Lambda^4_a\left( 1-\cos\left(Q^a_{i}\phi^i \right) \right)
\end{align}
with $Q^a_{j}\equiv U^{i}{}_j\mathcal{Q}^{a}_{i}$.

\noindent Having found the Hessian in the lattice basis, \ie
\begin{align}
  \mathcal{H}^\theta_{ij} = \frac{\partial^2 V(\theta)}{\partial\theta^i\partial\theta^j}
\end{align}
at a minimum of the potential, we require a basis transformation in order to extract the physical axion masses.  This results in the mass matrix $\mathcal{M}_{ij}$:
\begin{align}
  \begin{split}
    \mathcal{M}_{ij}:=\mathcal{H}^\phi_{ij} &= F^m{}_{p}U^{p}{}_{i}F^n{}_{q}U^q{}_{j}\mathcal{H}^\theta_{mn} \\
    &\Rightarrow \mathbf{F}\cdot\mathbf{U}^T\cdot\mathcal{H}^\theta\cdot\mathbf{U}\cdot\mathbf{F}
\end{split}
\end{align}
where $F^i{}_{j} = \frac{1}{M_{\mathrm{Pl}}}\cdot\left(\mathrm{diag}\left( 1/f_K \right)\right)^i{}_{j}$.
We can also transform the quartic coupling matrix, $\lambda_{ijkl}^\theta$ in a similar way,
\begin{align}
  \begin{split}
    \lambda_{ijkl}^\theta \equiv&\,\frac{\partial^4 V(\theta)}{\partial\theta^i\partial\theta^j\partial\theta^k\partial\theta^l}\to  \frac{\partial^4 \tilde{V}(\phi)}{\partial\phi^i\partial\phi^j\partial\phi^k\partial\phi^l}:=\lambda_{ijkl}^\phi\\
    \Rightarrow\lambda_{ijkl}^{\phi}=&\,F^p{}_{a} U^a{}_{i}\, F^q{}_{b}U^b{}_{j}\, F^r{}_{c}U^c{}_{k}\,F^s{}_{d} U^{d}{}_{l}\lambda_{pqrs}^\theta \\
    &\Rightarrow \mathbf{F}\cdot\mathbf{U}^T\cdot\left( \mathbf{F}\cdot\mathbf{U}^T\cdot\mathbf{\lambda}^\theta\cdot\mathbf{U}\cdot \mathbf{F}\right)\cdot\mathbf{U}\cdot\mathbf{F}
\end{split}
\end{align}
We then transform $\lambda_{ijkl}^{\phi}$ into the mass eigenbasis using $T^i{}_{j} = \mathrm{eigvec}\left( \mathcal{M} \right)^i{}_{j}$, which defines the mass eigenstates $\varphi^i=T^i{}_j\phi^j$ with masses $m_i$ given by the square roots of the eigenvalues of $\mathcal{M}_{ij}$ \ie
\begin{align}
  \mathbf{\lambda}^{\varphi}=\mathbf{T}^T\cdot\left(\mathbf{T}^T\cdot\mathbf{\lambda}^{\phi}\cdot\mathbf{T}\right)\cdot\mathbf{T}.
\end{align}
Finally, we define the perturbative decay constants, $f_{i,\mathrm{pert}}$, by
\begin{align}
f_{i,\mathrm{pert}} =\sqrt{\frac{m_i^2}{ |\lambda^\varphi_{iiii}|}}\, .
\end{align}
The definition of $f_{\rm pert}$ is by analogy to the case of a single axion with a cosine potential where:
\begin{equation}
V(\phi)=\Lambda^4[1-\cos(\phi/f)] = \frac{1}{2}\frac{\Lambda^4}{f^2}\phi^2-\frac{1}{4!}\frac{\Lambda^4}{f^4}\phi^4 +\mathcal{O}(\phi^6/f^6):= \frac{1}{2}m^2\phi^2-\frac{1}{4!}\frac{m^2}{f^2} \phi^4 +\mathcal{O}(\phi^6/f^6)\, .
\end{equation}

\section{Null Tests}\label{app:tests}

During the course of our analysis, we conducted many tests in order to cross-check our conclusions and ensure the robustness of our results.  We present these here for completeness.

\subsection{Statistics}\label{app:stats}
First we show various statistical tests that we conducted.

\begin{figure}[ht!]
\begin{center}
  \includegraphics[width=\textwidth]{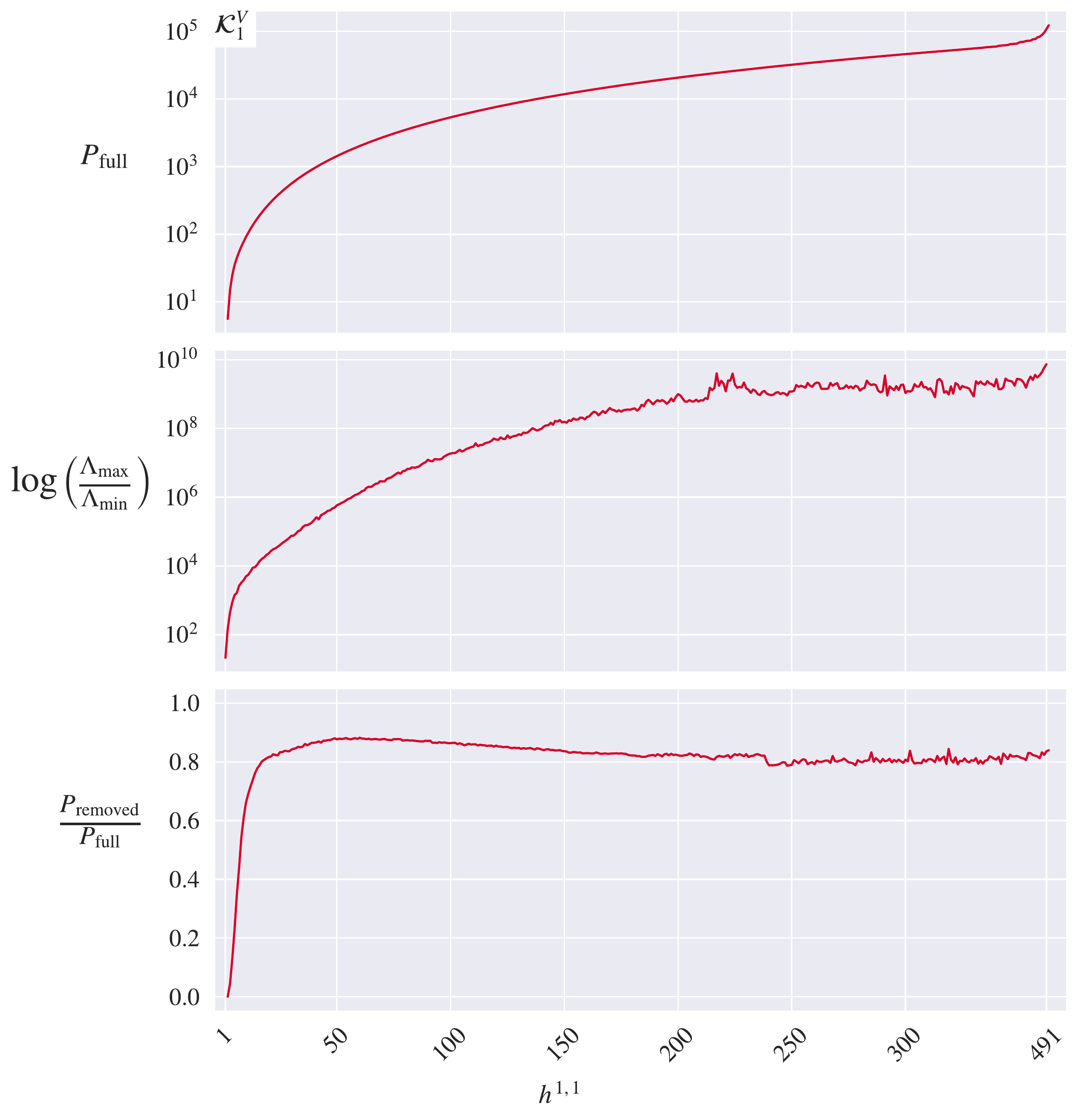}
  \caption{\textit{Top:} The total number of instanton contributions versus $h^{1,1}$. \textit{Middle:} The hierarchies of instanton scales versus $h^{1,1}$. \textit{Bottom:} The fraction of instanton contributions removed when applying $\Lambda_a\geq \Lambda_{\tt{num}} \simeq 10^{-53}$eV versus $h^{1,1}$.  These plots are fully explained in \S\ref{sec:Lambdas}.}
\label{fig:rankQ}
\end{center}
\end{figure}

\paragraph{Charge Matrix}  As discussed in \S\ref{sec:Lambdas}, applying a cutoff -- by using double precision tools -- reduces the number of contributing instantons, as shown in Fig. \ref{fig:rankQ}.  However, these represent contributions that are far below what is defined as effectively massless, \ie $m_a< H_0$, and so are rendered irrelevant for our analysis. In the same figure we also see the large disparity in cycle volumes and the quadratically-increasing number of terms in the potential with $h^{1,1}$. For further statistical relationships and details in the geometrical data, we refer readers to \cite{Demirtas:2018akl}.

As already pointed out in \cite{Demirtas:2018akl}, the additional cycle volumes included as $h^{1,1}$ increases get larger and larger, thus increasing the hierarchy within $\Lambda_a$.  In fact, for $h^{1,1}\gtrsim 100$,
\begin{align}
  \frac{\Lambda_{\mathrm{max}}}{\Lambda_{\mathrm{min}}}\gtrsim 10^6.
\end{align}

Retaining only those $\Lambda_a\geq \Lambda_{\tt{num}}$ results in a reduction in $P$ and therefore can affect the rank of $\mathcal{Q}$.  This results in changing $\mathcal{Q}$ shapes for the various triangulations, as shown in the bottom panel of Fig.~\ref{fig:rankQ}.  We note that $P_{\mathrm{removed}}/P_{\mathrm{full}}>0.5$ in more than 99\% of the geometries we studied, and $P_{\mathrm{removed}}/P_{\mathrm{full}}>0.75$ in more than 90\%.\footnote{However, in no CY$_3$ is $P_{\mathrm{removed}}/P_{\mathrm{full}}=1$.}

\begin{figure}[ht!]
\begin{center}
  \includegraphics[width=\textwidth]{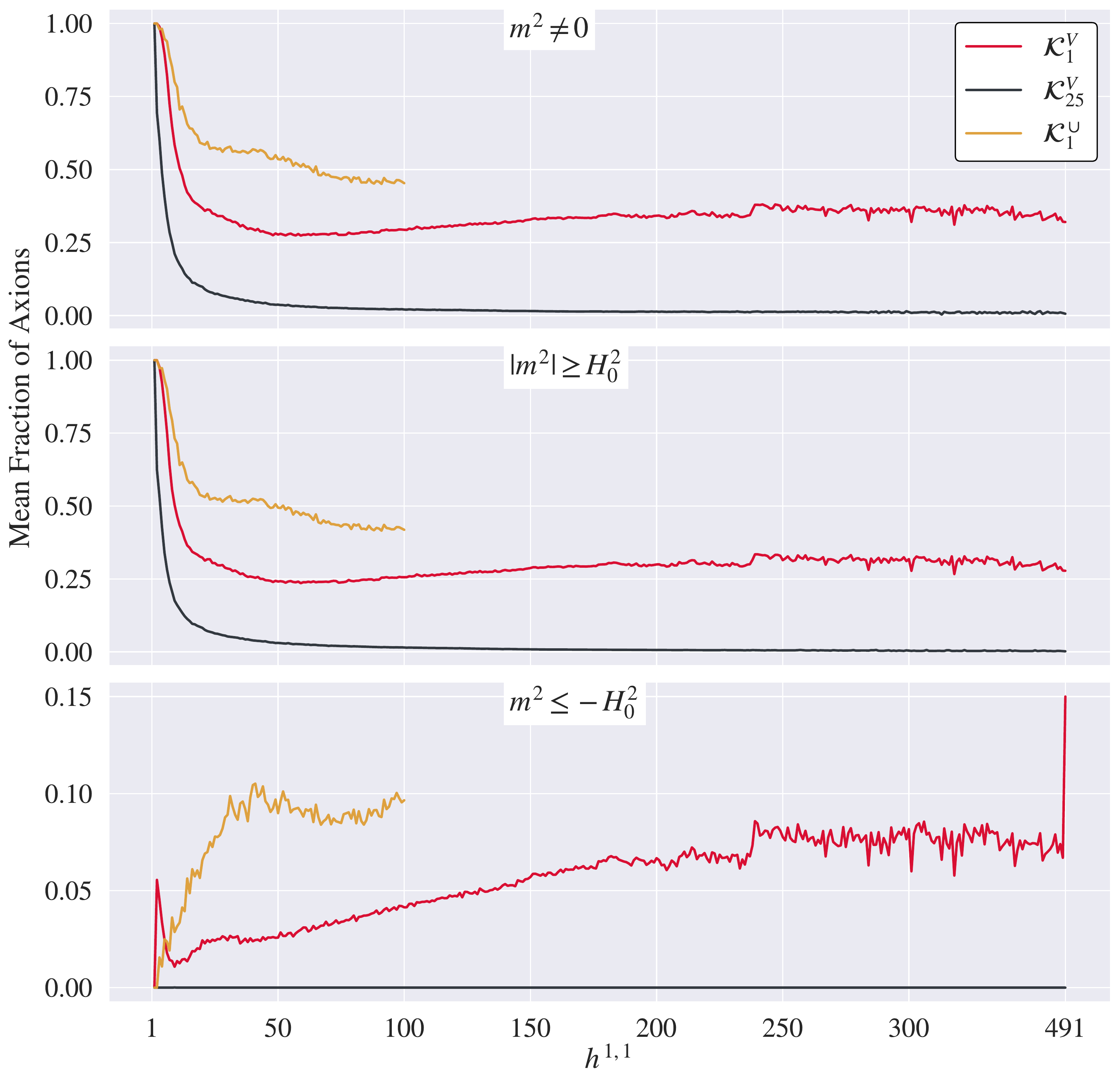}
  \caption{Proportion of massive/massless/tachyonic axions versus $h^{1,1}$}
\label{fig:tachyon}
\end{center}
\end{figure}

\paragraph{Axion Masses}  We then look at the proportions of Hessian eigenvalues that result in \textit{massless} axions, \ie below our working precision of $10^{-5000}$; \textit{effectively massless} axions, \ie below the Hubble scale, and also those that are \textit{negative}, \ie arising from the evaluation of the Hessian at a point other than a minimum\footnote{As shown in \S\ref{sec:axdata} and Figs. \ref{fig:local_mass20} \& \ref{fig:local_mass40}, the eigenvalue spectra are equivalent at different points in the field space, despite the change in sign.} in Fig. \ref{fig:tachyon}.

\begin{figure}[ht!]
\begin{center}
  \includegraphics[width=\textwidth]{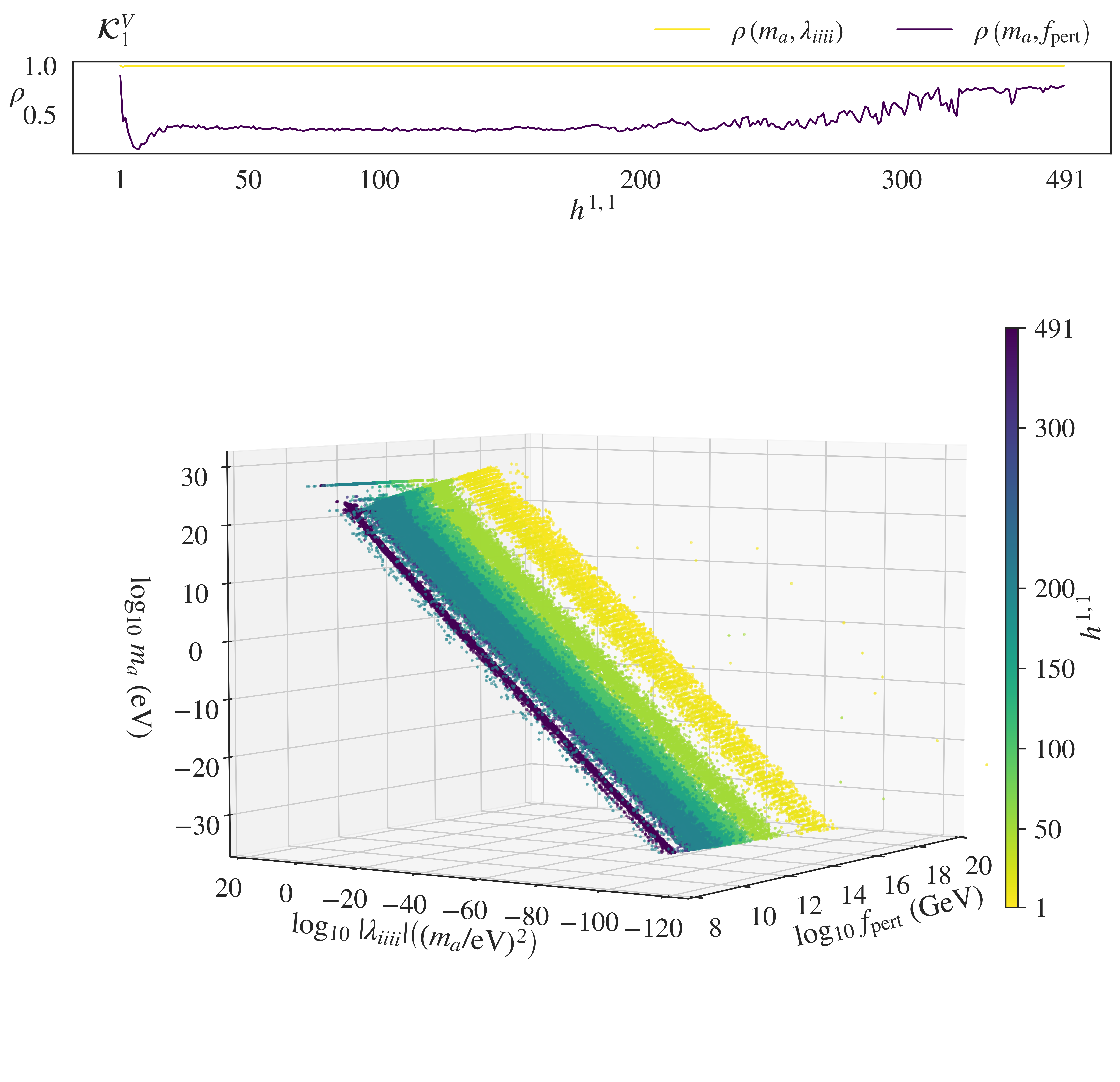}
  \caption{\textit{Top:} The Spearman rank coefficient, $\rho$, for comparisons of our physical quantities $m_a, \lambda_{iiii}, f_{\mathrm{pert}}$ from the $\mathcal{K}_1^V$ dataset.  As we see, $\rho\left( m_a,\lambda_{iiii} \right)\sim 1$.  The fact that $1-\rho$ is small but nonzero accounts for the spread in the $f_\mathrm{pert}$ distributions.
  \textit{Bottom:} 3d scatter-plot of $m_a,f_{\mathrm{pert}},$ and $\lambda_{iiii}$ for a sample of values of $h^{1,1}$.}
\label{fig:mfl_corr}
\end{center}
\end{figure}

\paragraph{Correlations}  Finally, in Fig. \ref{fig:mfl_corr}, we look at the Spearman rank correlations of the physical data, noting $\rho\left( m_a,\lambda_{iiii}\right)\simeq 1$ while $\rho\left( m_a,f_{\mathrm{pert}}  \right)\simeq \rho\left( \lambda_{iiii},f_{\mathrm{pert}} \right)\simeq 0.5$.

\begin{figure}[ht!]
\begin{center}
  \includegraphics[width=\textwidth]{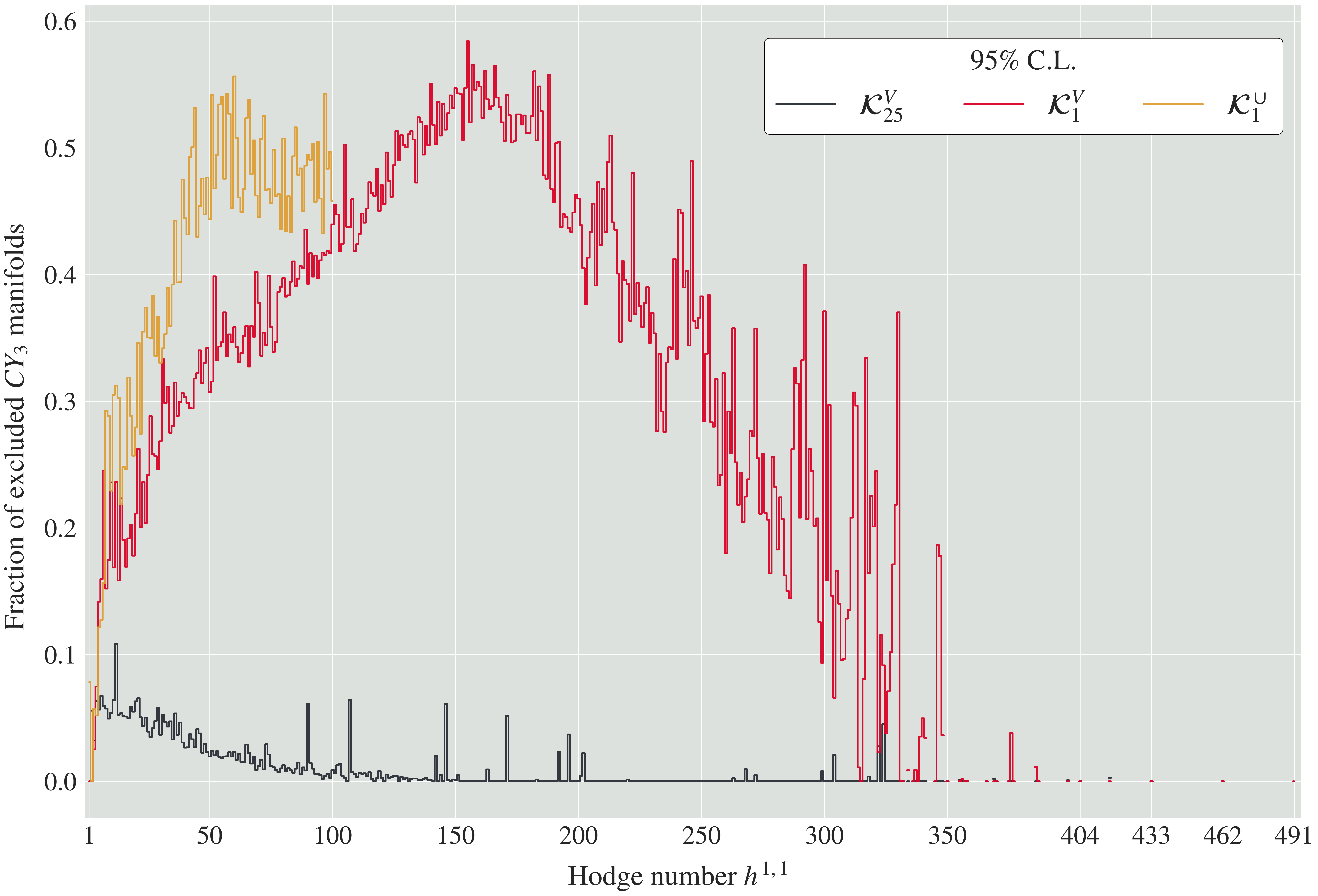}
  \caption{Fractions of geometries excluded by BHSR after resampling the masses and decay constants.}
\label{fig:10kredraws}
\end{center}
\end{figure}

\begin{figure}[ht!]
\begin{center}
  \includegraphics[width=\textwidth]{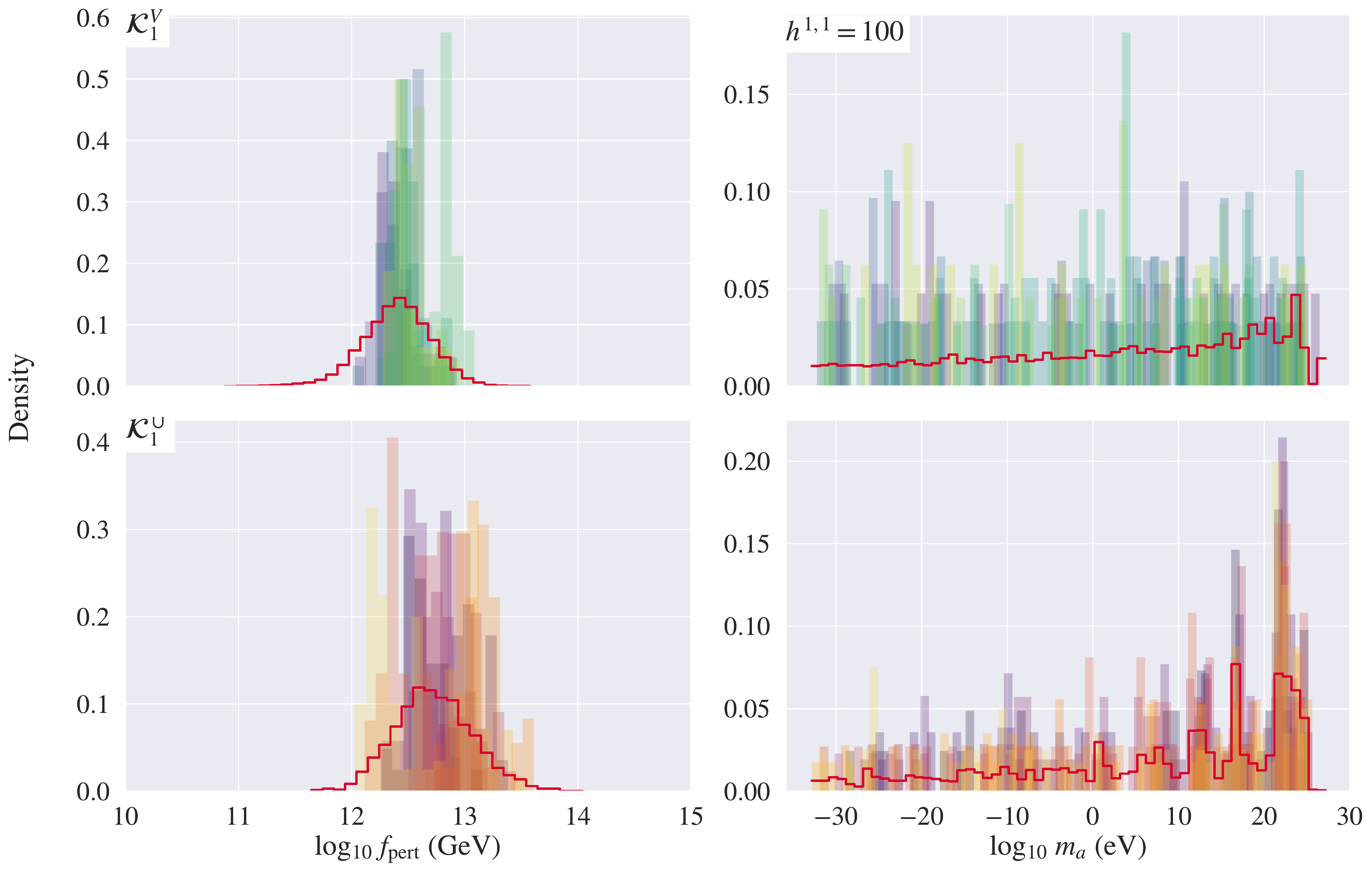}
  \caption{A sample of 10 individual distributions (\textit{coloured bars}) of $m_a$ and $f_{\mathrm{pert}}$ per CY$_3$ and their average distribution (\textit{red}) at $h^{1,1}=100$.}
\label{fig:sep_geoms}
\end{center}
\end{figure}

\paragraph{Resampling}  In Fig. \ref{fig:10kredraws} we calculate the fractions of geometries excluded by BHSR using the overall distribution of masses and decay constants, by concatenating the distributions of each geometry per $h^{1,1}$ and randomly drawing $\left( m_a,f_{\mathrm{pert}} \right)$ pairs. Comparing to Fig. \ref{fig:moneyplot} we see that using the overall distributions does not change the BHSR exclusions by an appreciable amount.  For illustration, we show a sample of the separate distributions of $m_a$ and $f_{\mathrm{pert}}$ per CY$_3$ at $h^{1,1}=100$ in Fig. \ref{fig:sep_geoms} and compare it to the distribution of their combination.

\subsection{Minimisation}\label{sec:minima}

The Hessian matrix and the axion self-interactions vary from point to point in axion field space.  It is therefore critical that we carry out our analysis at points where the axion vevs evolve slowly, or not at all, on cosmological timescales.
We therefore seek to identify \emph{local minima} of the potential, and to evaluate the axion couplings there.  In fact, \emph{critical points} in which all unstable directions have cosmologically-long lifetimes, \ie in which even the most tachyonic axions have $m_a^2 \gtrsim -H_0^2$, are equally suitable for this purpose.

As detailed in \S\ref{sec:KSstats}, there are vast hierarchies in the instanton mass scales, $\Lambda_a$, which inhibit many optimization routines.  Below we detail the tools and methodology we used to find minima and critical points in a subset of our geometries -- namely all those in our datasets up to $h^{1,1}=100$ -- and detail the various tests we carried out in order to ensure our obtained mass spectra and decay constants are robust. Having tested a variety of algorithms in the optimisation suite \texttt{pygmo2}, we found that \textit{differential evolution} was both the most effective and computationally efficient.

This optimization method stochastically populates a function and, at each pass, mutates each possible solution with other possible solutions to create a trial solution.  Should the trial solution be an improvement on others, it is added to the population of possible solutions, and otherwise it is discarded.  This process continues until an optimised solution is found or the maximum number of iterations is reached.  More precisely, we used the \textit{self-adapting} differential evolution (\texttt{sade}) algorithm in \texttt{pygmo2}, which provides a variety of strategies to build the trial solution. We chose \texttt{rand-to-best-and-current/2/bin} for our analysis.\footnote{We ran benchmark tests and found that \texttt{randtobest1bin} and \texttt{best1bin} were the most appropriate strategies for our problem.  However, natively these do not take advantage of the \textit{self-adaptive} qualities of the algorithm and thus, we chose \texttt{rand-to-best-and-current/2/bin}, which allows for some adaptation in mutation coefficients and crossover rate -- defined below.} For illustrative purposes, \texttt{best/1/bin} is the default strategy in \texttt{scipy}.  Here, two members of the population, $r_0$ and $r_1$, are randomly chosen and their difference is used to \textit{mutate}\footnote{With \textit{weighting factor}, or mutation coefficient, $F$.} the \texttt{best 1}, $t_0$,
\begin{align}\label{eqn:testvector}
  t^\prime = t_0 + F\left( r_0-r_1 \right).
\end{align}
The trial vector is then built and filled sequentially with parameters from $t^\prime$ or the original candidate -- this choice is made by generating a random number in $\left[0,1\right)$ in a \texttt{bin}omial distribution.  If this random number is lower than the \textit{crossover rate} ($CR$) then the parameter is taken from $t^\prime$, else it is taken from the original solution.\footnote{$CR$ and $F$ are input parameters that are adapted using the algorithm outlined in \cite{sade-pygmo}.  These can have a profound effect on the minimum found and thus it is often useful to automate their selection.}  The trial vector is then assessed for fitness: if it is an improvement on vectors already in the population, it is kept and the original candidate solution is discarded, otherwise $t^\prime$ is discarded and a new trial vector is built.  This procedure is conducted over a set domain and repeated until a solution is found.  We used $\mathcal{R}_{\mathrm{bound}}:= \pi\sqrt{h^{1,1}}\,\mathrm{max}(f_{K})$ as the diameter of the search domain for the algorithm.  We refer the reader to \cite{Demirtas:2018akl} for further details.

\begin{figure}[ht!]
\begin{center}
  \includegraphics[width=\textwidth]{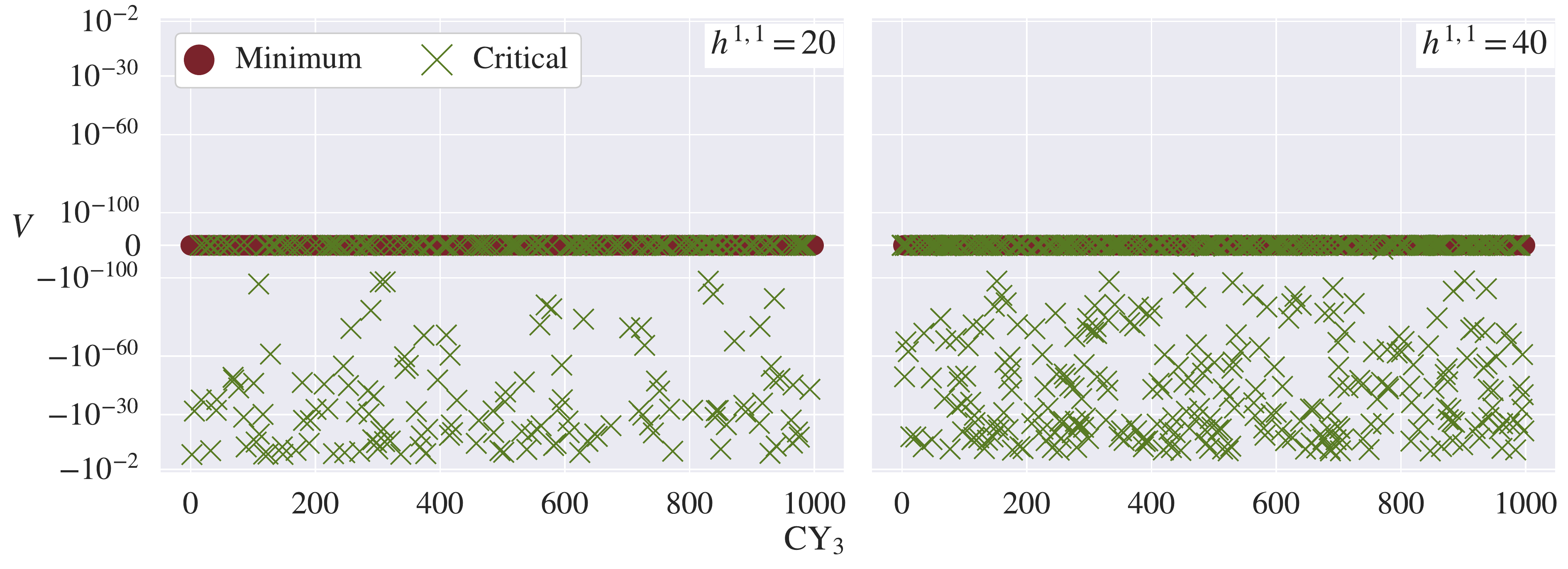}
  \caption{Values of the potential at the points found by the differential evolution algorithm, \ie $\theta=\theta_{\mathrm{min}}$, for the geometries at $h^{1,1}=20$ \& $h^{1,1}=40$.  Minima are defined as points where the entire eigenvalue spectrum of the Hessian matrix is positive semi-definite.}
\label{fig:minima2040}
\end{center}
\end{figure}

Note that we were unable to guarantee the \textit{global} minimum was found -- see Fig. \ref{fig:minima2040} -- due to the noisy nature of the potentials we studied.  In fact, in many cases the algorithm found a critical point on completion, rather than a minimum.

We therefore repeated the minima search for a random sample of triangulations with different initialisation points, and also varied the search domain in this sample.  We sought to verify that, even if the minimum found was not the global one, the physics would remain essentially unchanged --  \ie that at all minima in the potential, the masses and quartic self-couplings do not vary significantly. This comes about due to the vast hierarchies in $\Lambda_a$: in general, physically relevant masses, $m_a\geq H_0$, take very few of the elements of $\Lambda_a$ as dominant contributions and these would determine the overall distributions of physical quantities of interest.

\begin{figure}[ht!]
\begin{center}
  \includegraphics[width=\textwidth]{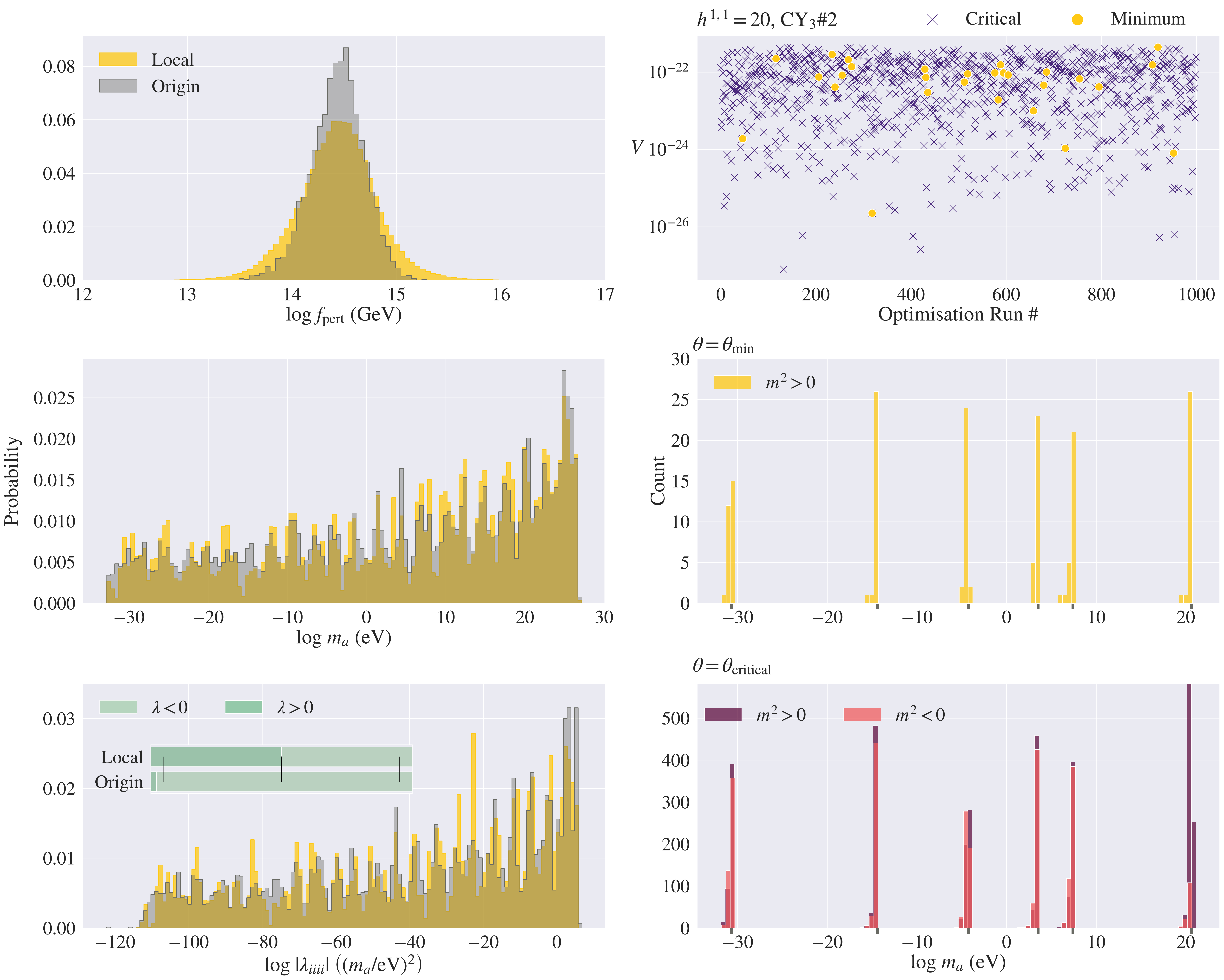}
  \caption{\textit{Left:} Mass, decay constant and quartic self-coupling distributions of 1000 local minima/critical points for a particular geometry at $h^{1,1}=20$. Here we compare the distributions obtained at the local critical points and the origin.  In the bottom plot, we show the fraction of positive/negative $\lambda_{iiii}$ in each distribution.  \textit{Right:} In the top plot we show the value of $V$ at the different critical points.  A minimum is defined as a point at which the evaluated Hessian eigenvalues are positive semi-definite.  We highlight that, due to the early stopping criteria on the optimiser, there are no $V<0$.  The next two plots show that, interestingly, the magnitude of the masses are unchanged at critical points vs minima \textit{despite} the change in sign, where the grey markers represent the masses found at the origin.}
\label{fig:local_mass20}
\end{center}
\end{figure}

\begin{figure}[ht!]
\begin{center}
  \includegraphics[width=\textwidth]{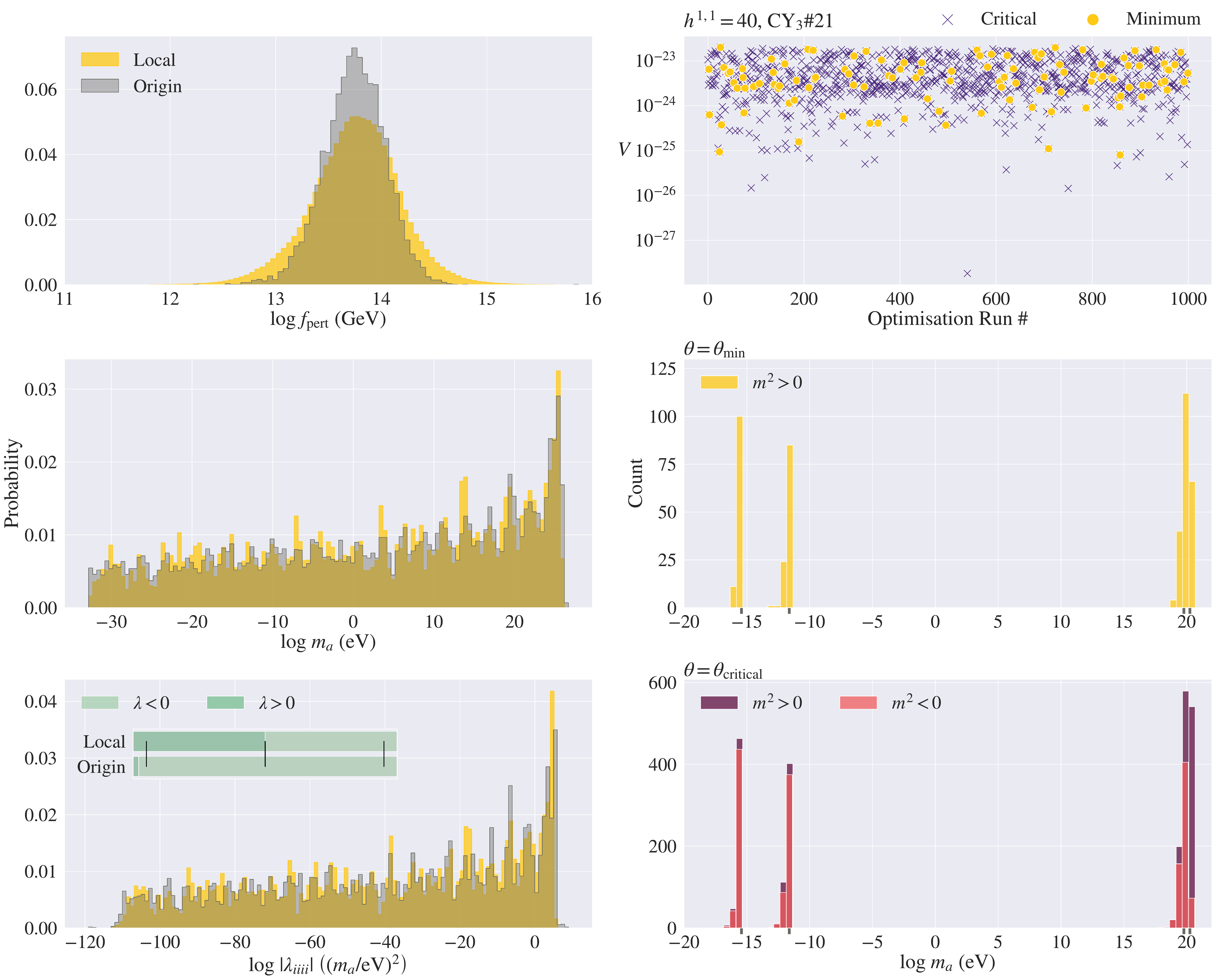}
  \caption{\textit{Left:} Mass, decay constant and quartic self-coupling distributions of 1000 local minima/critical points for a particular geometry at $h^{1,1}=40$. Here we compare the distributions obtained at the local critical points and the origin.  In the bottom plot, we show the fraction of positive/negative $\lambda_{iiii}$ in each distribution.  \textit{Right:} In the top plot we show the value of $V$ at the different critical points.  A minimum is defined as a point at which the evaluated Hessian eigenvalues are positive semi-definite.  We highlight that, due to the early stopping criteria on the optimiser, there are no $V<0$.   The next two plots show that, interestingly, the magnitude of the masses are unchanged at critical points vs minima \textit{despite} the change in sign, where the grey markers represent the masses found at the origin.}
\label{fig:local_mass40}
\end{center}
\end{figure}

We verified this hypothesis by finding multiple critical points for 100 different geometries at $h^{1,1}=20$ and $40$. The results are shown in Figs.~\ref{fig:local_mass20} and \ref{fig:local_mass40}. As can be seen in the figures, upon calculation of the Hessian eigenvalues at the various local critical points, we found little -- $\log\Delta \mathcal{H} \ll 1$ -- or no difference.\footnote{Some Hessian eigenvalues were found to be negative, however, both critical points and minima resulted in equivalent eigenvalue spectra \textit{by magnitude}.} Additionally, \textit{global} optimization was conducted for all geometries $1\leq h^{1,1}\leq 100$ and compared with the eigenvalue spectra at the origin\footnote{In fact, in the majority of cases, the origin was found to be the global mininum.}.  The overall distributions were found to be equivalent\footnote{A Kolmogorov-Smirnov 2-sample test gave $p\gtrsim 0.5$.} and so for $h^{1,1}>100$ optimization was not conducted, with evaluation of the Hessian and self-interactions at the origin.

\subsection{Phases}\label{app:phases}
\begin{figure}[ht!]
\begin{center}
  \includegraphics[width=\textwidth]{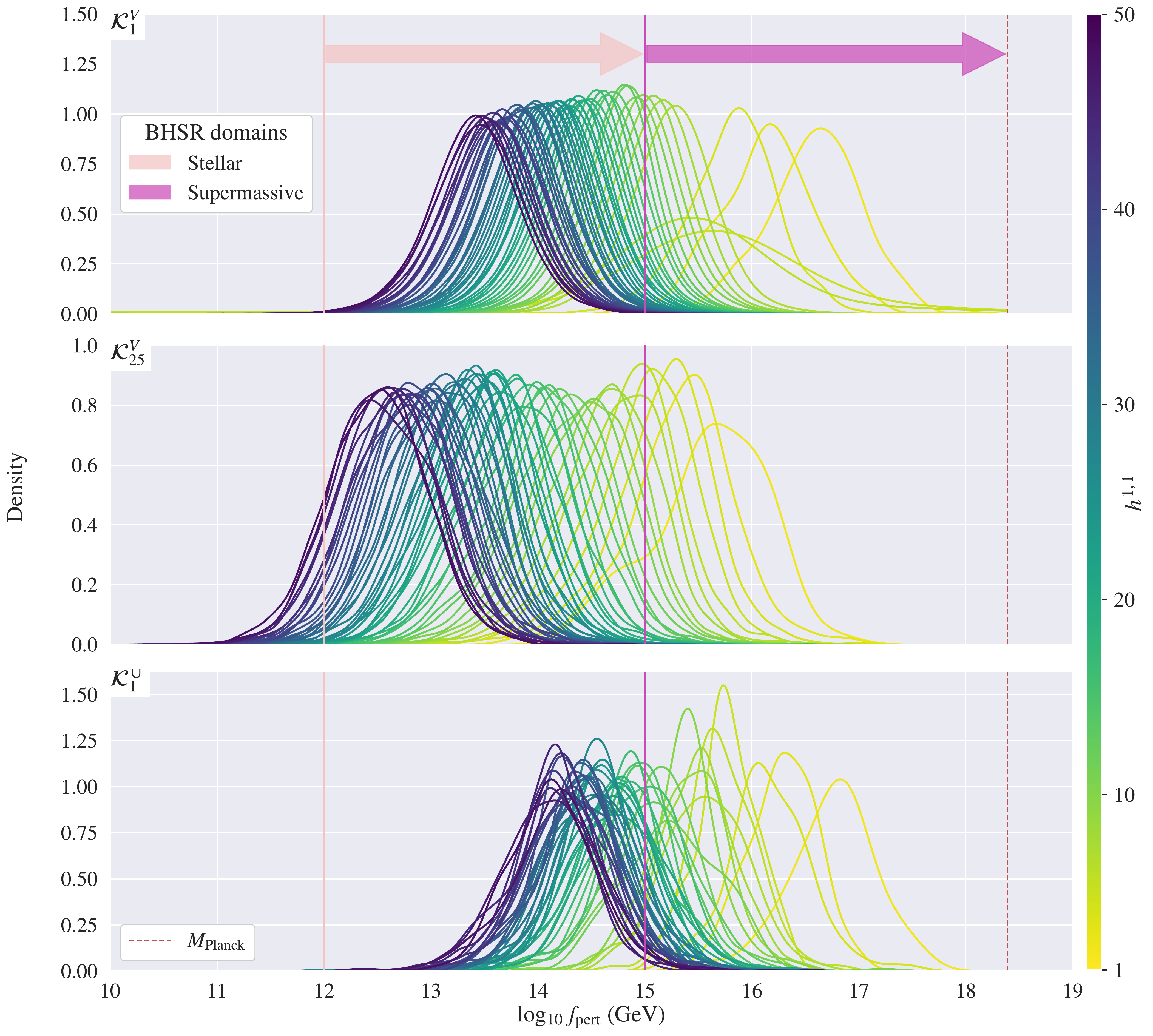}
  \caption{$f_\mathrm{pert}$ distributions when introducing a random phase, $\delta^a$, into the potential.  These reproduce the $f_{\mathrm{pert}}$ distributions at the origin for each dataset.}
\label{fig:fpert_phases}
\end{center}
\end{figure}

\begin{figure}[ht!]
\begin{center}
  \includegraphics[width=\textwidth]{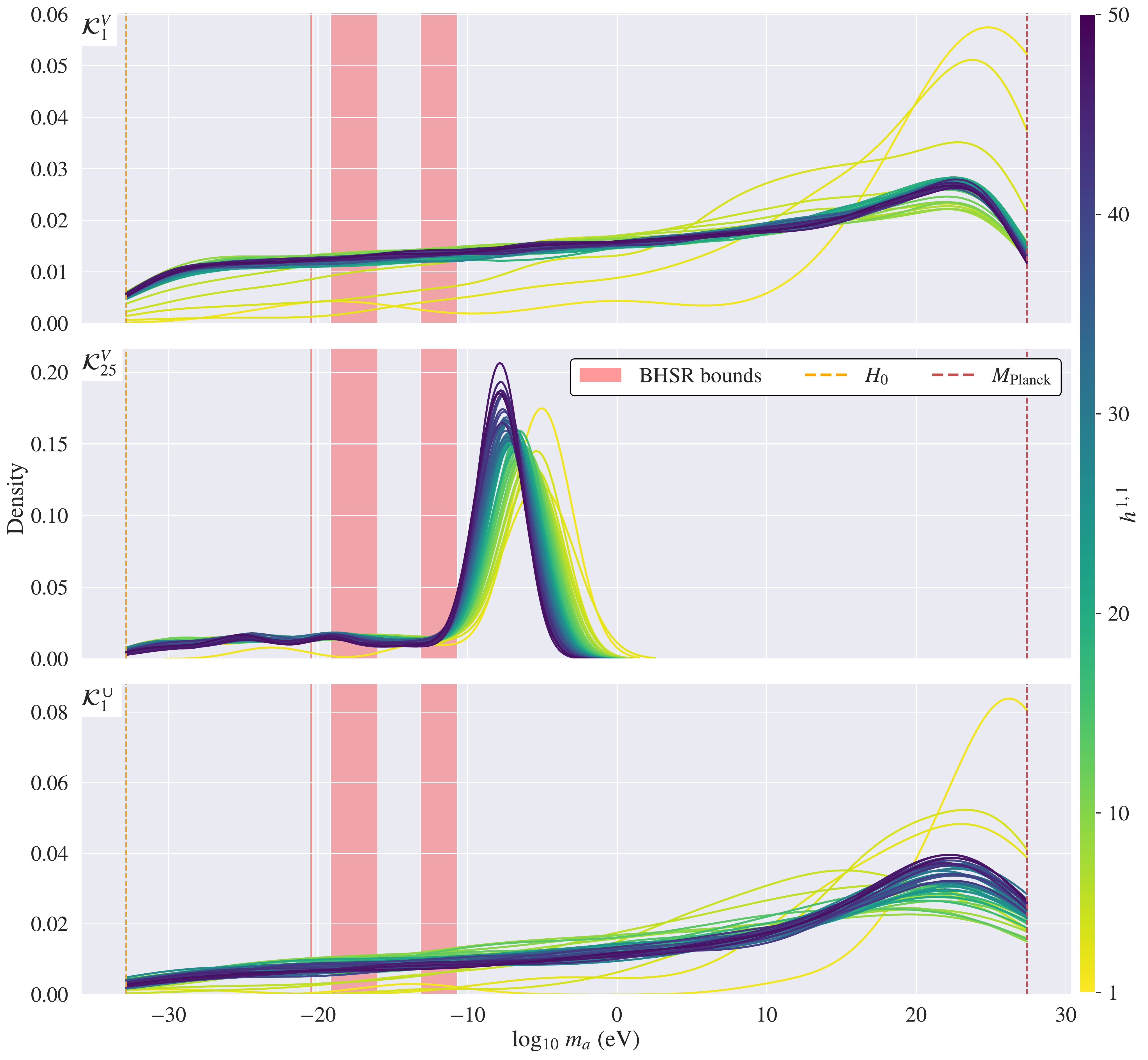}
  \caption{Distributions of masses when introducing a random phase, $\delta^a$, into the potential.  These reproduce the mass distributions at the origin for each dataset.}
\label{fig:mass_phases}
\end{center}
\end{figure}

\begin{figure}[ht!]
\begin{center}
  \includegraphics[width=\textwidth]{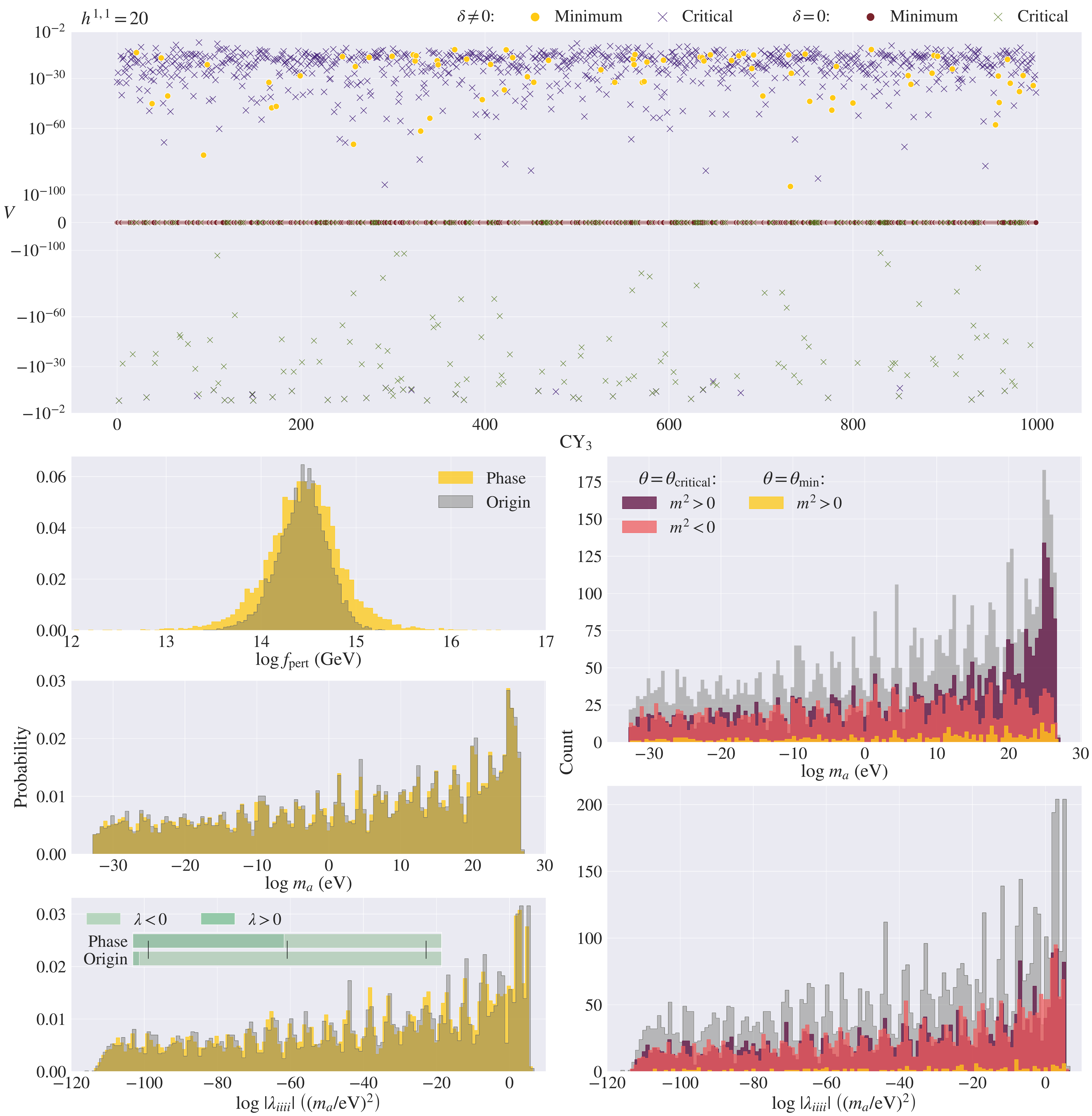}
  \caption{\textit{Top:}  The difference in $V$ at $\theta_{\mathrm{min}}^{\mathrm{phase}}$ and  $\theta_{\mathrm{min}}^{\mathrm{global}}$, where $\theta_{\mathrm{min}}$ indicates the field space coordinate of a minimum \textit{or} critical point. \textit{Left:} Mass, decay constant, and quartic self-interaction distributions for the 1000 geometries sampled at $h^{1,1}=20$ with an additional phase, $\delta^a$, compared with those at the origin. \textit{Right:} We show the distributions of masses and quartic self-couplings, indicating which are evaluated at minima, \ie with $\mathrm{eig}(H)>0$, and those evaluated at critical points, \ie with $\mathrm{eig}(H)\in\mathbb{R}$.  We note that the magnitude of the values is unchanged at critical points vs minima \textit{despite} the change in sign.}
\label{fig:phase_mass20}
\end{center}
\end{figure}

\begin{figure}[ht!]
\begin{center}
  \includegraphics[width=\textwidth]{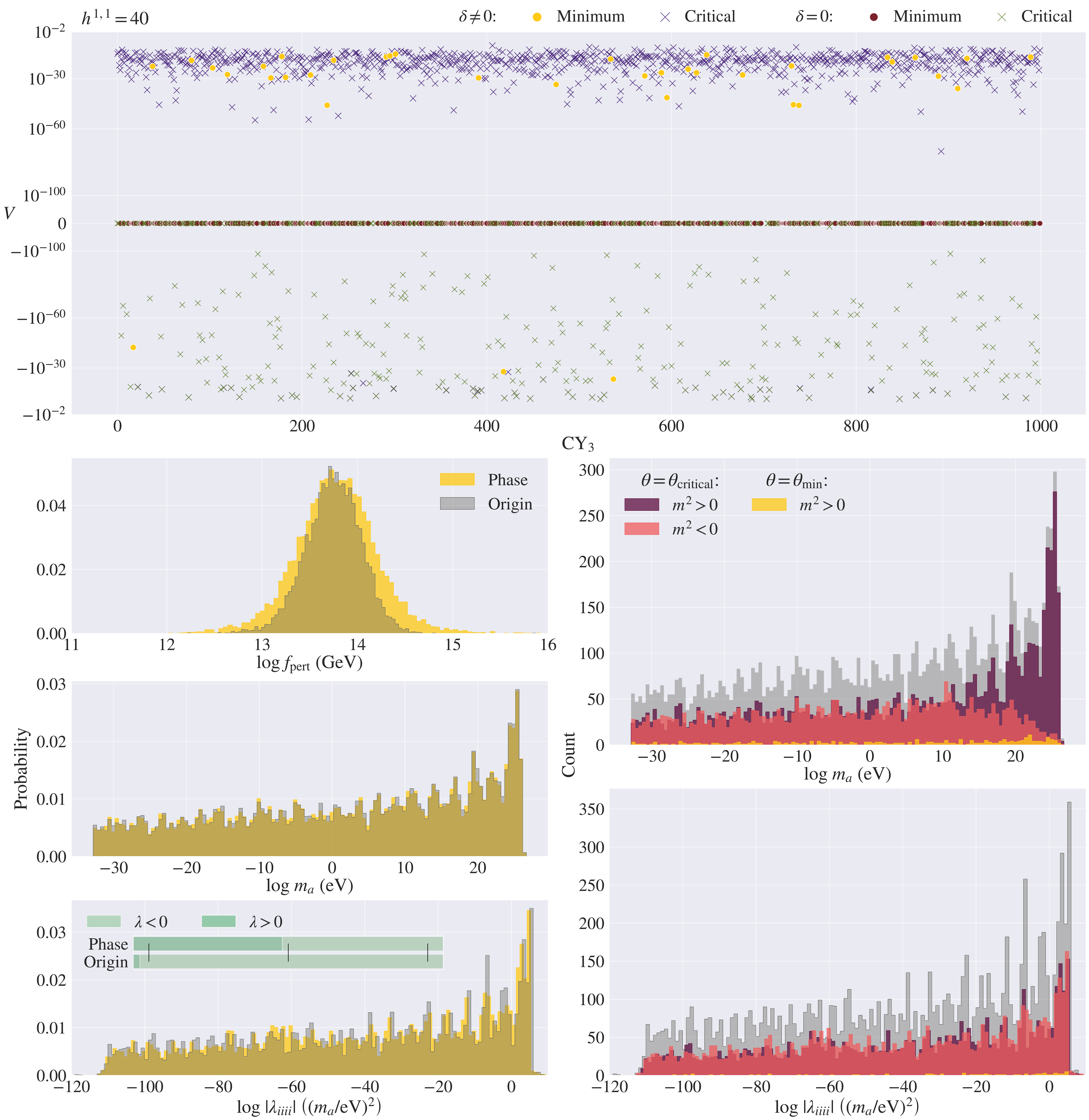}
  \caption{\textit{Top:}  The difference in $V$ at $\theta_{\mathrm{min}}^{\mathrm{phase}}$ and  $\theta_{\mathrm{min}}^{\mathrm{global}}$, where $\theta_{\mathrm{min}}$ indicates the field space coordinate of a minimum \textit{or} critical point. \textit{Left:} Mass, decay constant, and quartic self-interaction distributions for the 1000 geometries sampled at $h^{1,1}=40$ with an additional phase, $\delta^a$, compared with those at the origin. \textit{Right:} We show the distributions of masses and quartic self-couplings, indicating which are evaluated at minima, \ie with $\mathrm{eig}(H)>0$, and those evaluated at critical points, \ie with $\mathrm{eig}(H)\in\mathbb{R}$.  We note that the magnitude of the largest eigenvalues is unchanged at critical points vs minima \textit{despite} the change in sign.}

\label{fig:phase_mass40}
\end{center}
\end{figure}

In the main part of our analysis, we used $\delta^a=0$ in eq. \eqref{eq:full_lagrangian}.  Using current geometric tools, this factor is currently inaccessible.  However, it is reasonable to assume that in some cases it will take approximately uniformly distributed values $\delta^a\in\left[0,2\pi\right)$.  We thus conducted a smaller analysis to investigate the physical consequences of such a random phase.  We show the distributions of decay constants and masses, for all geometries in our datasets up to $h^{1,1}=50$, in Figs. \ref{fig:fpert_phases} and \ref{fig:mass_phases}, respectively.  We found that the phase had a negligible effect on the overall mass distributions, though it did give a slight broadening of the distribution of decay constants.  In some cases the similarities can be expected to arise from the phases simply shifting minima along flat directions.  The minor differences in masses and quartic self-interactions -- shown in Figs.~\ref{fig:phase_mass20} and \ref{fig:phase_mass40} -- then result in a slightly weaker correlation coefficient, and thus a broader decay constant distribution.

Interestingly, we see an effect on the optimization: the addition of the phase lifts most minima and critical points to $V>0$. Some, however, are shifted to more negative values. The minima with $V>0$ should not be interpreted as string theoretic constructions of de Sitter vacua, as we have not explicitly stabilized the K\"ahler moduli, and have added an arbitrary constant to the potential.

%\subsection{Tests}

\begin{figure}[ht!]
\begin{center}
  \includegraphics[width=\textwidth]{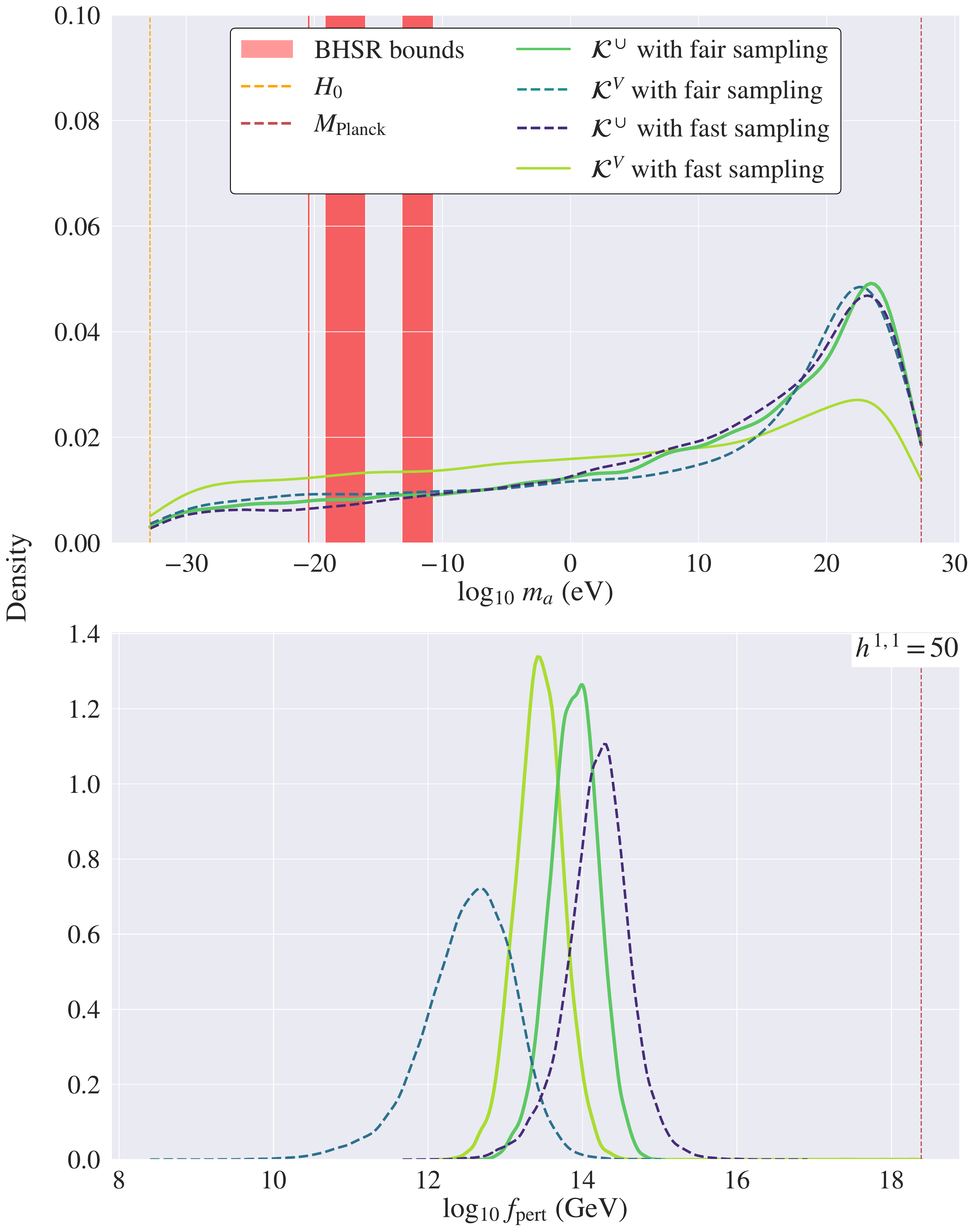}
  \caption{A comparison of mass and decay constant distributions varying the construction of the datasets as detailed in \S\ref{sec:KS}.  The solid lines represent data used in our analysis.}
\label{fig:stringtests}
\end{center}
\end{figure}

\FloatBarrier
\newpage
\bibliography{KSAxiverseBHSR}
\end{document}